\documentclass[twocolumn,aps,prl,superscriptaddress]{revtex4-1}

\usepackage{color,amsmath,amssymb,epsfig,graphicx}
\usepackage{bm}
\usepackage{mathptmx, courier, pifont}
\usepackage[scaled=0.92]{helvet}
\usepackage[T1]{fontenc}
\usepackage{textcomp}
\usepackage[colorlinks=true,linkcolor=blue,filecolor=blue,urlcolor=blue,citecolor=blue]{hyperref}

\begin{document}
\author{S.~Frauendorf}
\email{sfrauend@nd.edu}
\affiliation{Department of Physics and Astronomy, University of Notre Dame, Notre Dame, Indiana 46556,  USA}
\title{ Wobbling motion in triaxial nuclei }
\date{\today}
\begin{abstract}
The experimental evidence for the collective wobbling motion of triaxial nuclei is reviewed. The classification into transverse and longitudinal in the presence
of  quasiparticle excitations is discussed. The description  by means of the quasiparticle+triaxial rotor model is discussed in detail.
 The structure of the states is analyzed using the spin-coherent-state and spin-squeezed-state representations of the reduced density matrices of the total and particle angular
 momenta, which distill   the corresponding  classical precessional motions. Various approximate solutions of the quasiparticle+triaxial rotor model are evaluated.
 The microscopic studies of wobbling in the small-amplitude random phase approximation are discussed.  Selected studies of wobbling by means
 of the triaxial projected shell model are presented, which focus on how this microscopic approach removes certain deficiencies of the semi-microscopic quasiparticle+triaxial rotor model.

\end{abstract}
\maketitle

\section{Introduction}
At the  Conference "Chirality and Wobbling in Atomic Nuclei" (CWAN'23) in Huizhou, China, July 10-14, 2023 I presented a review talk on "Wobbling motion in triaxial nuclei".
This article presents   my perspective on the  wobbling mode in triaxial nuclei, which has been an active field of both experimental and 
theoretical research in the recent couple of decades. The article is organized as follows.

 Section "SEMICLASSICAL ANALYSIS OF THE EIGENSTATES OF THE TRIAXIAL ROTOR" discusses the triaxial rotor (TR) model. 
  Based on the correspondence with the rotation of the classical TR, the TR eigenstates are classified as "wobbling"  and "flip" states. 
The appropriateness  of this classification is justified in the subsections: 
"Restrictions by symmetry", 
"Correspondence with the classical orbits", 
"Spin Coherent States" (a tool to distill the proxies of the orbits from the quantal states), 
"Discussion of the Triaxial Rotor Dynamics", 
"Alternative terminology".
The subsection "Experimental evidence for triaxial rotor states in even-even- nuclei" discusses the few examples found so far.
 
Section "PARTICLES COUPLED TO THE TRIAXIAL ROTOR" covers the work on wobbling excitations in the framework
of "The particle plus triaxial rotor model" (PTR), introduced in the subsection by this name. "Wobbling in triaxial strongly deformed nuclei" is devoted
to the discovery of the wobbling mode in odd-A nuclei and the inconsistencies  of the  pertaining PTR interpretation. These are removed  in "Transverse and longitudinal wobbling"
with classification of the PTR eigenstates as "transverse wobbling" (TW) and
"longitudinal wobbling"  (LW) based on the topology of the precession cones of the corresponding classical the angular momentum  vectors
with respect to the principal axes. 
The subsection "Transverse wobbling in triaxial normal deformed nuclei" reviews the work on this type of wobbling.
Three selected nuclei are discussed in detail, where
the controversy about the TW interpretation is addressed. 
The instability of the TW mode with increasing angular momentum   and the transition to LW via the  flip regime are discussed in 
"Transition from transverse to longitudinal wobbling". 
 An alternative to illustrate the dynamics of the particle rotor system is introduced in "Spin squeezed states".
 The evidence for LW  is reviewed in subsection "Longitudinal wobbling", where one case is exposed in  more detail.
Subsection "Approximate solutions of the PTR model" presents my perspective on the various approximate solutions of the PTR model.
The whole section is concluded with the summary "Virtues and limits of the PTR".

 Section 
"MICROSCOPIC DESCRIPTIONS OF WOBBLING" reviews the studies of the wobbling mode by means  microscopic  methods based on the 
"The pairing-plus-quadrupole-quadrupole Hamiltonian", some relevant details of which are given in this subsection.
The small-amplitude approach (RPA) is discussed in subsection "Random Phase Approximation".
Subsection "Triaxial projected shell model" gives a short introduction to the microscopic description of wobbling in the frame work of the TPSM.
 It presents its application to two nuclei. Additional TPSM results are presented in the examples of section "PARTICLES COUPLED TO THE TRIAXIAL ROTOR".   
 
  Section "SOFT CORE" addresses  wobbling in nuclei with a soft triaxial shape.

\section{Semiclassical analysis of the eigenstates of the  triaxial rotor}\label{s:TR}

 The triaxial rotor Hamiltonian is given by 
\begin{equation}\label{eq:HTR}
 H_{\textrm{TR}}=\sum_{i=1,2,3}A_i\hat{J}_i^2,~~~A_i=\frac{1}{2{\mathcal{J}_i}},
\end{equation}
where   $\hat{J}_i$ are the three angular momentum projections onto the principal axes of the triaxial density distribution.  
The model parameters are the three moments of inertias (MoI's) $\mathcal{J}_i$. In the case of molecules,  they  take the
classical values of an ensemble of point masses located at their positions in the molecular skeleton. 
Usually all eigenstates of $ H_{\textrm{TR}}$ are observed because the atoms are non-identical.

The eigenstates are determined by the asymmetry parameter of the MoI's
\begin{equation}\label{eq:asym}
\kappa=\frac{2A_1-A_2-A_3}{A_2-A_3},~~ A_2\leq A_1\leq A_3.
\end{equation}
The inertia  ellipsoid changes from  a prolate spheroid  for $\kappa=-1,~A_1=A_3$ to  an oblate spheroid for   $\kappa=1,~A_1=A_2$ with
maximal triaxiality at $\kappa=0$. 

\begin{figure}[t]
\center{\includegraphics[width=\linewidth,trim=0 0 0 0 ,clip]{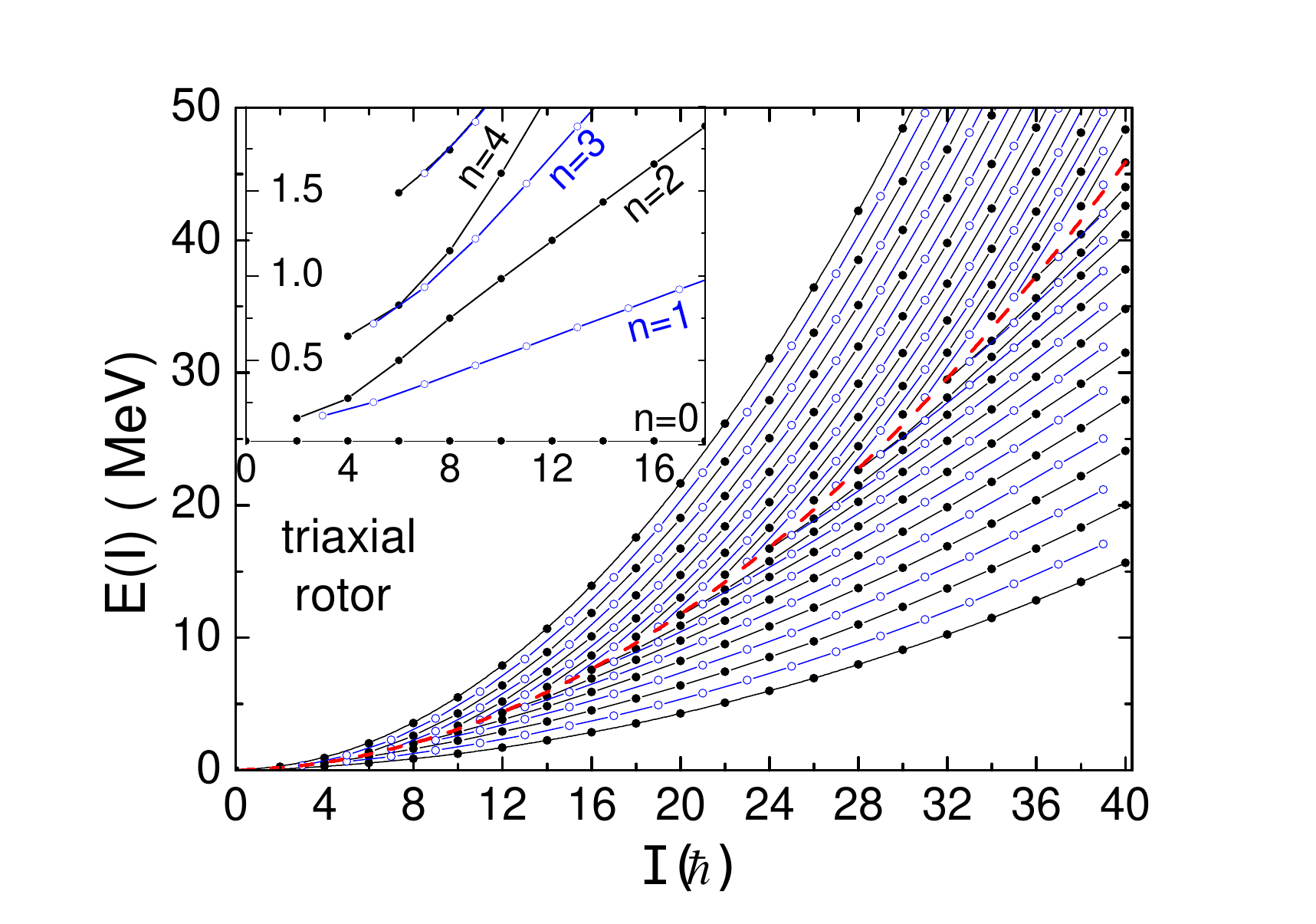}}
 \caption{\label{f:TRM_energy} Energies of the triaxial rotor the asymmetry parameter $\kappa=0.24$ as 
 functions of the angular momentum $R$. 
 The dashed line 
 represents the energy of the separatrix $I(I+1)/(2\mathcal{J}_s)$.  The inset displays  only the completely symmetric representations of the D$_2$ point group.
 The states are labelled 
 by the wobbling quantum number $n$.  }
\end{figure}

 Fig.~\ref{f:TRM_energy} shows the TR energies  
with the MoI's $\mathcal{J}_m=30~\hbar^2$/MeV, 
$\mathcal{J}_s=10~\hbar^2$/MeV, $\mathcal{J}_l=5~\hbar^2$/MeV, which correspond to $\kappa=0.24$, which is close to 0 of maximal  asymmetry oft the inertia ellipsoid. 

\subsection{Restrictions by symmetry}\label{s:sym}

In contrast to molecules, 
 nuclei are composed of two species identical fermions, the protons and neutrons. This implies that the orientation of the nuclear density with respect to a symmetry axis cannot be specified,  
and collective rotation about this axis does not exist. The rational bands of deformed axial nuclei reflect the collective rotation about the axis perpendicular to the symmetry axis.
The axial rotor Hamiltonian is simply $(\hat R^2-\hat R_3^2)/2{\cal J}$ with 3 being the symmetry axis and ${\cal J}$ the MoI.
For reflection-symmetric nuclei the rotational band built on the ground state is the sequence $I(I+1)/2{\cal J}$, $I$= 0, 2, 4, .... The odd $I$ do not appear because a rotation ${\cal R}_\bot(\pi)$
about an axis perpendicular to the symmetry axis leaves the intrinsic state of the rotor invariant. In their textbook \cite{BMII}, Bohr and Mottelson discussed in detail the rotational features of 
the axial rotor when the intrinsic  state contains excited quasiparticles. 
 This restricts the 
choice of the Moi's  by the fundamental condition that collective rotation about a symmetry axis is not allowed.  
More specifically,  the more the density of a principal axis deviates from rotational symmetry the larger is its MoI, 
which requires the order $\mathcal{J}_m>\mathcal{J}_s>\mathcal{J}_l$.
In the following, we associate the axes $i$= 1, 2, 3, with  the short, medium, 
and long principal axes of the density distribution, respectively, using the short-hand notation $s-$, $m$- and $l$- axes.

\begin{figure}[t]
\center{\includegraphics[width=\linewidth,trim=0 0 0 0 ,clip]{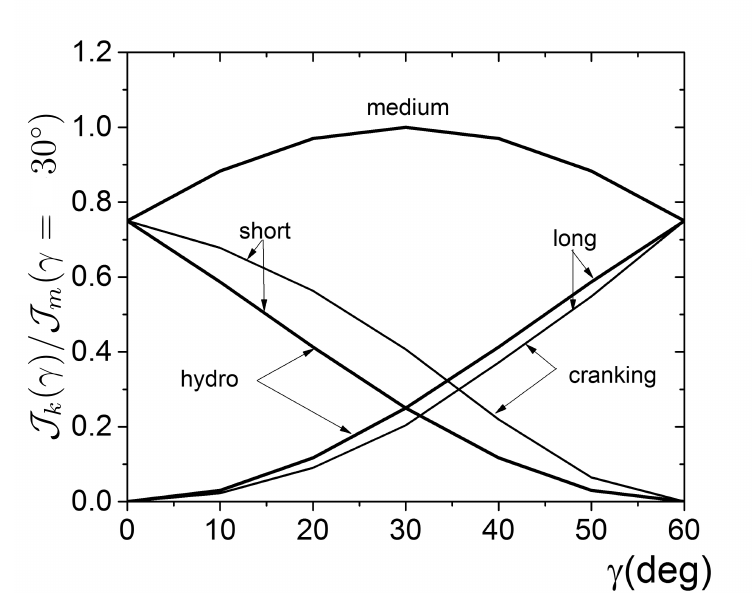}}
 \caption{\label{f:Jgam} 
 Thick curves: Irotational flow MoI's as functions of the triaxiality parameter $\gamma$  (Copenhagen convention).
 Thin curves: Microscopic cranking calculations for $Z=68$, $N=96$, $\Delta_p=1.1$ MeV, $\Delta_n=1.0 $ MeV,  $\varepsilon=0.25$.
 Reproduced with permission from Ref.~ \cite{Frauendorf2018PRC} (where the Lund convention with the opposite sign of $\gamma$ was used).}.
\end{figure}
\begin{figure}
\center{\includegraphics[angle=-90,width=0.8\linewidth,trim=0 40 0 40 ,clip]{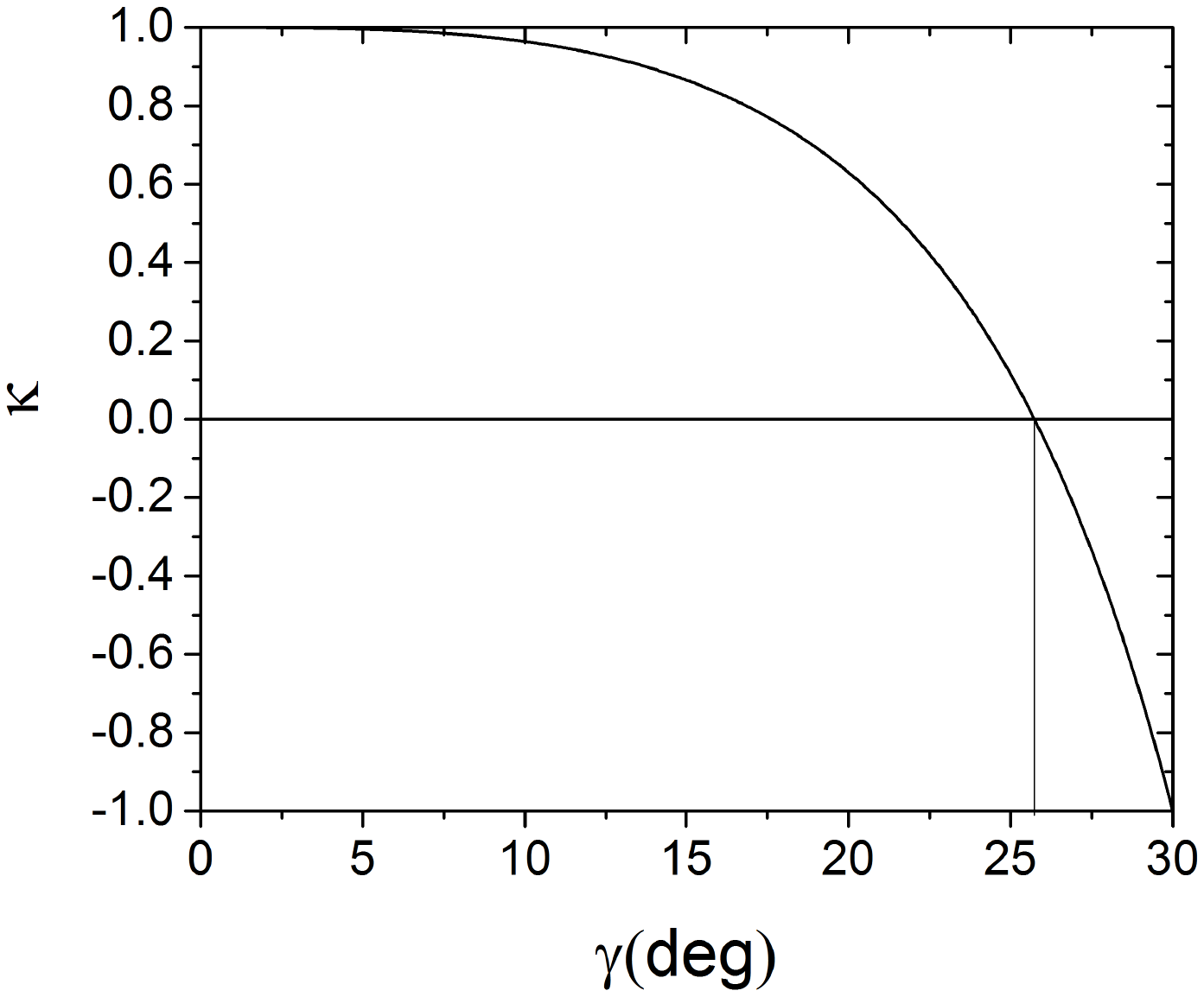}}
 \caption{\label{f:kap-ga} 
 The asymmetry parameter $\kappa(\gamma)$ Eq. (\ref{eq:asym}) for irrotational flow MoI's.}
\end{figure}

For triaxial shape, specified by the deformation parameters $\beta$ and $\gamma$, (see \cite{BMII}) collective rotation about the third axis becomes possible. 
The new degree of freedom leads to a new class of collective excitations - the "wobbling" mode,
which will be discussed in the following.
The condition that $\mathcal{J}_2$ is the largest is obeyed by assuming, like  Bohr and Mottelson ~\cite{BMII} Eq. (6B-17), that  the ratios between MoI's 
agree with the ones of irrotational 
flow of an ideal liquid,
\begin{equation}\label{eq:momiIF}
 \mathcal{J}_i(\beta,\gamma)=\mathcal{J}_0(\beta)\sin^2\left(\gamma-\frac{2\pi i}{3}\right),
\end{equation}
where  $\mathcal{J}(\beta)$ sets the scale. Fig. \ref{f:Jgam} shows the  irrotational-flow MoI's and Fig. \ref{f:kap-ga} the pertaining asymmetry parameter as 
functions of $\gamma$.   Below $15^\circ$, where ${\cal{J}}_3$ is small, the inertia ellipsoid does not change much.
 For $\gamma=25.76^\circ$ it is maximal triaxial.  Above,  it  rapidly develops  to a prolate  spheroid  with ${\cal{J}}_2>>{\cal{J}}_1\approx{\cal{J}}_3$. 

The analysis of of the rotational energies and E2 matrix element from COULEX experiments ~\cite{Allmond2017PLB} 
demonstrated that the dependence of the MoI's on the
triaxiality parameter $\gamma$ is well accounted for by Eq.~(\ref{eq:momiIF}).
Microscopic calculations  by means of the cranking model~\cite{Frauendorf2018PRC}
demonstrated that the MoI's approach the ratios 
(\ref{eq:momiIF}) for strong pairing correlations. Fig. \ref{f:Jgam} includes the MoI's for realistic pairing.  For weak pairing substantial deviations appear. 
However the deviations do not indicate an approach to the ratios of rigid body flow, because these contradict 
the no-rotation-about-a-symmetry-axis restriction of the nuclear many body system. The moment of the medium axis is always the largest because
the density distribution maximally deviates from axial symmetry.

\begin{figure*}[t]
\center{\includegraphics[width=\linewidth,trim=0 0 0 100 ,clip]{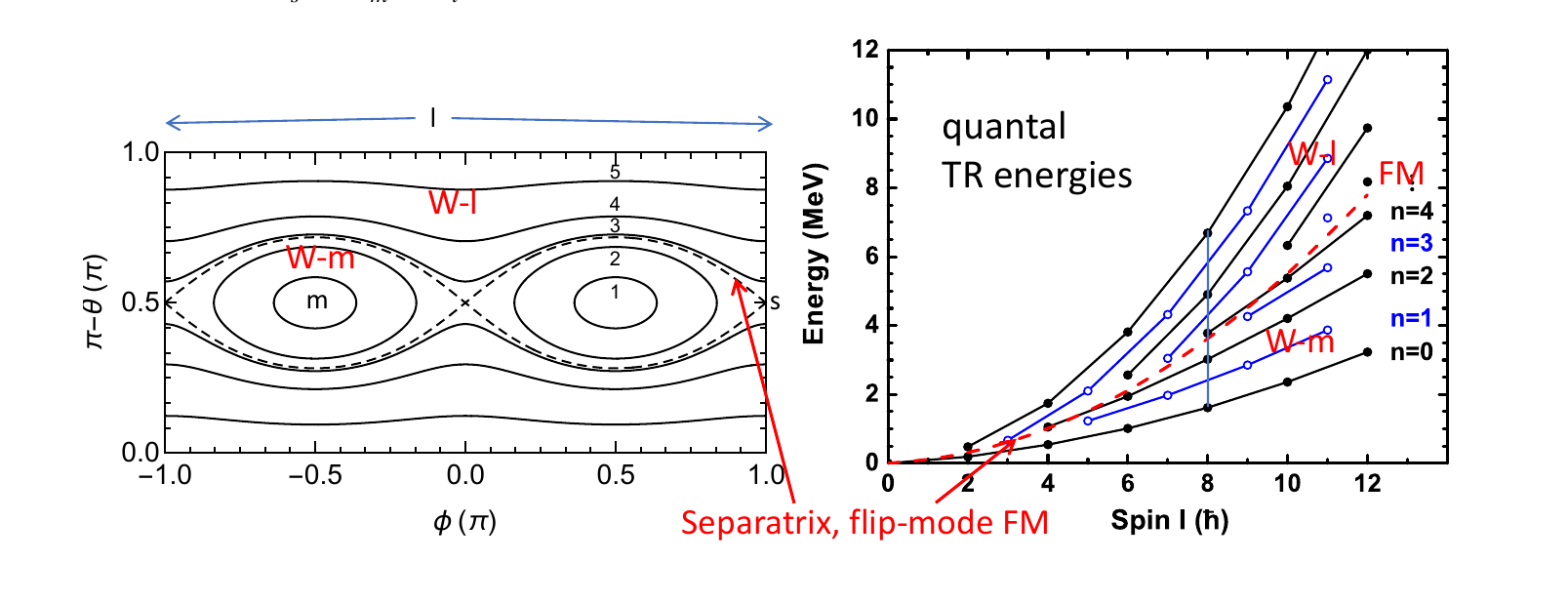}}
 \caption{\label{f:TRM_energy_orbits} Left panel: Classical orbits of the triaxial rotor states that are connected by the vertical line at $I=8$ the right panel. 
 The energies are equal to their quantal
  energies in the right panel, labelled  in ascending order. Right panel: Energies of the triaxial rotor as 
 functions of the angular momentum $R$. The states are labelled 
 by the wobbling quantum number $n$. The dashed line 
 represents the energy of the separatrix $I(I+1)/(2\mathcal{J}_s)$. Reproduced with permission from  Ref.~\cite{Chen_Frauendorf2022EPJA} and \cite{Frauendorf2014PRC} and augmented. }
\end{figure*}

Not all solutions of the triaxial rotor (TR) Hamiltonian are accepted for nuclei. Restricting the discussion  to the case that the intrinsic triaxial ground state 
is invariant with respect to the three rotations ${\cal R}_i(\pi)$ only the TR solutions that are symmetric with respect to the three 
rotations are rotational excitations of the intrinsic ground state (completely symmetric representations of the D$_2$ point group) are allowed.
The inset of Fig. \ref{f:TRM_energy} and the right panel of Fig.~\ref{f:TRM_energy_orbits} show the selection of these states.

 The TR Hamiltonian is diagonalized 
in the basis of the axial rotor states  $\vert IMK\rangle$, where $M$ is the angular momentum  projection 
on the laboratory $z$-axis and $K$ the projection  the 3-axis
axes of the rotor. The TR eigenstates
\begin{equation}\label{eq:rotorstate}
\vert IM\nu\rangle=\sum_{K=-I}^{I} C_{IK}^{(\nu)}\vert IMK\rangle
\end{equation}
are given by the amplitudes $C_{IK}^{(\nu)}$, which depend only on $K$.
the angular momentum  projection on one of the body-fixed principal axes. 
The symmetry restricts 
$K$ to be even and requires $C_{I-K}^{(\nu)}=(-1)^IC_{IK}^{(\nu)}$. 

\subsection{Correspondence with the classical orbits}\label{s:correspondence}

 It is instructive to classify the states using quasiclassic correspondence. 
The left hand panel shows the classical orbits of the angular momentum vector $\mathbf{R}$ with respect to the principal axes of the triaxial shape, where 
\begin{equation}
R_1=R\sin\theta\cos\phi,~R_2=R\sin\theta\sin\phi,~R_3=R\cos\theta.
\end{equation}

The orbits are determined by the conservation of the angular momentum 
\mbox{$R^2=I(I+1)$} and the energy $E_{TR}(I)$, Eq. (\ref{eq:HTR}), which is set equal to the quantal energies in right panel of Fig.~\ref{f:TRM_energy_orbits}. 
The area $\oint R\cos\theta(\phi)d\phi$ between the orbits takes the value $2\pi \hbar$, which is the semiclassical quantization condition.

The figure shows the orbits for the states $I_\nu=8_\nu$ which are connected by the vertical line at $I=8$ in the right panel.
There are two types of orbits, which are
separated by the separatrix shown as the dashed orbit in the left panel. For the  states inside the separatrix the angular momentum revolves around the axis with 
the maximal  MoI, and for the ones outside it revolves around the $l$-axis with 
the minimal MoI. The orbits are numbered by their energy in ascending order. For the lowest states the motion is harmonic as seen by the elliptic shape of the orbit.

 Using semiclassical correspondence the states are classified by their topology. On page 190 of their textbook Nuclear Structure II \cite{BMII}
Bohr and Mottelson introduced the name "wobbling" as 
"...the precessional motion of the axes with respect to the direction of I; for small amplitudes this motion has 
the character of a harmonic vibration... ". This clearly describes the states that correspond to the orbits inside the separatrix. Hence it is appropriate to call 
these states wobbling excitations. Accordingly,  the states labelled by $n=$ 0, 1, 2, ....are called zero, single, double, ...  wobbling excitation, respectively.
In Ref. \cite{Lawrie2020}  the authors suggested restricting the name "wobbling" to the states with elliptic (harmonic) precession cones and to call the other tilted precession (TiP) states
(see the detailed discussion below).
In my view this is an unnecessary complication. It does not comply with Bohr and Mottelson, who associate wobbling with a general precession and then add 
harmonic vibration as a special case. I recommend using the simple topological classification with respect to the separatrix. 

It is instructive  indicating the axis about which the angular momentum vector precesses. In the TR model one has the $m$-axis 
wobbling states which correspond to the orbits inside the separatrix and the $l$-axis wobbling states which correspond to the orbits outside it.

Classically, the separatrix represent the unstable rotation about the s-axis with the intermediate MoI. The orbits in its vicinity represent a number of revolutions about the
unstable axis and then a rapid change to rotation about the opposite orientation of the unstable axis for another number of revolutions and back to the original orientation, and so on. 
This behavior of the triaxial top is also known as the Dzhanibekov effect after the Russian cosmonaut  who observed it when he unscrewed a wing nut,  which flipped  off his hand and floated 
around. Instructive videos can be found in Wikipedia. In Ref.~\cite{Chen_Frauendorf2022EPJA}, Chen and Frauendorf suggested the name "flip mode" (FM) for the  states corresponding to orbits close to the separatrix. 
The right panel of Fig. \ref{f:TRM_energy_orbits}, demonstrates how new FM states are born at the energy of the separatrix while  the lowest wobbling states become more and 
more harmonic when the angular moment increases. The anharmonicity of the mode increases with the wobbling number $n$. There is no sharp borderline between the wobbling and flip regimes.

\begin{figure}[t]
\center{\includegraphics[width=0.7\linewidth,trim=0 0 0 0 ,clip]{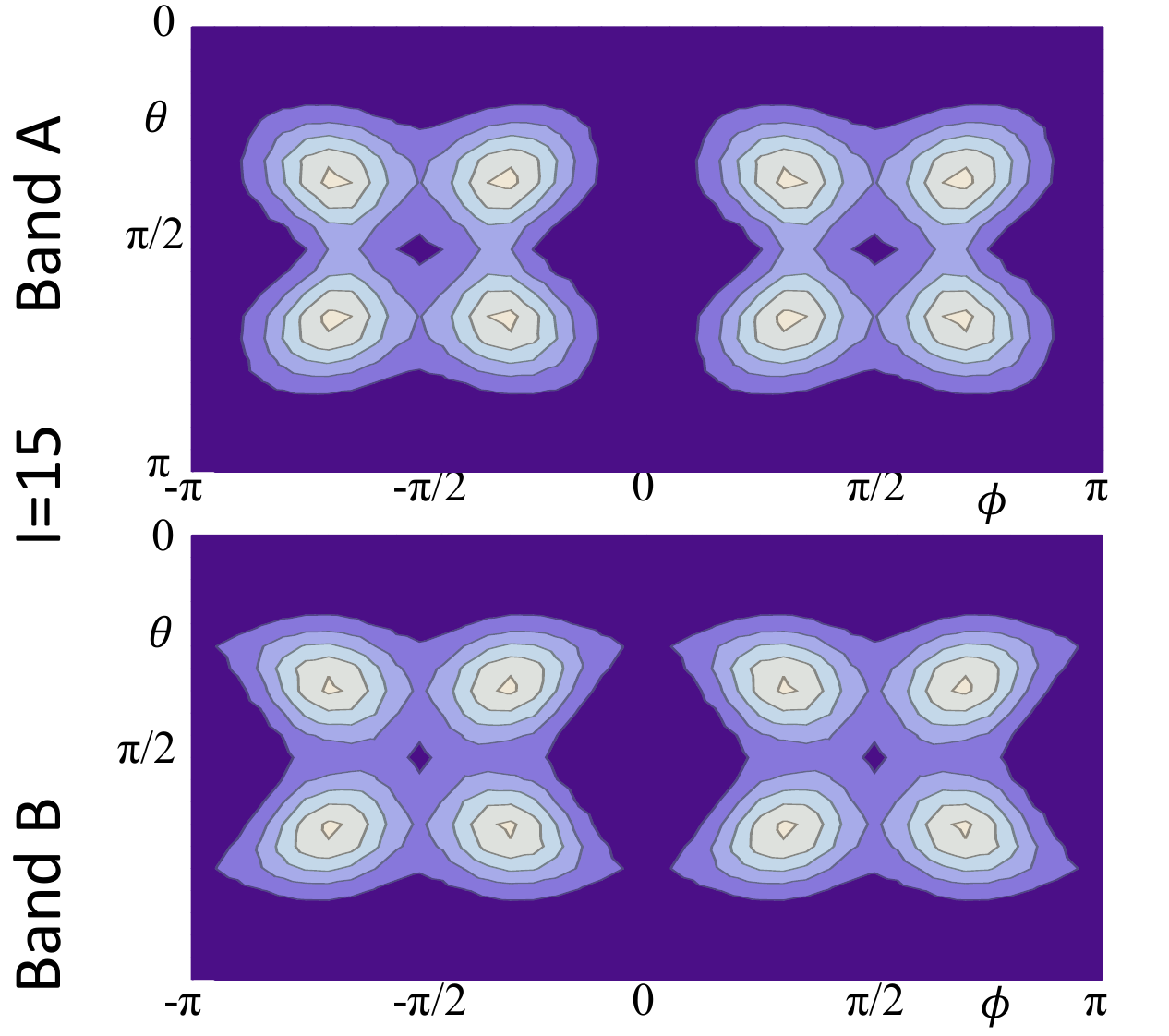}}
 \caption{\label{f:firstSCS} Probability distribution of the angular momentum orientation $ P(\theta \phi)_{I\nu} $  
 generated from the reduced density matrix of the particle-hole plus TR calculation  for $^{134}$Pr in Ref. \cite{Frauendorf-Meng97}. 
 The SCS map shows that the distributions of the chiral partner bands are almost the same at $I=15$ where the two bands are nearly degenerate. 
 }
\end{figure}

\subsection{Spin Coherent States}\label{s:SCS}
The Spin Coherent State (SCS) representation  distills the semiclassical features from the TR states. The probability of the angular momentum orientation is given by
\begin{align}\label{eq:SCS}
 P(\theta \phi)_{I\nu} &=\frac{2I+1}{4\pi}\sin\theta\notag\\
 & \times \sum\limits_{K,K'=-I}^I D^{I*}_{IK}(0,\theta,\phi)
 \rho^{(\nu)}_{KK'}D^{I}_{K'I}(0,\theta,\phi).
 \end{align} 
 In the case of the TR the density matrix is
 \begin{equation}\label{eq:dmTR}
   \rho^{(\nu)}_{KK'}=C^{(\nu)}_{IK}C^{(\nu )*}_{IK'}
 \end{equation}
   with $\nu=n$ being the wobbling number.
 Plots of $ P(\theta \phi)_{I\nu}$ (SCS maps) have first been produced for the two-particle triaxial 
rotor model ~\cite{Frauendorf2015Conf}  
 in order to visualize the appearance of the chiral geometry of the quantal results.  
 In these cases $ \rho^{(\nu)}_{KK'}$ is the reduced density matrix.
 Fig. \ref{f:firstSCS} shows an example of the SCS maps from Ref.  ~\cite{Frauendorf2015Conf}.
  Subsequently Chen {\it et al.}~\cite{F.Q.Chen2017PRC} published  the visualization technique
 under the name  ``azimuthal plots" for  the  projected shell model. 
  Later on it has been used quite a bit to illustrate the chiral and wobbling geometry.

\begin{figure}[ht]
\center{\includegraphics[width=\linewidth]{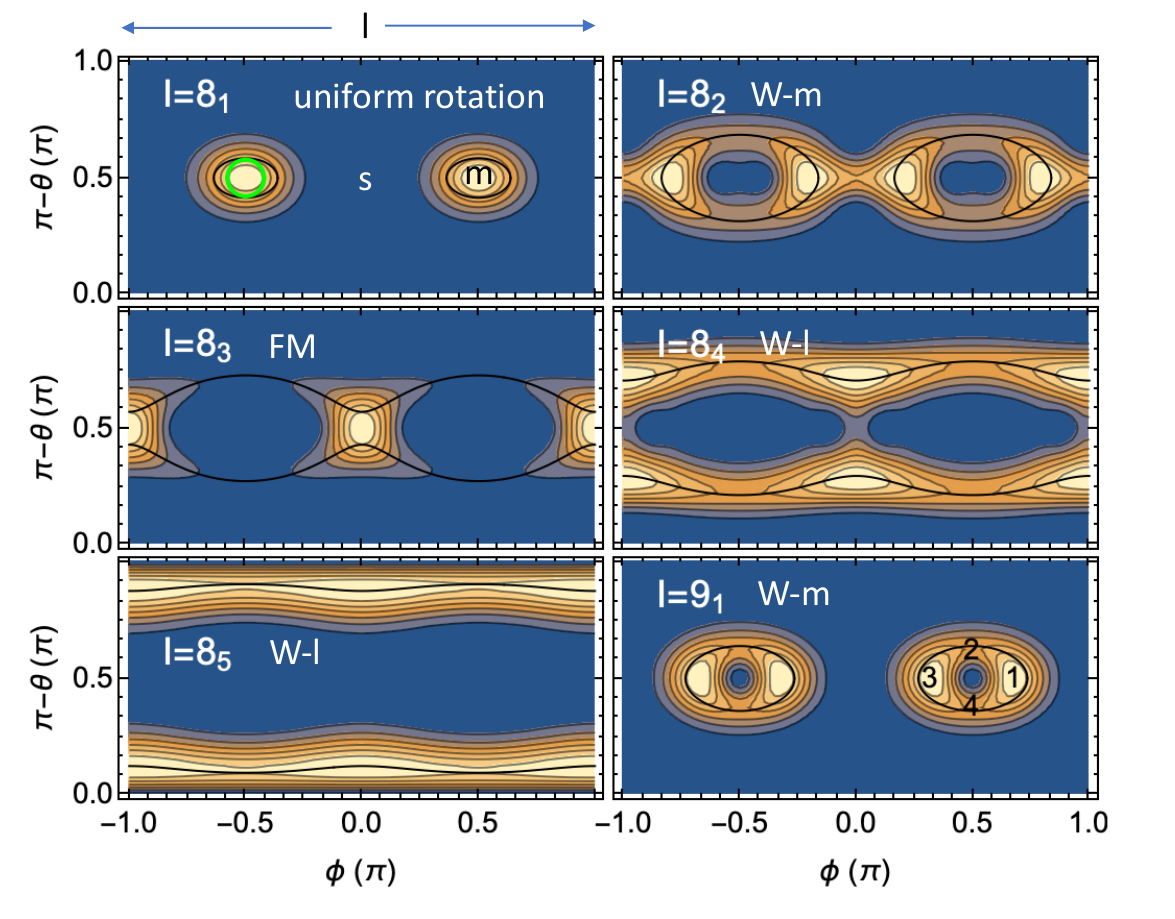}}
 \caption{ \label{f:TRmaps1} SCS probability densities of the angular momentum orientation $ P(\theta \phi)_{I\nu} $  of the triaxial rotor states $I_\nu=8_1$, ... $8_5$ and $9_1$ 
shown  in  Fig.~\ref{f:TRM_energy_orbits}. The states are labelled by $I_\nu$ in direction 
 of ascending energy. 
 The color sequence with increasing probability is
 dark blue $-$ zero, light blue, dark brown, light brown, white. 
 The densities are normalized. The classical orbits are shown as 
 black full curves. The energies are equal to their quantal energies in
 Fig.~\ref{f:TRM_energy_orbits}. The numbers indicate the turning points.
 The green circle indicates the finite resolution
 of the plot caused by the non-orthogonaliy of the SCS basis. 
 Reproduced with permission from  Ref.~\cite{Chen_Frauendorf2022EPJA}. }
\end{figure}

 The SCS map  in Fig. \ref{f:TRmaps1} shows that the distributions  are fuzzy rims tracing the classical orbits, which are included as black curves. 
 As discussed in detail in Ref.~\cite{Chen_Frauendorf2022EPJA}, the quantal values of $ P(\theta \phi)_{I\nu} d\theta \vert \cos\theta\vert d\phi $ correspond  to the 
 ratio $dt/T$ of the time $dt$ the classical orbit spends within the area $d\theta \vert \cos\theta\vert d\phi $ and $T$ the time for one revolution.
 
 \subsection{Discussion of the Triaxial Rotor Dynamics}\label{s:TRdyn}
 
 Inspecting Figs. \ref{f:TRM_energy_orbits} and  \ref{f:TRmaps1}, it is instructive to remember  
  the physical pendulum, for which the elongation angle $\varphi$ and the momentum $p=\mathcal{J}d\varphi/dt$ are canonical variables.
 In analogy, the angle $\phi$ and $R_3=R\cos\theta$ are the conjugate position and momentum of the classical TR.
  For small energy the motion is  harmonic (mathematical pendulum), and  the phase space curve is an ellipse. 
 With increasing energy the curve deforms. Along with this, the time near the two turning points increases and the time near $\phi=\pi/2$ or $\varphi=0$ deceases.
 At the critical energy of the separatrix the pendulum takes the unstable position $\varphi=\pm\pi$ and the TR the position $\phi=0, ~\pi$ corresponding to the unstable rotation about the s-axis.
 For energies slightly below the pendulum stays a long time at the turning points to swing rapidly through $\varphi=0$. For energies slightly above the pendulum very slowly passes over 
 $\varphi=\pi$ and keeps going, i. e. it rotates about its axis. The relative passage time through $\varphi=\pi$ decreases with a further increase of the energy. In analogy, 
for energies above the critical value the TR angular momentum precesses around the $l$-axis instead about the $m$-axis as for the energies below. 
 For energies near the critical value the TR angular momentum vector stays very long  near the s-axis  and moves rapidly through $\phi=\pi/2$,  
 like the rapid passage of $\varphi=0$ of the pendulum.

 The state $8_3$ has an energy somewhat above the critical value. Its angular momentum stays long near the s-axis $\phi=0$, rapidly moves (flips) to the negative s-axis $\phi=\pm\pi$,
 stays long, flips back  to the positive s-axis $\phi=0$, and so on. Accordingly it is called the "flip mode" and labelled by FM in Fig. \ref{f:TRM_energy_orbits}, where 
 the lower states are labelled as W-m and the higher as W-l according to their precession axes.
 
 \begin{figure}[t]
\center{\includegraphics[width=\linewidth,trim=0 0 300 0 ,clip]{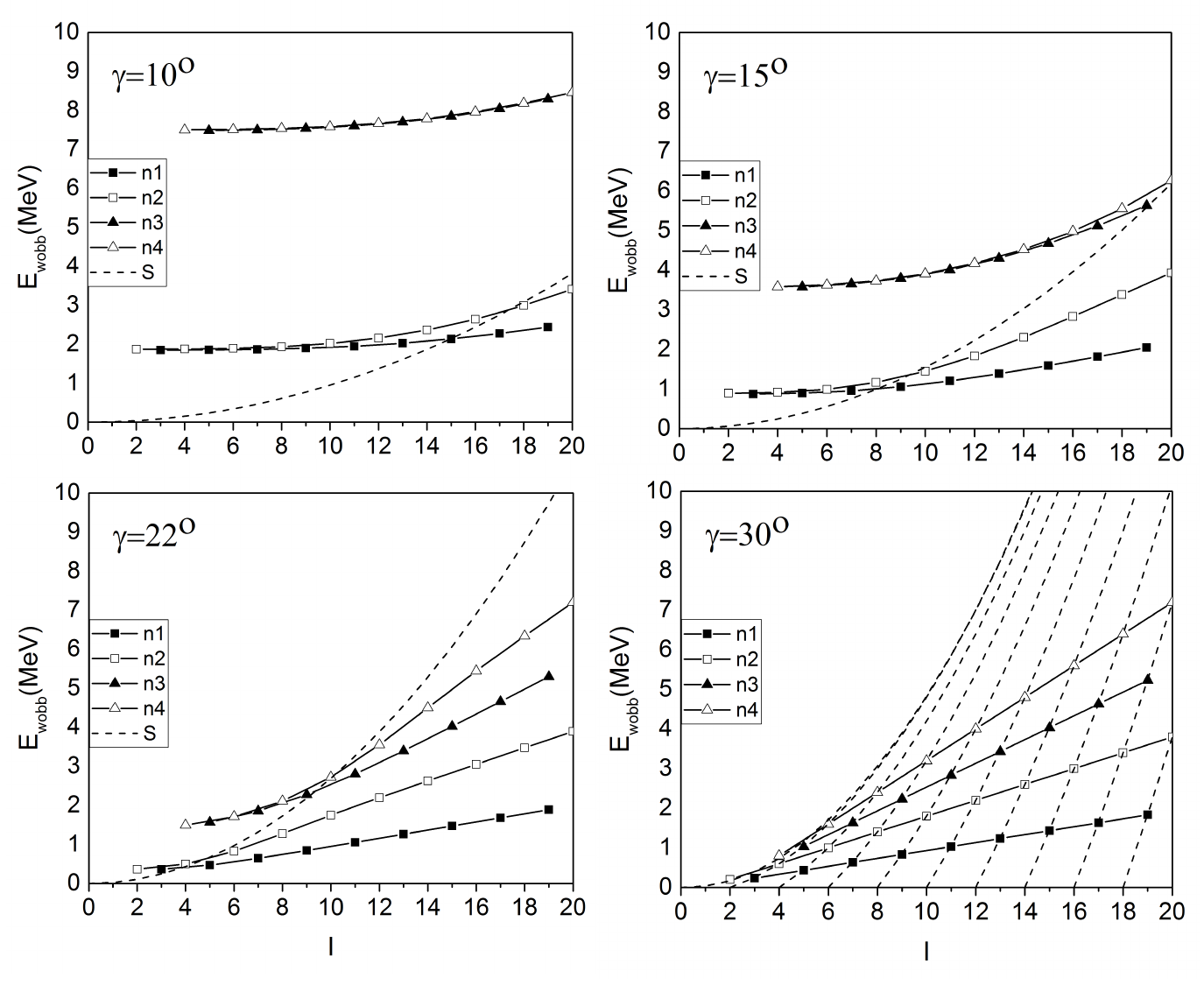}}
 \caption{\label{f:EWTRga} 
 Experimental wobbling energies Eq. (\ref{eq:Ewexp}) for  $\gamma=10^\circ,~15^\circ,~22^\circ,~30^\circ$ assuming irrotational-flow MoI's Eq.(\ref{eq:momiIF}).
 The respective inertia asymmetry parameters are  $\kappa=$ 0.96, 0.87, 0.47, -1. The energies are normalized to $E(2^+_1)$. 
 The dashed curves show the separatrix, except for $\gamma=30^\circ$, where they connect the sequences with the same projection $K$ on the $m$-axis.}
\end{figure}

 "Experimental" wobbling energies of the TR spectrum are defined as the excitation energies above the yrast states, 
\begin{align}\label{eq:Ewexp}
E_w(I)=E(I)-\left(E(I-1)_{yrast}+E(I+1)_{yrast}\right)/2
\end{align}

Fig. \ref{f:EWTRga} shows $E_w(I)$ of the  TR with irrotational-flow MoI's for different $\gamma$.
 For the  lowest states and sufficiently high angular momentum, located far enough below the separatrix, the harmonic 
 approximation  \cite{BMII}, p. 190 ff, \cite{Frauendorf2014PRC} becomes valid.
 The TR energies are given by 
 \begin{equation}\label{eq:EW-harm}
E_n(I)= A_3I(I+1) + \left( n + \frac{1}{2} \right)\hbar\omega_w, ~~A_i=1/2\mathcal{J}_i
\end{equation}
where $n$ is the number of wobbling quanta and
the wobbling frequency $\hbar\omega_w$ is equal to 
\begin{equation}\label{eq:omW-harm}
\hbar\omega_w = 2I\,[(A_1 -A_2)(A_3-A_2)]^{1/2}.
\end{equation}
Reduced transition probabilities are given in Refs.  \cite{BMII} and \cite{Frauendorf2014PRC} (where the different axes assignment  $A_1\geq A_2\geq A_3$ was used).
An example is the case $\gamma=22^\circ$ with $\kappa$=0.47 in  Fig. \ref{f:EWTRga}, where  for $ I > 15$ the  distance between lowest wobbling excitations $n$=1, 2, 3 is about the same. 

 For $\gamma=30^\circ$  the  irrotational-flow MoI's have the ratios $\mathcal{J}_m:\mathcal{J}_s:\mathcal{J}_l=4:1:1$. 
The TR is a symmetric top  (Meyer-ter-Ven limit \cite{MtV1975}) with the energy expression 
\begin{equation}
E_n(I)= I(I+1)A_s+K^2(A_m-A_s),
\end{equation}
where $K=I-n$ is  the  angular moment projection on the $m$-axis, which takes  even values between $I$ and $-I$.
The expressions for the transition probabilities can be obtained from  Ref.  \cite{BMII} by relabelling the axes. 

The states group into sequences of fixed $K$, which are displayed by the 
dashed curves in Fig. \ref{f:EWTRga}. The full lines display the lowest bands to be seen experimentally. As usual, these states are connected by strong E2 transitions.
Their corresponding classical motion is a precession of the angular momentum around the $m$-axis, where the orbit in the $s$-$l$-plane is a circle with the radius $n=I-K$, which is
a special case of wobbling.   When $\gamma$ increases toward $30^\circ$ the elliptical orbits develop into circles and the separatrix approaches the $K=0$ sequence.

 In Fig.  \ref{f:EWTRga}, $\gamma=10^\circ$  illustrates the other case of small  asymmetry of the inertia tensor $\kappa=0.96$, 
with the MoI's ratios $\mathcal{J}_m:\mathcal{J}_s:\mathcal{J}_l=29:19:1$. The separatrix is close to the yrast line, such that the $n=1$ wobbling 
state appears only for $I\geq 17$. For smaller $I$ the lowest excited states form a doublet, which correspond to a narrow precession cone around the  $l$- axis. 
The SCS maps show maxima at $\theta=\pi/2, ~\phi=0,~\pm\pi$ for even $I$ and  $\theta=\pi/2, ~\phi=\pm\pi/2$, which represent
the near-axial  density distribution $\propto \vert D^I_{I,2}(0,\theta,\phi)+(-1)^I D^I_{I,-2}(0,\theta,\phi)\vert^2$.
The doublet sequence has the characteristics of the $\gamma$ band of axial nuclei. As already pointed in Ref. \cite {BMII}, the properties 
of the $\gamma$ band are well accounted for by a slightly asymmetric TR. Examples  are the experimental information on the E2 matrix elements in $^{168}$Er 
 (TR $\gamma=7^\circ$) \cite{Frauendorf2018} and the work in Ref. \cite{Allmond2017PLB}. The $\gamma$ vibration represents a wave  that travels around the $l$- axis
 of the axial nucleus, which is equivalent with the $K=2$ excitation of a TR with  weak triaxiality.

 \begin{figure}[ht]
\begin{minipage}{0.49\textwidth}
\center{\includegraphics[angle=-90,width=\textwidth,trim=0 0 0 0 ,clip]{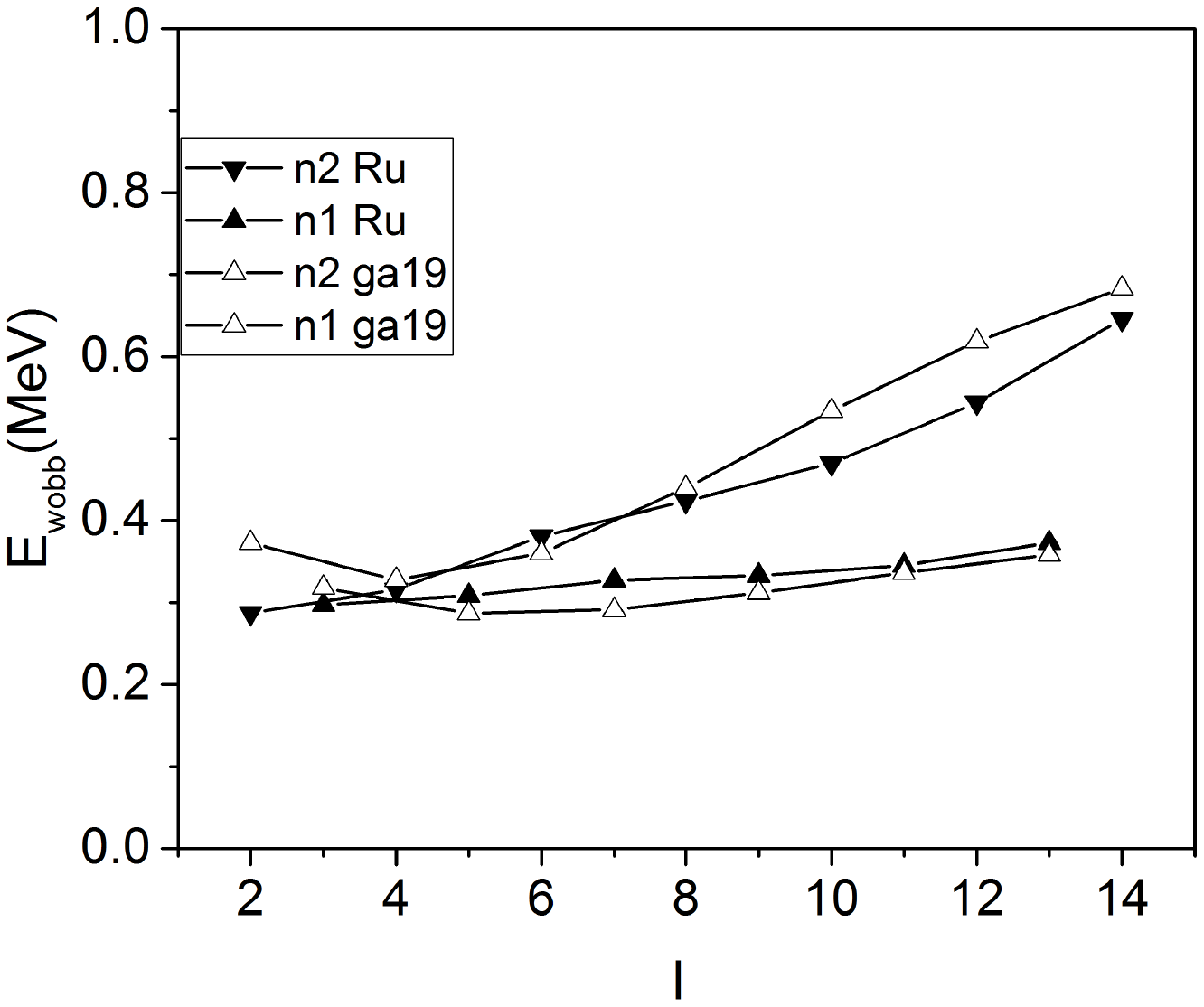}}
\end{minipage}
\hfill\mbox{}\hfill
\begin{minipage}{0.49\textwidth}
 \caption{\label{f:Ruga19} 
 Experimental wobbling energies Eq. (\ref{eq:Ewexp}) for $^{112}$Ru (full symbols) compared with the TR calculation (open symbols) using 
  irrotational-flow MoI's for \mbox{$\gamma=19^\circ$ and ${\cal J}_0=30(1+0.2I)\hbar^2/\mathrm{MeV}$. }}
\end{minipage}
\end{figure}

\subsection{Experimental evidence for triaxial rotor states in even-even- nuclei}\label{s:TRexp}
Fig. \ref{f:Ruga19} compares  the $n=1,~2$ wobbling bands in $^{112}$Ru \cite{ensdf} with the a TR calculation assuming 
irrotational flow, Eq. (\ref{eq:momiIF}) with $\gamma=19^\circ$, and the variational moment of inertia (VMI) ${\cal J}_0=30(1+0.2~I)\hbar^2/\mathrm{MeV}$.
With increasing $I$ the structure changes  from the a flip mode to anharmonic wobbling. The use of the VMI is decisive for the good
agreement with  the  experimental wobbling energies. However, it renders too low energies for the ground band. It is not possible to adjust 
$\gamma$ and  ${\cal J}_0$ such that good agreement for both the wobbling and yrast energies is achieved.

 The case of $^{112}$Ru together with $^{76}$Ge \cite{Toh13} and $^{192}$Os\cite{ensdf}
are the scarce examples of wobbling  excitations on the ground state band. There is a larger number of even-even nuclei for which the
even-I sequence of the quasi $\gamma$ band is lower than the odd-I one, while for wobbling the order is opposite. This "staggering of the $\gamma$ band"
indicates whether respectively, the deviation from axial shape is dynamic or static or, as commonly said, the nuclei are $\gamma$ soft or rigid (see e. g. Refs.
\cite{Frauendorf15,Ruoof2024}). In Refs. \cite{Ruoof2024,SJ21,Na23} the authors demonstrated  that the microscopic triaxial projected shell model (TPSM) \cite{JS99}
very well describes the experimental energies and E2 matrix elements of both $\gamma$ rigid and soft nuclei, provided the coupling to the quasiparticle excitations 
is taken into account.
This microscopic approach removes the problem with the yrast energies of the phenomenological TR description of $^{112}$Ru (See Figs. 5 and 7 and Tab. 3 of Ref. \cite{Ruoof2024})
It would be  interesting to investigate  the angular momentum geometry of $\gamma$ soft nuclei.  I will return to the topic in section "Soft Core".

\subsection{Alternative terminology}\label{s:terminology}

Lawrie {\it et al.} \cite{Lawrie2020}  introduced a different terminology for the classical orbits. They restrict the name "wobbling" to the near-elliptic orbits, which can be 
approximated as harmonic oscillations, and  call all other orbits  "Tilted Precessions" (TiP). The qualitative properties of the TiP orbits are described as the ones of  
the $\gamma=30^\circ $symmetric top, and the authors use  the
angular momentum projection on the $m$-axis, $K$ to label the TiP bands. The TR Hamiltonian (\ref{eq:HTR}) is more general, it encompasses
different MoI's for all three axes for which the TiP approximation does not hold.  On the other hand, the harmonic approximation for wobbling 
by Eqs. (\ref{eq:EW-harm}, \ref{eq:omW-harm}) and the pertaining expressions for the transition probabilities hold up to small terms of the order of $1/I$
for the lowest bands of the $\gamma=30^\circ$ TR (see Fig.1  of Ref. \cite{Lawrie2020}), which makes the TiP terminology inconsistent.
Moreover "tilted precession", as meant by the authors, is a tautology, because in a precession cone the angular momentum vector is tilted with respect 
to the axis it revolves by definition. 

In my view, it is better keeping the terminology simple and call all the  states that correspond to the orbits which revolve around the $m$-axis wobbling states, irrespective
if the precession is harmonic or not.
The name is quite appropriate, because the change of the rotational axis of the earth, which is quite irregular, and the tumbling  of the rotational axis of a baseball
are both called wobbling motion. It is also consistent with the way Bohr and Mottelson introduced the name into nuclear physics (see above). The first phrase indicates a general 
precessional mode, the second phrase indicates the special case of  small amplitude motion, which is worked out in the book.

\section{Particles coupled to the triaxial rotor}\label{s:PTR}

\subsection{The particle plus triaxial rotor model }

The experimental evidence for wobbling is clearer when one (or more) high-j particles are coupled to the TR, because the mutual interaction stabilizes the triaxial shape.
The high-j particles act like gyroscopes, which change the wobbling motion. The Particle plus Triaxial Rotor (PTR) model \cite{MtV1975}  describes the coupled system and has become
a standard tool for interpretation. 
The corresponding Hamiltonian is~\cite{MtV1975,BMII}
\begin{align}\label{eq:HPTR}
 H_{\textrm{PTR}}=\sum\limits_{i=1,2,3}\frac{(\hat J_i-\hat j_i)^2}{2{\cal J}_i} +h_{p}(\gamma),
  \end{align}
  where $\hat{J}_i=\hat{R}_i+\hat{j}_i$ is the total angular momentum, 
$\hat{j}_i$ the angular momentum of the particle, $\hat{R}_i$ the one of the triaxial rotor,
and $h_{p}(\gamma)$ describes the coupling of the particle with the triaxial potential. 
If necessary, the pair correlations are taken into account by applying the BCS approximation to $h_p$. In this case the model
is called the quasiparticle triaxial rotor (QTR) model. In the following I will refer to both PTR and QTR even when the pair correlations are unimportant.
The coupling of the high-j particles $f_{7/2},~h_{11/2}~i_{13/2},~j_{15/2}$ is often described by the simplified Hamiltonian \cite{Hamamoto1986PLB}
which assumes that their angular momentum is conserved 
\begin{align}
 \label{eq:hproton}
 h_p(\gamma)=\frac{\kappa}{j(j+1)}\left[\left(3j_3^2-j(j+1)\right)\cos\gamma
     +\sqrt3\left(j_1^2-j_2^2\right)\sin\gamma\right].
\end{align}
  The coupling strength $\kappa$ is determined by adjusting the single particle energies generated by $h_p(\gamma)$ to the 
  ones of a realistic triaxial nuclear potential of the same triaxiality parameter $\gamma$. As the shape of the neutron and proton densities are very similar 
  the same $\gamma$ value determines the intrinsic charge quadrupole moments and the $E2$ transition probabilities as well.
  The inclusion of pairing 
 and the generalization to quasiparticles is straightforward  \cite{Hamamoto1986PLB}.
 The MoI's are input parameters which are confined by the conditions of the TR discussed in section \ref{s:TR}.
 In some studies the irrotational-flow ratios (\ref{eq:momiIF}) for the same value of $\gamma$ as for the particle potential are used, and only ${\cal J}_0$
 is adjusted to the rotational energies. In most studies  some deviation of the inertia asymmetry $\kappa$ (not to be confused with $\kappa$ in Eq. (\ref{eq:hproton}))
 from the irrotational-flow value is accepted as long as the general conditions ${\cal J}_m>{\cal J}_s>{\cal J}_l$  are met. 

The PTR eigenstates are represented in the basis
$\vert IIK\rangle\vert jk\rangle$, where $\vert IIK\rangle$ are the 
rotor states for half-integer $I$ and $\vert jk\rangle$ the high-$j$ 
particle states,
\begin{equation}
 |II\nu\rangle=\sum_{K,k} C_{IKk}^{(\nu)} \vert IIK\rangle\vert jk\rangle.
\end{equation}
The coefficients $C_{IKk}^{(\nu)}$  are restricted by the
requirement that rotor-core states must be symmetric 
representations of the D$_2$ point group, which implies
\begin{equation}\label{eq:D2symmetry}
C_{I-K-k}^{(\nu)}=(-1)^{I-j}C_{IKk}^{(\nu)},~~K-k~~\mathrm{even}.
\end{equation}
Expression for the transition probabilities are given for example in Ref. \cite{Chen_Frauendorf2024}.
The reduced density matrices
\begin{equation}\label{eq:TWrhoj}
 \rho_{kk'}^{(\nu)}=\sum_K C_{IKk}^{(\nu)}C_{IKk'}^{(\nu)*}
\end{equation}
and
\begin{equation}\label{eq:TWrhoJ}
 \rho_{KK'}^{(\nu)}=\sum_k C_{IKk}^{(\nu)}C_{IK'k}^{(\nu)*}
\end{equation}
are inserted into Eq. (\ref{eq:SCS}) to generate SCS maps, which illustrate the corresponding motion of  the
particle  $\mathbf{j}$ and  total angular momentum $\mathbf{J}$, respectively. 

\subsection{Wobbling in  triaxial strongly deformed nuclei}\label{s:TWSD}

Clear evidence for wobbling in  triaxial strongly deformed (TSD) nuclei  was established by the Copenhagen group. 
They found in $^{161-167}$Lu \cite{161Lu,163Lu1,163Lu2,165Lu,167Lu} and $^{167}$Ta \cite{167Ta} very regular bands in the spin range $I\approx 15-40$, which are
based on the $i_{13/2}$ proton orbital and a deformation of $\beta\approx 0.4$. The bands TSD2 $\rightarrow$ TSD1  and TSD3 $\rightarrow$ TSD2 are interconnected by 
strong $I\rightarrow I-1$ collective $E2$ transitions, which are the hallmark of the 
wobbling motion of the charge density of the whole  nucleus. The bands TSD1, TSD2, TSD3 were assigned to carry $n=$0, 1,2 wobbling quanta, respectively.
Hamamoto and Hagemann \cite{163Lu1,hamamoto1,hamamoto2}  carried out PTR calculations.
They assumed $\gamma=20^\circ$, irrotational flow MoI's and adjusted $\kappa{\cal J}_0$ to account for the average wobbling energy. 
However, the authors exchanged the MoI such that ${\cal J}_s> {\cal J}_m> {\cal J}_l$, in
 contrast to the  order implied by the principles of spontaneous symmetry breaking. 
  
\begin{figure}[ht]
\center{\includegraphics[width=0.7\linewidth,trim=0 0 0 0 ,clip]{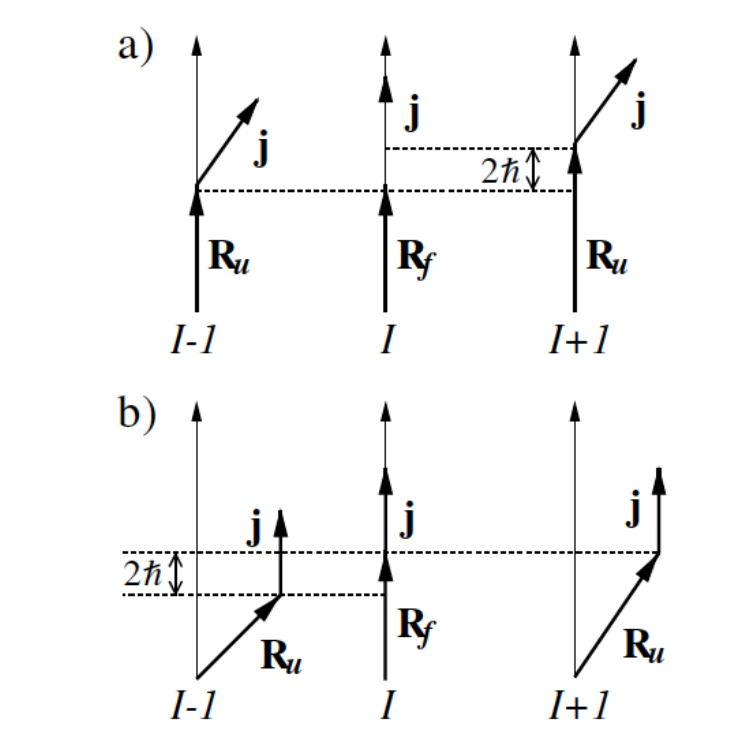}}
 \caption{\label{f:wob-crank} 
 Schematic coupling scheme of the particle and core
angular momenta in the favored ($I$) and unfavored ($I \pm 1$) states 
 for (a) the cranking regime and (b) the wobbling mode ($n_w = 1$). Reproduced with permission from Ref. \cite{163Lu1}.}
\end{figure}

 The PTR system has the particle degrees of freedom in addition to the orientation of the total angular momentum.
  The authors illustrated the  pertaining types of excitations in Fig. \ref{f:wob-crank}.  The $i_{13/2}$ proton has the lowest energy when its angular momentum $\mathbf{j}$
aligns with the s-axis, because this orientation generates the best overlap with the triaxial potential (See the discussion of particle alignment in Ref.  \cite{Frauendorf2018}).
In the  yrast band (TSD1) the TR angular momentum $\mathbf{R}$ aligns with the s-axis as well. In panel (a) the proton  $\mathbf{j}$ is tilted from the s-axis 
about which it precesses
while  $\mathbf{R}$ stays aligned. This is the excitation type that the cranking model describes as a signature partner band (The signature is defined as $\alpha=I+2n$.)
In panel (b) the proton $\mathbf{j}$ stays aligned while the $\mathbf{R}$ is tilted from the s-axis about which it precesses. This the first wobbling excitation. It generates strong E2 radiation,  
which is not the case for the signature partner band. The difference  is used to identify the bands as wobbling or cranking excitations.

\begin{figure}
\center{\includegraphics[angle=0,width=\linewidth,trim=0 0 0 0 ,clip]{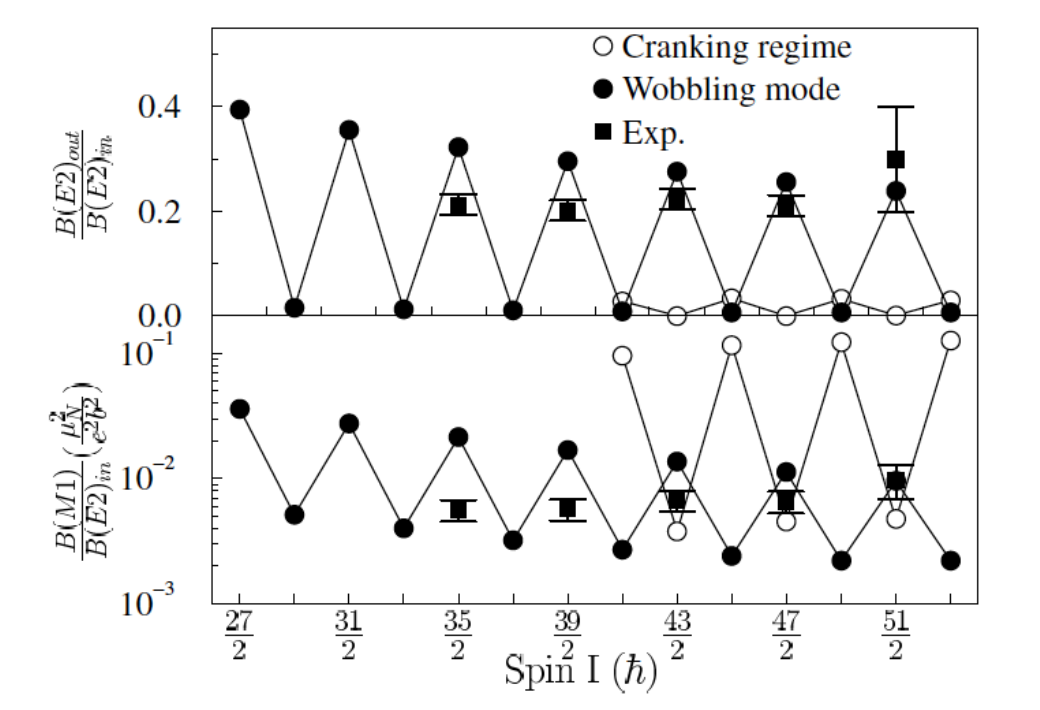}}
 \caption{\label{f:163LuBE2M1} 
Ratios $\frac{B(E2,I\rightarrow I-1)_{out}}{B(E2,I\rightarrow I-2)_{in}}$ and 
$\frac{B(M1,I\rightarrow I-1)_{out}}{B(E2,I\rightarrow I-2)_{in}}$ between the bands TSD2 and TSD1
compared with the PTR calculation in Refs. \cite{163Lu1,hamamoto1,hamamoto2}.
Reproduced with permission from Ref. \cite{163Lu1}.}
\end{figure}

The PTR $\frac{B(E2,I\rightarrow I-1)_{out}}{B(E2,I\rightarrow I-2)_{in}}$ ratios in  Fig. \ref{f:163LuBE2M1} agree well with the experimental ones. 
The ratio of 0.2 indicates a strong collective enhancement of the inter band E2 transitions, which the hallmark of the wobbling mode. 
However the PTR energy difference between the $n_w=0$ yrast band the $n_w=1$ wobbling
bands increases with $I$, which is in contrast to the decrease of the distance between the bands  TSD2 and TSD1. The authors noticed that the natural order  ${\cal J}_s<{\cal J}_m$
leads to a decrease of the wobbling energy, but dismissed it, because it was too rapid. 

The studies of $^{161,165,167}$Lu \cite{161Lu,165Lu,167Lu} rendered similar wobbling behavior as $^{163}$Lu. The accordance of 
$\frac{B(E2,I~n=2 \rightarrow I-1,~n=1)_{out}}{B(E2,I\rightarrow I-2)_{in}}$  ratios with the PTR calculations is the same as for $^{163}$Lu.


\begin{figure}[ht] 
\center{\includegraphics[width=\linewidth,trim=0 0 0 0 ,clip]{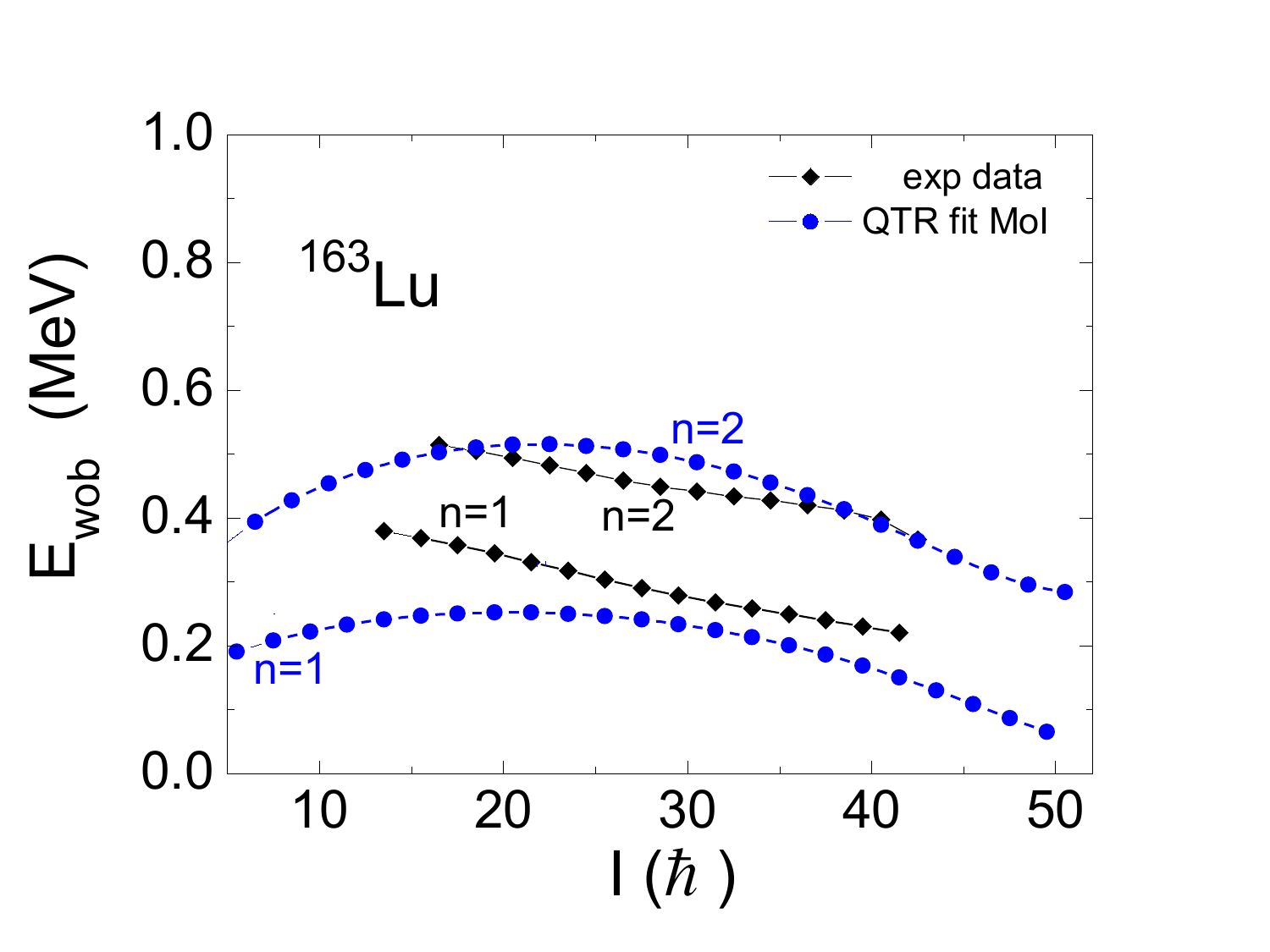}}
 \caption{\label{f:163LuEwob} 
Experimental wobbling energies Eq. (\ref{eq:Ewexp}) of the bands TSD2 and TSD3 in ${163}$Lu compared with the PTR (labelled by QTR) calculations in Ref. \cite{Frauendorf2014PRC}
 using the parameters $\varepsilon=0.4,~\gamma=20^\circ,~{\cal J}_{m,s,l}=( 64, 56, 13)\hbar^2$/MeV, respectively.
Reproduced with permission from Ref. \cite{Frauendorf2014PRC}.}
  \end{figure}

\begin{figure}
\center{\includegraphics[angle=0,width=\linewidth,trim=0 0 0 0 ,clip]{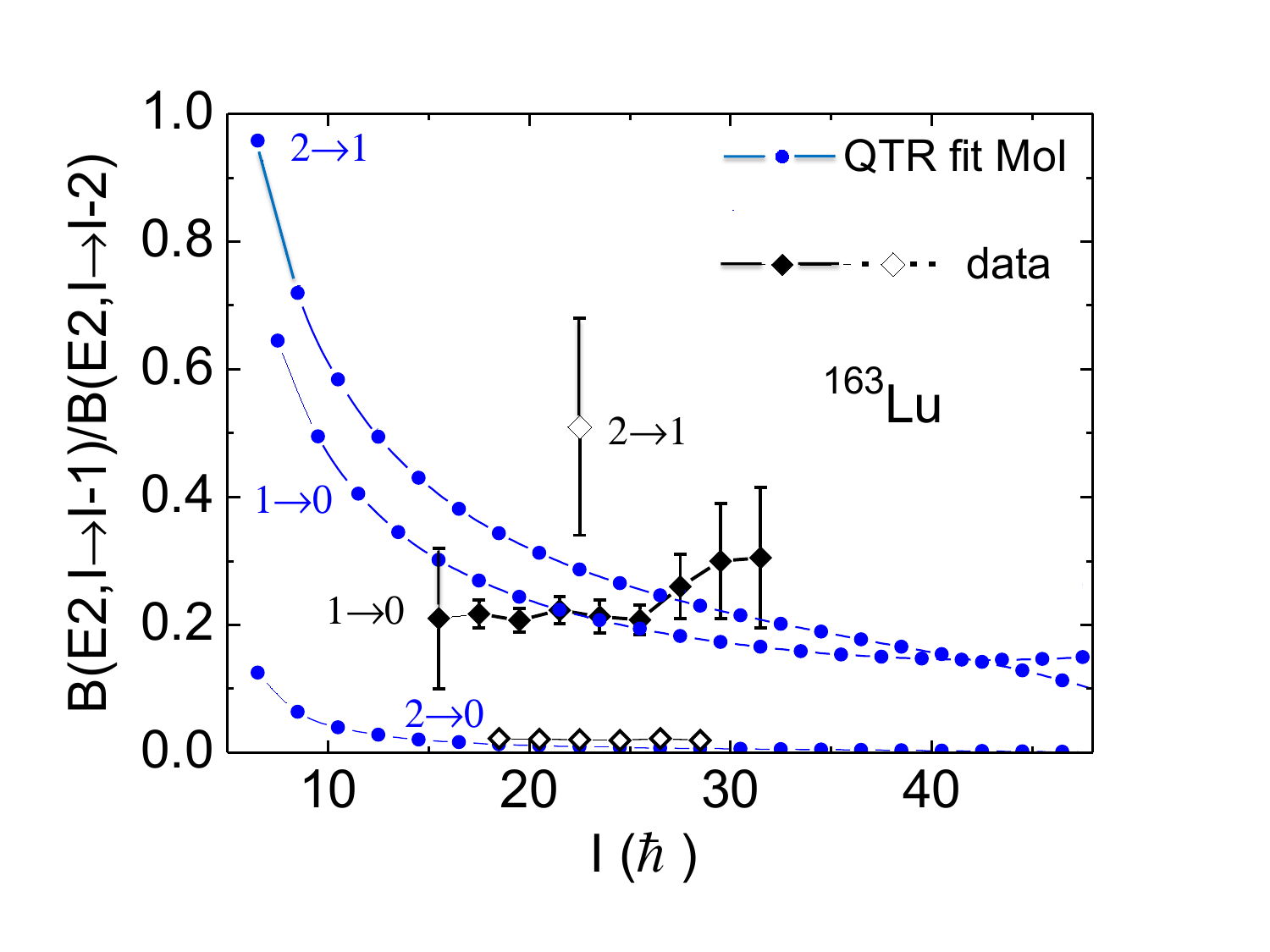}}
 \caption{\label{f:163LuBE2} 
Ratios $\frac{B(E2,I\rightarrow I-1)_{out}}{B(E2,I\rightarrow I-2)_{in}}$ and 
$\frac{B(M1,I\rightarrow I-1)_{out}}{B(E2,I\rightarrow I-2)_{in}}$ between the bands TSD2 and TSD1 compared with the QTR  calculations in Ref. \cite{Frauendorf2014PRC}.
Reproduced with permission from Ref. \cite{Frauendorf2014PRC}.}
\end{figure}

\subsection{Transverse and longitudinal wobbling}

The exchange of the two MoI's appears to be a serious problem because it violates a basic principle of spontaneous symmetry breaking. 
Analyzing their  PTR calculations, Frauendorf and Meng  discovered  that  the rotational motion of
odd-odd nuclei may attain a chiral character  \cite{Frauendorf-Meng97} (c.f. Figs. 2b and 5 of Ref. \cite{Frauendorf-Meng97}).
The chiral geometry emerges as the combination of the angular momenta of a proton particle, a neutron hole  and
the TR core aligned with the s- axis, the $l$-axis  and the $m$-axis, respectively.  That is,  chirality  appears because the $m$-axis  has the larges MoI. Fig. 2a of Ref. \cite{Frauendorf-Meng97}
demonstrates that the rotational spectrum is analogous when both the proton and the neutron are holes  with their angular momentum aligned with the $l$-axis.
The distance between the lowest even and odd $I$ bands decreases with $I$. 
 The same kind of decrease is expected in the  spectrum of the TR combined with  one proton with its  angular momentum aligned with the s-axis, which is observed
 in the Lu isotopes. In order to arrive at an interpretation that is consistent  with  the existing work on chirality, Frauendorf and D\"onau \cite{Frauendorf2014PRC}
 reanalyzed the  TSD bands of the Lu isotopes by means of the QTR, requiring that the $m$-axis has the largest MoI. 
 
 In Ref.  \cite{Frauendorf2014PRC} the authors carried out QTR calculations 
 with the same deformation parameters $\varepsilon=0.4$ and $\gamma=20^\circ$ as in Refs. \cite{163Lu1,hamamoto1,hamamoto2}. (The actual calculations were carried out in the framework of the core-quasiparticle-coupling model (QCM), which for the TR core is just the PTR or QTR  reformulated in the laboratory frame.)
 Three sets of MoI's were investigated which obey the requirement that ${\cal J}_m$ is the largest. Figs. \ref{f:163LuEwob} and \ref{f:163LuBE2} 
 compare the energies and $\frac{B(E2,I\rightarrow I-1)_{out}}{B(E2,I\rightarrow I-2)_{in}}$ ratios with the ones from TSD2 and TSD3. The ratios between the TR MoI's were adjusted to obtain 
 the best agreement of the wobbling energies with experimental ones.  The  corresponding  inertia asymmetry of $\kappa=0.93$ is close to
 $\kappa=0.91$ for the microscopic MoI's obtained by the cranking model, which gave results very similar  to the ones in Figs. \ref{f:163LuEwob} and \ref{f:163LuBE2}.
 The irrotational flow MoI's for $\gamma=20^\circ$ correspond to   $\kappa=0.64$. The enhanced asymmetry of the inertia ellipsoid causes a too steep 
 decrease of the wobbling energy with $I$. Thus, the features of the TSD wobbling bands are well accounted for by the QTR applying it in consistency with the studies of chirality.
 The only caveat is an overestimate of the $\frac{B(M1,I\rightarrow I-1)_{out}}{B(E2,I\rightarrow I-2)_{in}}$ ratios.
 
In order elucidate the physics behind the decreasing wobbling energy D\"onau and Frauendorf  \cite{Frauendorf2014PRC}
drastically simplified  the PTR by assuming that the proton $\mathbf{j}$ is rigidly aligned with the s-axis,
which they called the frozen alignment (FA) approximation. 
\begin{eqnarray}\label{FAham}
H_{FA} = A_1(\hat J_3 - j)^2 + A_2 \hat J_1 ^{2} + A_3\hat J_{2}^{2},
\end{eqnarray}where $j$ is a number.

Applying the small-amplitude approximation 
$$\hat J_3=(\hat J^2-\hat J^2_1-\hat J^2_2)^{1/2}\approx J-\frac{\hat J_1^2}{2J}-\frac{\hat J_2^2}{2J}, ~J=\sqrt{I(I+1)}$$
 to the FA Hamiltonian the  harmonic FA  Hamiltonian (HFA) is obtained,
\begin{eqnarray}\label{HFAham}
H_{HFA}=A_1(J-j)^2+(A_2-\bar A_1) \hat J_2 ^{2}+(A_3-\bar A_1) \hat J_3 ^{2},\nonumber\\\bar A_1=A_1(1-j/I).
\end{eqnarray}
Replacing $A_1$ by $\bar A_1$, the HFA Hamiltonian  agrees with the one in Ref. \cite{BMII} 
from which Bohr and Mottelson derived the  harmonic wobbling energies and transition probabilities.
The	 wobbling frequency (\ref{eq:omW-harm}) becomes  
\begin{equation}\label{eq:omTW-harm}
\hbar\omega_w = 2I\,[(A_2 -\bar A_1(I))(A_3-\bar A_1(I))]^{1/2}.
\end{equation}
The expressions for the transition probabilities are given in Refs. \cite{BMII,Frauendorf2014PRC}.

The linear increase of $\hbar\omega_w$  with $I$ is counter acted by the  decrease of the  product under the square root caused by the increase of $A_3(I)$, which results in the 
HFA curve shown in Fig. \ref{f:163LuEwob}.
At the critical spin 
\begin{equation}\label{eq:Ic}
I_c=j{\cal J}_2/({\cal J}_2-{\cal J}_1)
\end{equation}
 the HFA
becomes unstable because  $A_2-\bar A_1(I))=0$, the wobbling frequency is zero, and the HFA breaks down.
 
The presence of the  particle $\mathbf{j}$  aligned with the s-axis 
leads  to the replacement of the rotational parameter $A_1$ by the effective one $\bar A_1=A_1(1-j/I)$, which is the smallest as long as  $0<\bar A_1<A_2,~A_3$.
  The wobbling motion represents  the precession of the total angular momentum $\mathbf{J}$
  about the s- axis, which is  perpendicular to the $m$-axis with the largest TR MoI. 
  Frauendorf and D\"onau suggested the name "transverse wobbling" (TW) in order to indicate that the particle 
  $\mathbf{j}$ and the axis of the precession cone are aligned with the s-axis, which is perpendicular to the $m$-axis with the maximal MoI.  
  In the FA approximation the orientation of the precession cone agrees with the one of $\mathbf{j}$. The axis of the $\mathbf{J}$ precession cone provides  
  general classification criterion that encompasses the case when  $\mathbf{j}$ is no longer narrowly aligned with the s-axis (see below).

The presence of a hole $\mathbf{j}$  aligned with the $l$-axis leads  to the replacement of the rotational parameter $A_3$ by the effective $\bar A_3=A_3(1-j/I)$, which is another case 
of transverse wobbling with the hallmark of decreasing wobbling energies. A midshell quasiparticle has the tendency to align its  $\mathbf{j}$ with the $m$-axis. 
The HFA limit corresponds to the replacement of $A_2$ by $\bar A_2=A_2(1-j/I)$, which remains to be the smallest.
Frauendorf and D\"onau classified this case as "longitudinal  wobbling" because $\mathbf{j}$ and the axis of the precession cone have the direction of the $m$-axis with the maximal MoI. 
The wobbling frequency (\ref{eq:omW-harm}) replaced with $\bar A_2$ increases more rapidly with $I$ than for constant $A_2$, which is the hallmark of LW.

It should be stressed that Frauendorf and D\"onau introduced the notations LW and TW in order to  {\it classify the exact PTR} results. Contrary to the claim by Lawrie {\it et al.}
 \cite{Lawrie2020} they did not intend to restrict LW and TW to the $I$ range where the HFA is a good approximation. Below I will be expound how to use the classification scheme  
beyond the $I$ range where  the HFA  is valid.  

For $^{163}$Lu the  critical spin is $I_c\approx 50$, which is  just reached by the QTR calculations.  The QTR $n=1$ wobbling energy is small there as seen in Fig. \ref{f:163LuEwob}.
Although the  calculation shows  the hallmark of TW, the decrease of the wobbling energies at large $I$,  it does not reproduce the almost linear  decline seen in experiment.
Fig. \ref{f:163LuBE2} shows that the QTR ratios $\frac{B(E2,I\rightarrow I-1)_{out}}{B(E2,I\rightarrow I-2)_{in}}$ monotonically decrease with $I$ while the experimental ones first stay constant
to increase at high spin. The calculation underestimates the experimental $\frac{B(E2,I~n=2 \rightarrow I-1,~n=1)_{out}}{B(E2,I\rightarrow I-2)_{in}}$  ratio of about 
two times the value for the $n=1\rightarrow 0$ transitions, which seems to point to a more harmonic nature of the wobbling mode. 
The QTR ratios $\frac{B(M1,I\rightarrow I-1)_{out}}{B(E2,I\rightarrow I-2)_{in}}$ are about two orders of magnitude larger than in experiment.
As demonstrated in subsection "{Triaxial projected shell model", the discrepancies of the PTR with the experiment go away  for the microscopic TPSM calculations.

\subsection{Transverse wobbling  in triaxial normal deformed nuclei}\label{s:TWND}

\subsubsection{The nucleus $^{135}$Pr}

 The nucleus $^{134}$Pr was presented as the first example for chirality \cite{Frauendorf-Meng97}. Having in mind the similarity with TW, Frauendorf and D\"onau \cite{Frauendorf2014PRC} carried out 
 QTR calculations for $^{135}$Pr, which has only one odd $h_{11/2}$ proton.  They used the mean field equilibrium deformation parameters 
 $\varepsilon =0.16,~\gamma=26^\circ$  and fitted  the MoI to the observed band energies. 
 Fig. \ref{f:135PrEwobPRC} shows that  observed wobbling energies decrease until $I=29/2$ and then increase. 
 The QTR $n=1$ curve has the same shape with a less pronounce minimum around the critical spin $I_c=14.4$, where the HFA becomes unstable. 
 The authors  interpreted the results as follows.
 On the down side  the nucleus is in the TW regime and on the up side it is in the LW regime.
 
  The fitted MoI's correspond to an inertia asymmetry of $\kappa=0.71$. 
 For $\gamma=27^\circ$ irrotational flow gives the smaller value of $\kappa=-0.17$, which causes a too early instability of the TW regime. Deviations from the irrotational flow
 ratios may have various reasons. There are shell effects which modify the MoI's (see Fig. \ref{f:Jgam}). Fluctuations of $\gamma$ may result in a smaller effective value.
  In  Ref. \cite{Shimizu08} it was noted that the $\gamma$ value of the modified oscillator potential corresponds to a smaller
  $\gamma$ of the triaxial  density distribution, which determines the MoI's. The fitted ratio  $\kappa=0.71$ is close to the irrotational flow value $\kappa=0.69$ 
 for  $\gamma=19^\circ$, which is smaller than  $\gamma=26^\circ$ used for the potential in the QTR calculation.(See Ref. \cite{Shimizu08}.) 

\begin{figure}[t] 
\begin{minipage}{0.49\textwidth}
\center{\includegraphics[width=\textwidth,trim=0 0 0 0 ,clip]{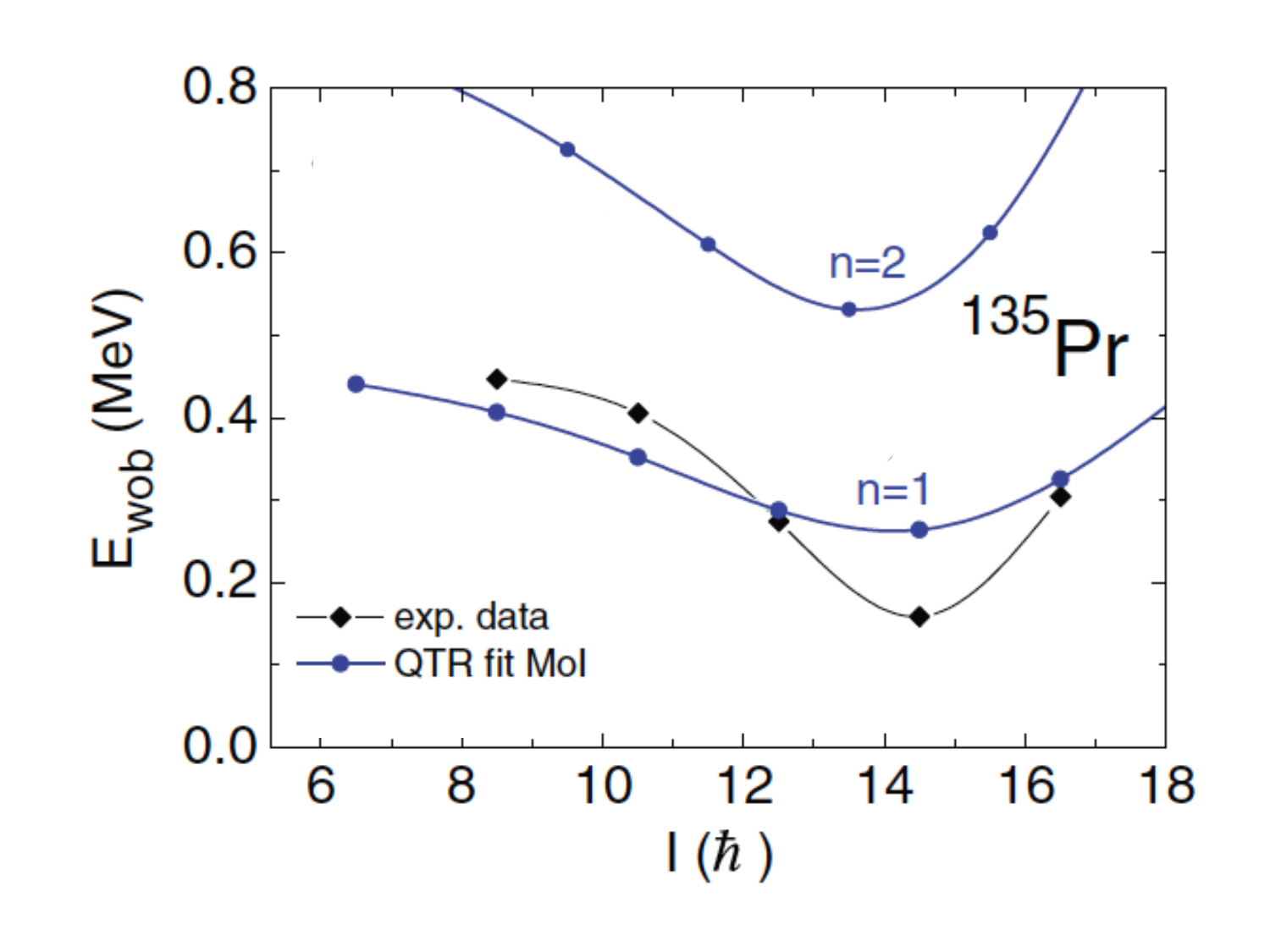}}
\end{minipage}
\hfill\mbox{}\hfill
\begin{minipage}{0.49\textwidth}
 \caption{\label{f:135PrEwobPRC} 
Experimental wobbling energies Eq. (\ref{eq:Ewexp}) of $^{135}$Pr compared with the QTR calculations in Ref. \cite{Frauendorf2014PRC}
 using the parameters $\varepsilon=0.16,~\gamma=26^\circ,~{\cal J}_{m,s,l}=( 21, 13, 4)\hbar^2$/MeV, respectively.
Reproduced with permission from Ref. \cite{Frauendorf2014PRC}.}
\end{minipage}
\end{figure}

\begin{figure}[t]
\center{\includegraphics[width=\linewidth,trim=0 0 0 0 ,clip]{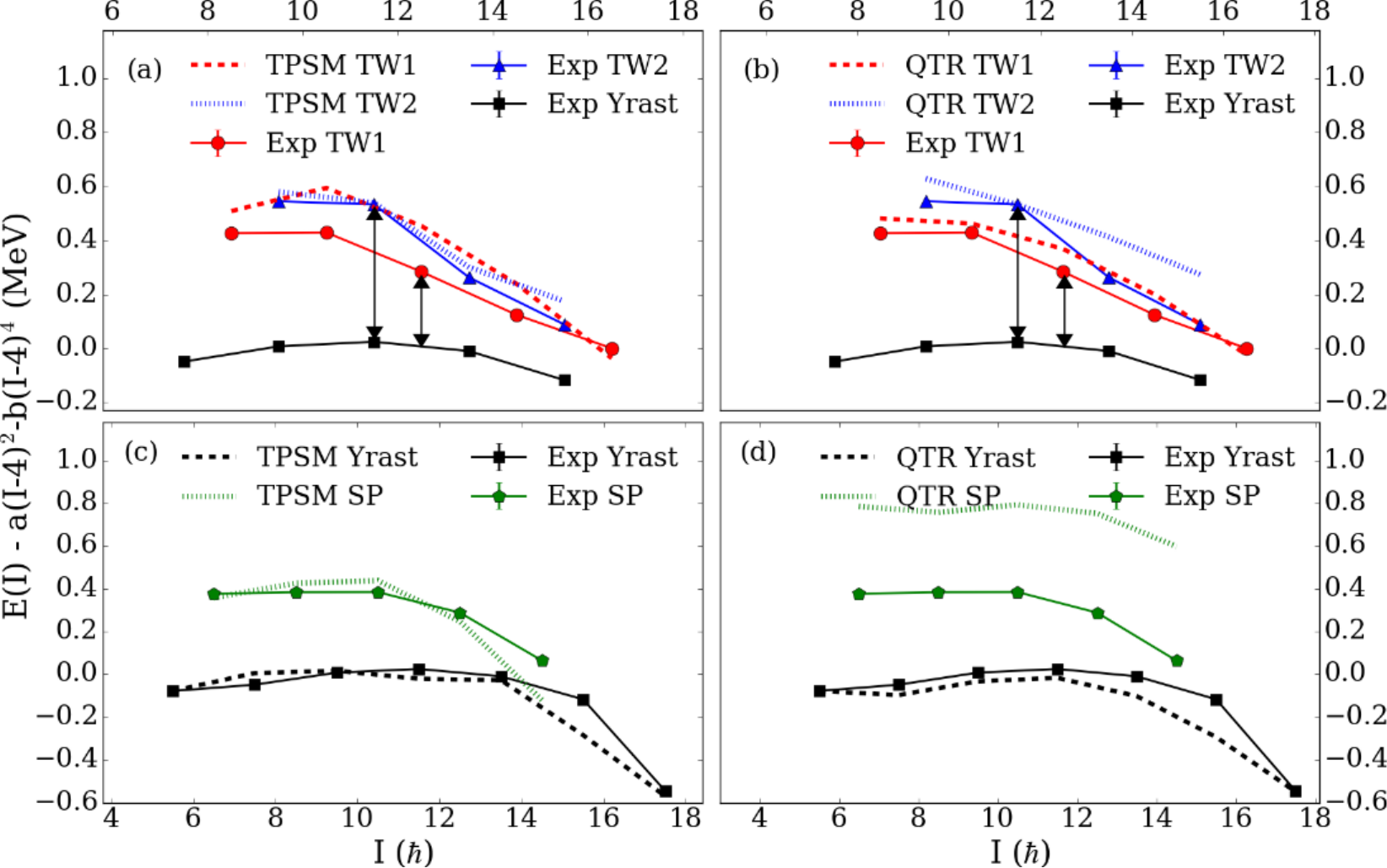}}
 \caption{\label{f:135PrEwob} 
  Experimental level energies minus a rotor contribution \textit{vs.} spin, for the TW1, TW2, Yrast and SP bands in $^{135}$Pr  compared with the corresponding  TPSM and QTR 
 (labelled  as QTR) calculations.  The double pointed arrows between the experimental TW2, TW1 and Yrast points indicate the wobbling energy for the TW2 and the TW1 bands, respectively.
  Reproduced with permission from Ref. \cite{135pr-2019PLB}.}
\end{figure}

\begin{figure}
\center{\includegraphics[angle=0,width=\linewidth,trim=0 0 0 0 ,clip]{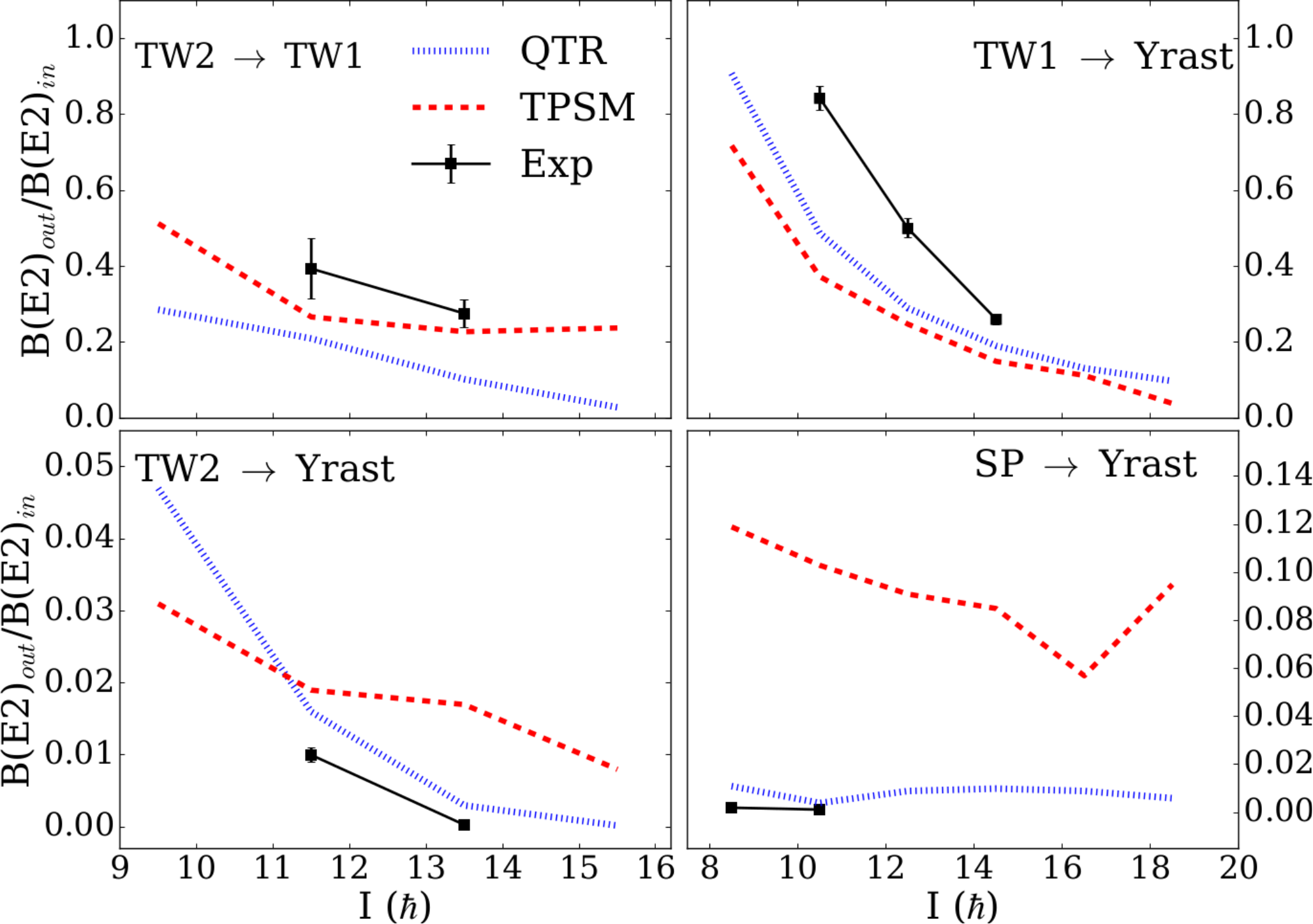}}
 \caption{\label{f:135PrBE} 
  B(E2)$_\text{out}$/B(E2)$_\text{in}$ \textit{vs.} spin for the TW2 $\rightarrow$ TW1, TW1 $\rightarrow$ Yrast, TW2 $\rightarrow$ Yrast and the SP $\rightarrow$ Yrast transitions. (TPSM (red dashed line), QTR (dotted line) and experiment (squares connected by 
  solid line).  Reproduced with permission from Ref. \cite{135pr-2019PLB}.}
\end{figure}

The calculations motivated U. Garg's group to measure the mixing ratios of the $I\rightarrow I-1$ transitions
 between the suggested wobbling  bands \cite{135pr-2015PRL,135pr-2019PLB,135pr-erratum}. 
 Figs. \ref{f:135PrEwob} and \ref{f:135PrBE} compare the results of the measurements  with a slightly modified QTR  calculation. 
 As in  Fig. \ref{f:135PrEwobPRC},  the excitation energy of the first (TW1) wobbling state  decreases below and increases above $I_c$. 
The distance between TW1 and the second (TW2) wobbling state is much smaller than the excitation energy of TW1, which indicates strong anharmonicities. 
The   ratios \mbox{ $\frac{B(E2,I\rightarrow I-1)_{out}}{B(E2,I\rightarrow I-2)_{in}}$} show the enhancement that signifies  the collective nature of the wobbling mode.
The ratios \mbox{$\frac{B(E2, TW2\rightarrow TW1)}{ B(E2, TW1\rightarrow yrast )}\sim 1$} indicate strong anharmonicities as well. 
The ratios \mbox{$\frac{B(E2, TW2\rightarrow yrast)}{B(E2,TW1\rightarrow yrast)} <0.05$} are small as expected.
The QTR calculations also describe  the  excitation of the odd particle, the signature partner (SP) band. As it should be, its $\frac{B(E2,I\rightarrow I-1)_{out}}{B(E2,I\rightarrow I-2)_{in}}$  
  ratios are very small like in the experiment.
However, the QTR places the SP band at too high  energy.

  \begin{figure}[t]
\center{\includegraphics[width=\linewidth,trim=0 0 0 0 ,clip]{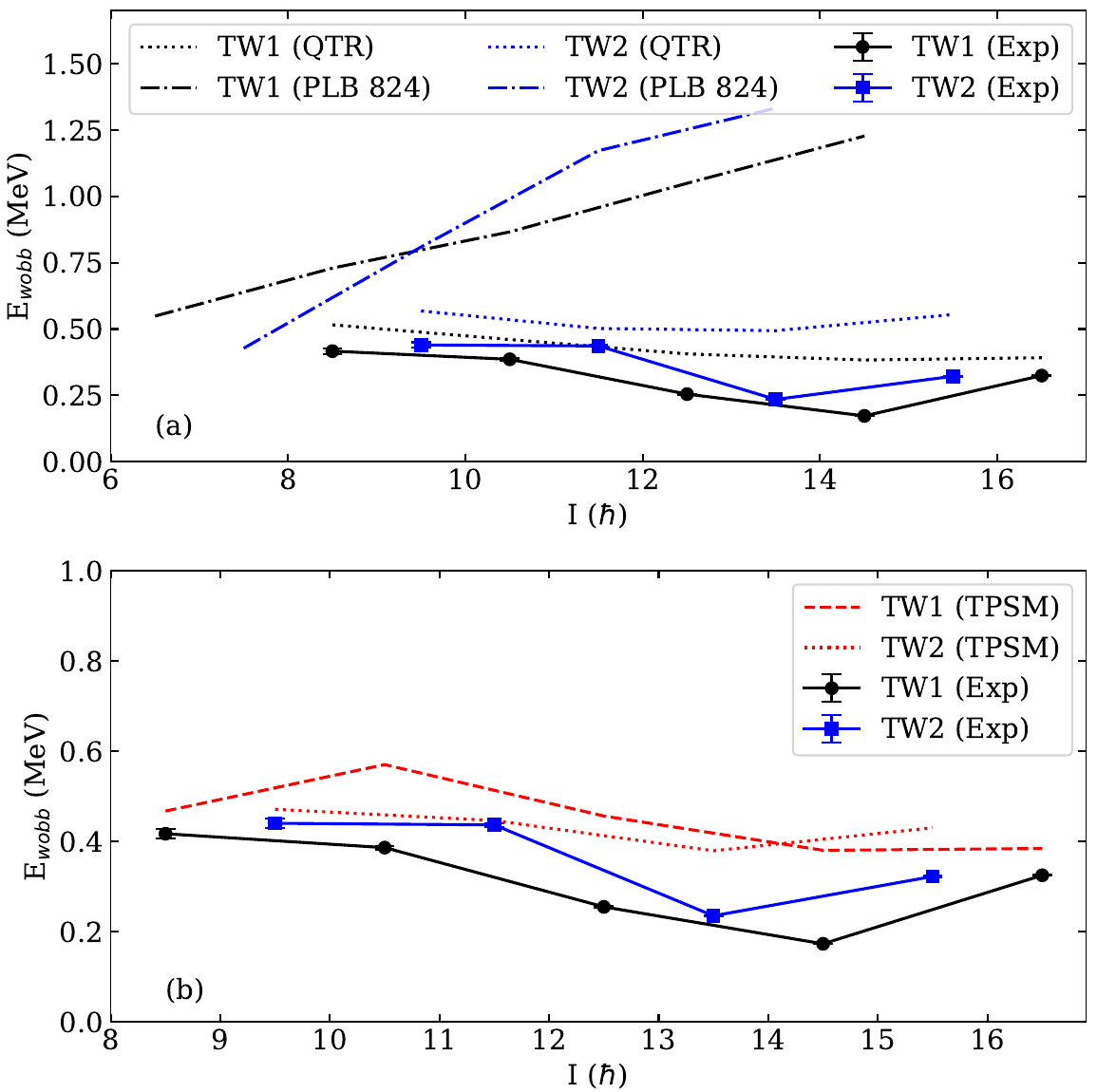}}
 \caption{\label{f:135PrEwobb} Wobbling energies $E_W(I)$ for $^{135}$Pr compared with calculations. The labels TW1 and TW2 refer to the first and second band above the yrast band. 
 QTR labels the calculations of Ref. \cite{135pr-2019PLB}, which uses fitted MoI's. Ref. \cite{135pr-Lv}. PLB 824 labels the QTR calculations of Ref. \cite{135pr-Lv}, which used irrotational
 flow MoI. The pertaining calculated small ratios $\frac{B(E2,I\rightarrow I-1)_{out}}{B(E2,I\rightarrow I-2)_{in}}$ ratios indicate that TW1  has the character of a signature partner band. TPSM labels the triaxial projected shell model calculations from Ref. \cite{135pr-2019PLB}.
Reproduced with permission from Ref. \cite{Sensharma2024}.}
\end{figure}

 Lv {\it et al.} \cite{135pr-Lv}  reinvestigated  $^{135}$Pr. In contrast to Refs. \cite{135pr-2015PRL,135pr-2019PLB,135pr-erratum} they determined small  mixing ratios $\vert \delta \vert <1$,
 which indicate small   $\frac{B(E2,I\rightarrow I-1)_{out}}{B(E2,I\rightarrow I-2)_{in}}$ ratios and large $\frac{B(M1,I\rightarrow I-1)_{out}}{B(E2,I\rightarrow I-2)_{in}}$ ratios.
 They presented the results as "Evidence against the wobbling nature of low-spin bands in
 $^{135}$Pr", because large collective $E2$ transitions  between the bands are the decisive signature for TW. 
  In measuring the mixing ratios the challenge is to decide between the two minima of  $\chi^2$ from the fit to the angular distributions.  The authors of Ref. \cite{135pr-Lv} 
  chose the minimum with small $\vert \delta \vert$ the authors of Refs. \cite{135pr-2015PRL,135pr-2019PLB,135pr-erratum} chose the one with large $\vert \delta \vert$.
  In Ref. \cite{Sensharma2024} Sensharma {\it et al.}
 present additional details of their fit procedure which support  their choice of the large $\vert \delta \vert$. 
 
 The QTR calculations in  Ref. \cite{135pr-Lv} account for the  small $\frac{B(E2,I\rightarrow I-1)_{out}}{B(E2,I\rightarrow I-2)_{in}}$ ratios, which reassign the TW1 band
  as a signature partner band.
 The new assignment is supported   by composition of the QTR states discussed in supplementary material of Ref. \cite{135pr-Lv}. The difference between the two
 QTR calculations consists in  the  inertia asymmetry, which is $\kappa=0.71$ for the fitted MoI's of Refs. \cite{135pr-2015PRL,135pr-2019PLB,135pr-erratum} and
  $\kappa=-0.17$ for the irrotational-flow MoI's of Ref. \cite{135pr-Lv}. The small $\kappa$  causes a very early instability of the TW regime  which moves the SP band below the wobbling band.
  The SP band is characterized 
  by the steady increase of $E_W(I)$ seen in the upper panel of Fig. \ref{f:135PrEwobb}. In my view, the stark contrast  with the  observed decrease  $E_W(I)$ makes the SP assignment
  unbelievable. (Private communication of the QTR energies by E. Lawrie is acknowledged.) 
  
   The TSPM calculations in the lower panel of Fig.~\ref{f:135PrEwob} evaluate the rotational response microscopically. 
   There is no freedom in choosing the MoI's.
  The similarity between the QTR of Refs. \cite{135pr-2015PRL,135pr-2019PLB,135pr-erratum} and the TSPM results lends credibility to the adjusted inertia asymmetry ratio of $\kappa=0.71$.
  
 If  the TW1 band were the
  SP band then it would be difficult to explain the presence of a second nearby band with SP properties,  which was  observed in Refs. \cite{135pr-2015PRL,135pr-2019PLB,135pr-erratum}.
  Finally  there is the far-reaching  analogy between $^{135}$Pr and $^{105}$Pd (see next paragraph), for which the 
 carefully measured mixing ratios indicate large $\frac{B(E2,I\rightarrow I-1)_{out}}{B(E2,I\rightarrow I-2)_{in}}$  ratios, which are consistent with the ones of 
 Refs. \cite{135pr-2015PRL,135pr-2019PLB,135pr-erratum}.
 
 In my opinion all the mentioned facts favor the TW interpretation of the bands. A new experiment would be needed to safely establish the experimental $\delta$ values.

 \begin{figure}[t]
\center{\includegraphics[width=\linewidth,trim=0 0 0 0 ,clip]{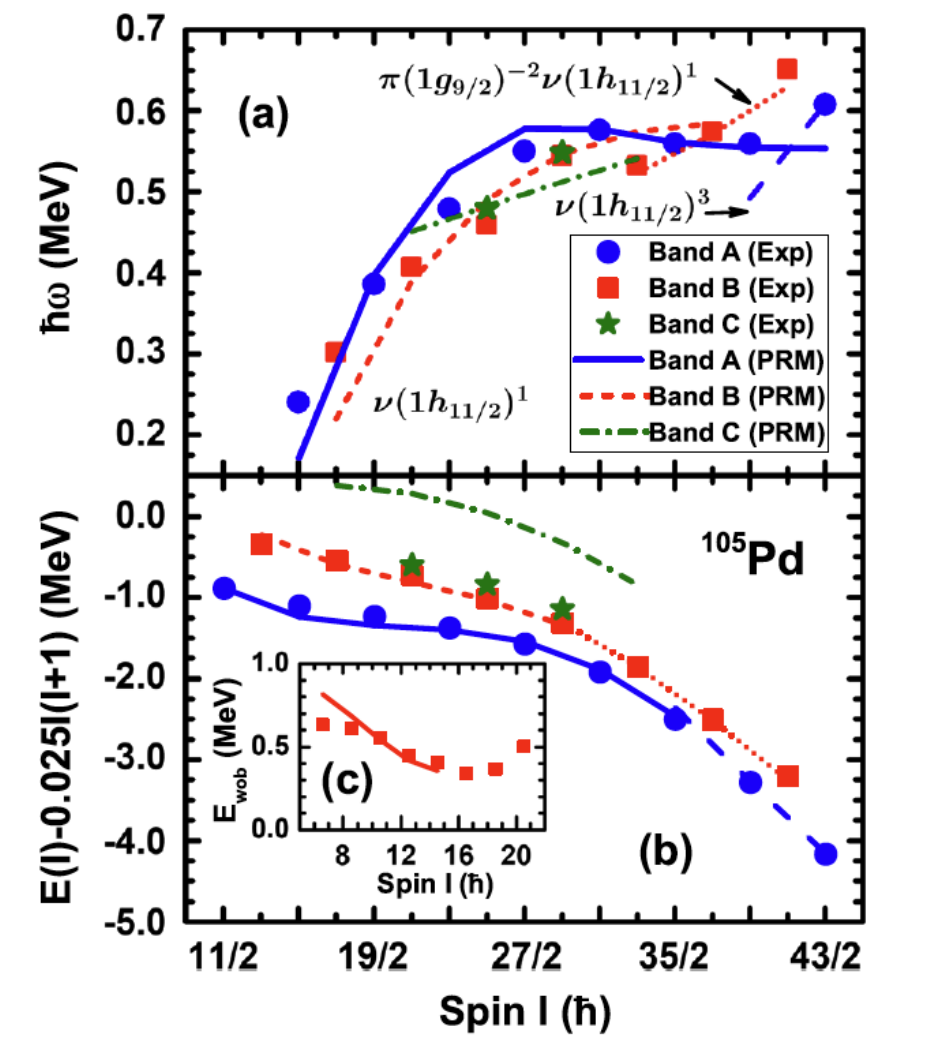}}
 \caption{\label{f:105Pd} 
 Experimental values derived from the energies of lowest bands in $^{105}$Pd compared with the QTR (labelled as PRM)  calculations:
 (a) rotational frequency,  (b) energies minus a rotor reference with  the wobbling energies $E_W(I)$ as the inset (c). The bands A, B and C are assigned
  to, respectively, $n=0$ TW,  $n=1$ TW and SP.
Reproduced with permission from Ref. \cite{Timar1}. 
  }
\end{figure}

\subsubsection{Other examples of transverse wobbling with one excited quasiparticle}

Another example for TW is $^{105}$Pd, which was studied in Ref. \cite{Timar1} . The authors carried out QTR (called PRM) calculations using a triaxial deformation of $\gamma=25^\circ$
derived from mean field calculations and fitted MoI's, which correspond to an inertia asymmetry of $\kappa=0.69$.
Fig. \ref{f:105Pd} compares the experimental results with the QTR calculations, where
the rotational frequency is defined as
\begin{equation}
\hbar \omega(I-1/2)=(E(I)-E(I-2))/2.
\end{equation}
The experimental (PRM) $\frac{B(E2,I\rightarrow I-1)_{out}}{B(E2,I\rightarrow I-2)_{in}}$  ratios for the $n=1 \rightarrow 0$ trrasitions are
 $I=$17/2: 0.66$\pm$0.18 (0.736), 
 $I=$21/2: 0.60 $\pm$0.09 (0.465), 
$I=$25/2:  0.34$\pm$ 0.07 (0.329).

From the PTR perspective the nucleus is much like $^{135}$Pr, just that the odd $h_{11/2}$ proton is
 replaced by the odd $h_{11/2}$ neutron. The slightly lower inertia asymmetry as compared to $^{135}$Pr is reflected by the somewhat higher value of $I_c=33/2$ of the 
 minimum of the wobbling energy. The QTR predicts energy of the SP bands too high as in the case of   $^{135}$Pr. 
 
 A fourth band was reported in Ref. \cite{Timar2}. The experiment
 could not substantiate the nature of the band. Karmakar {\it et al.} \cite{Karmakar2024} measured mixing ratios for the transitions  connecting it with the TW1 band, which 
 did not indicate a strong $E2$ component. The finding is in contrast to  $^{135}$Pr where the connecting transitions of the analog band with TW1  show a strong $E2$ component, 
 which allowed the authors of Ref. \cite{135pr-2019PLB} to identify it as the second wobbling excitation TW2. In $^{105}$Pd band 4 represents a TW excitation on top of the SP band.
 The reason for the difference is the lower energy of the SP band in $^{105}$Pd than in $^{135}$Pr.  As a consequence,
    in  $^{105}$Pd  the TW excitation on the SP band has a smaller energy than the second wobbling excitation, while in $^{135}$Pr  
    the second wobbling excitation has a smaller energy than  the TW excitation on the SP band. 
The microscopic 
TPSM calculations \cite{Karmakar2024,SJB24} account for the different nature of band 4, while the QPR predicts TW2 for band 4 in both nuclei. 
The latter is not surprising because the QTR
 overestimates the energy of the SP band in both nuclei.

Mukherjee {\it et al.} \cite{151Eu} identified in $^{151}$Eu a TW band built on the proton $h_{11/2}$ yrast band together with the pertaining SP band. See section "Soft Core" for an analysis 
in the framework of the TPSM.

Nandi {\it et al.} \cite{183AuTW} identified in  $^{183}$Au a TW band built on the proton $h_{9/2}$ band and another TW band built on the proton $i_{13/2}$ band. 
The wobbling energies from the $h_{9/2}$ configuration decrease rapidly corresponding to a $I_c\approx 14$. The wobbling energies from  the $i_{13/2}$ configuration 
slowly increase 
like the PTR $E_W$ in Fig. \ref{f:163LuEwob} far below $I_c\approx 30$. The authors used different MoI's to reproduce the  different  $I_c$ and the resulting 
$I$-dependences of $E_W(I)$, which also provided the
correct $\frac{B(E2,I\rightarrow I-1)_{out}}{B(E2,I\rightarrow I-2)_{in}}$  ratios. Remarkably, the microscopic TPSM reproduces these quantities of both TW bands
with nearly the same set of input deformations. See Fig. 12.11 and 12.14 of Ref. \cite{SJB24}.

Rojeeta Devi {\it et al.} \cite{133BaTW} confirmed the prediction \cite{Frauendorf2014PRC} that TW appears for high-hole states as well, where the $\mathbf{J}$ precesses about the $l$-axis.
They identified in $^{133}$Ba the first and second TW bands built on the neutron $h_{11/2}$ hole band and carried out a qualitative analysis in HFA approximation.

 \begin{figure}[h]
\center{\includegraphics[width=\linewidth,trim=0 0 0 0 ,clip]{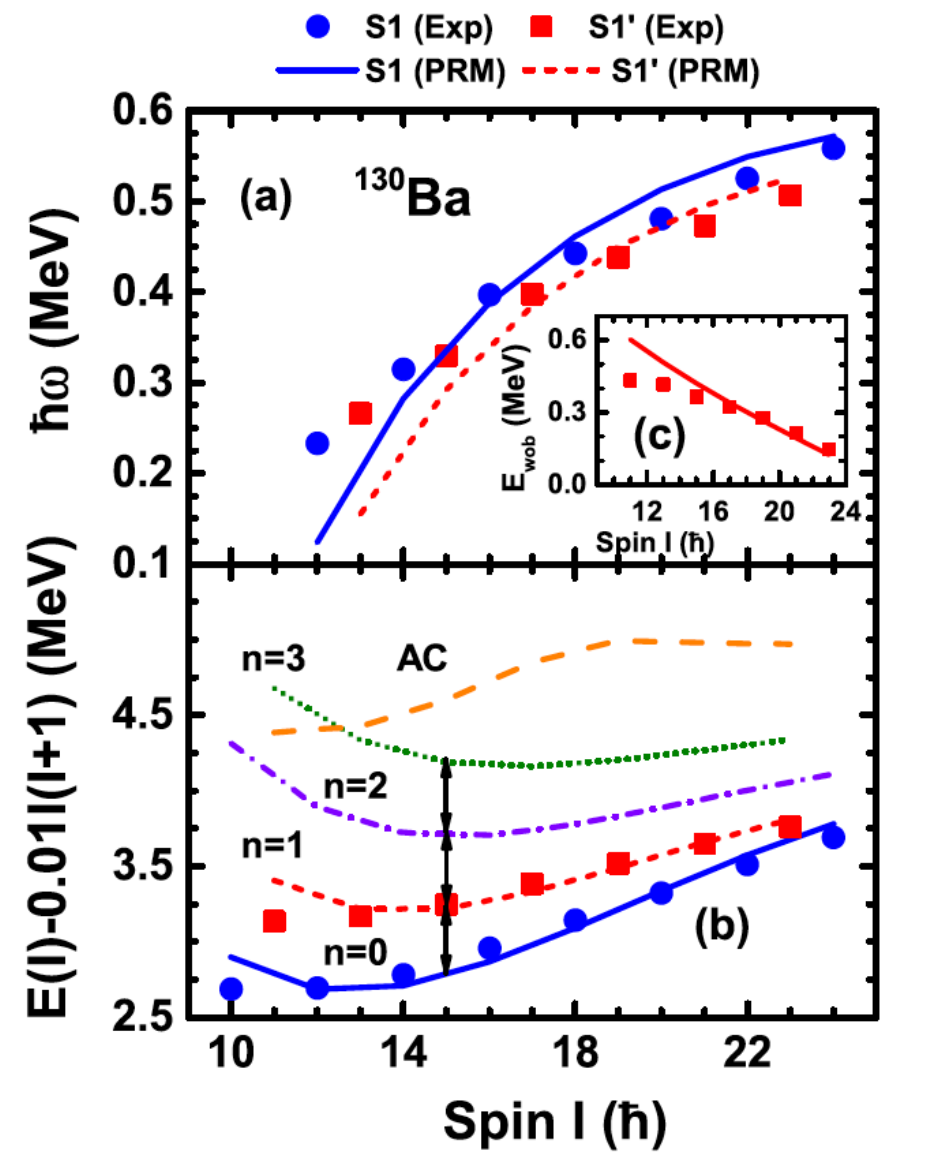}}
 \caption{\label{f:130BaEwob} 
 Experimental values derived from the energies of the bands S1, S1' and AC in $^{130}$Ba compared with the QTR (labelled as PRM)  calculations:
 (a) rotational frequency,  (b) energies minus a rotor reference with  the wobbling energies $E_W(I)$ as the inset (c).
 The QTR calculations used the parameters 
$\varepsilon=0.24,~\gamma=21^\circ,~{\cal J}_{s}/{\cal J}_{m}/{\cal J}_{l}=( 0.73,1,0.43)\hbar^2$/MeV, respectively.
 Reproduced with permission from Ref. \cite{130BaTW}.}
\end{figure}

 \subsubsection{ Transverse wobbling with two excited quasiparticles}

The TW regime is more stable when two particles align their angular momentum with the s-axis, as the  value of $j$ in $I_c$ doubles.
The first example of TW on a  $\pi(h_{11/2})^2$ configuration was reported by Chen {\it et al.} \cite{130BaTW} for $^{130}$Ba. 
The even-$I$ band S1, which is AB in CSM notation, was interpreted as the $n=0$ TW band. The odd-$I$ band S1'  was assigned to the $n=1$ TW band. 
The  second odd-$I$ band, called AC in CSM notation, represents the SP band, because the quasiproton B is replaced by  C with the opposite signature.
  They compared the experimental data 
with QTR (labelled as PRM) calculations using  the deformation parameters from a mean field calculation and adjusting the MoI's to fit the energies of the $n=0$ and 1 bands.

Fig. \ref{f:130BaEwob} compares  the rotational frequencies and the 
 energies of the  bands S, S' and AC with the PRM results, which agree rather well.
 The experimental mixing ratios of the $I\rightarrow I-1$ transitions between the $n=1$ and $n=0$ bands  allowed the authors 
determining    $\frac{B(E2,I\rightarrow I-1)_{out}}{B(E2,I\rightarrow I-2)_{in}}$ ratios between 0.3 and 0.4, which signify the wobbling character of the $n=1$ band.
They are consistent with the
 PRM  ratios $\frac{B(E2,I\rightarrow I-1)_{out}}{B(E2,I\rightarrow I-2)_{in}}$=0.51,~0.42, 0.35, 0.29, 0.25 for $I$=13, 17, 19, 21, respectively. The $E2$ component of the transition from the AC band to the $n=0$ (AB) band was found to be small as expected for the SP band. 

The wobbling energy in Fig. \ref{f:130BaEwob} shows that that the TW regime is stable within the displayed spin range ($I_c$=37). 
Fig. \ref{f:130BaSCS} displays the SCS probability distributions for $I=14$ and 15. The $n=0$ and AC states correspond to uniform rotation about the s-axis. 
The $n=1$ and $n=2$ states show the precession orbits around the s-axis, which characterize the TW regime.

\begin{figure}[h]
\center{\includegraphics[angle=0,width=\linewidth,trim=0 0 0 0 ,clip]{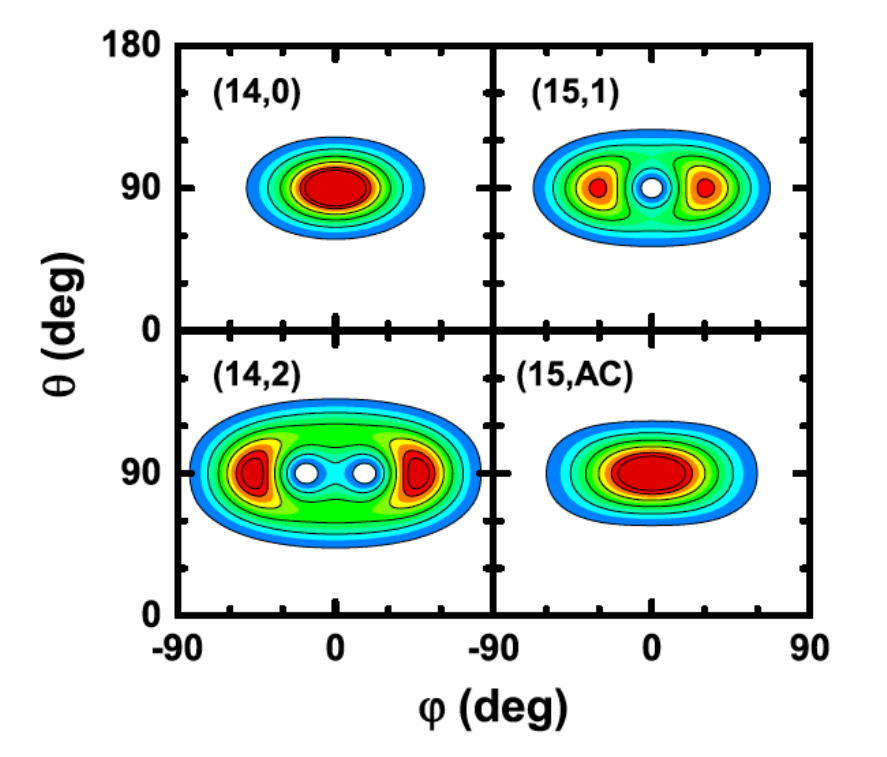}}
 \caption{\label{f:130BaSCS} 
 SCS  probability distributions $P(\theta,\phi)$ 
of the angular momentum orientation $\mathbf{J}$ with respect to the body-fixed frame 
 for the four lowest bands in Fig. \ref{f:130BaSCS}. 
The color sequence with increasing probability is dark blue - green - dark brown. 
   In accordance with Fig. \ref{f:130BaEwob}, the panels are labeled by $(I,n)$ with $n$ being  TW wobbling number or AC.
Reproduced with permission from Ref. \cite{130BaTW}.}
\end{figure}

 A second example for TW on the  $\pi(h_{11/2})^2$ configuration is $^{136}$Nd, which was studied by Chen and Petrache \cite{136NdTW}.

\subsection{Transition from transverse to longitudinal wobbling}\label{s:TWLWND}

 The FA approximation provides  only a  first qualitative TW-LW classification. The  $\mathbf{j}$ of the odd particle  reacts to wobbling motion of the nucleus,
which  must be taken into account to understand the region where the TW becomes unstable and beyond.  For this purpose Chen and Frauendorf 
\cite{Chen_Frauendorf2022EPJA} invoked the classical adiabatic energy. The classical energy is defined by replacing in the PTR Hamiltonian
 (\ref{eq:HPTR},\ref{eq:hproton})  the operators by their corresponding  classical numbers,
\begin{eqnarray}
\hat J_3 \rightarrow J\cos\theta,~~\hat J_1 \rightarrow J\sin\theta\cos\phi,~~\hat J_2 \rightarrow J\sin\theta\sin\phi,\\
\hat j_3 \rightarrow j\cos\vartheta,~~\hat j_1 \rightarrow j\sin\vartheta\cos\varphi,~~\hat j_2 \rightarrow j\sin\vartheta\sin\varphi.
\end{eqnarray}

The  adiabatic classical energy $E_{ad}(\theta,\phi)$  is obtained by minimizing the classical energy $E_{class}(\theta,\phi,\vartheta,\varphi)$ 
with respect to $\vartheta$ and $\varphi$ for given $\theta,~\phi$, that is, let the particle react to the Coriolis force. It is displayed Fig. \ref{f:135PrEad} for the PTR Hamiltonian of $^{135}$Pr. 
The energy contours represent the classical orbits when $\mathbf{j}$ adiabatically follows $\mathbf{J}$,
which is the case when the energy scale of
the particle is large compared to the scale of the wobbling mode. As demonstrated in Ref. \cite{Chen_Frauendorf2024}, adiabticity is a reasonable approximation 
for the lowest three bands.   See Figs. \ref{f:135PrSSS} and \ref{f:130BaSSS} below.

\begin{figure}[t]
\center{\includegraphics[angle=0,width=\linewidth,trim=0 0 0 0 ,clip]{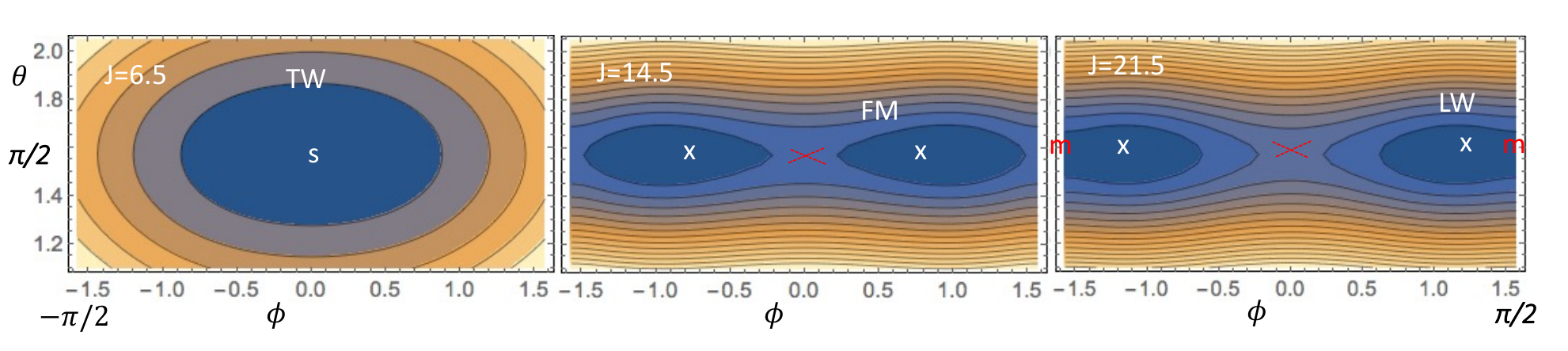}}
 \caption{\label{f:135PrEad} Adiabatic classical energy $E_{ad}(\theta,\phi)$ in $^{135}$Pr for the TW, FM and LW regimes.
 The contours of constant energy represent the classical orbits 
 with the corresponding energy and the indicated angular momentum. 
 Minima are marked by white crosses and saddles by red crosses. 
 Minima and saddles are marked by crosses. 
 The letters s and m indicate the location of the s-and $m$-axes.
}
\end{figure}

\begin{figure}[h]
\center{\includegraphics[angle=0,width=\linewidth,trim=0 0 0 0 ,clip]{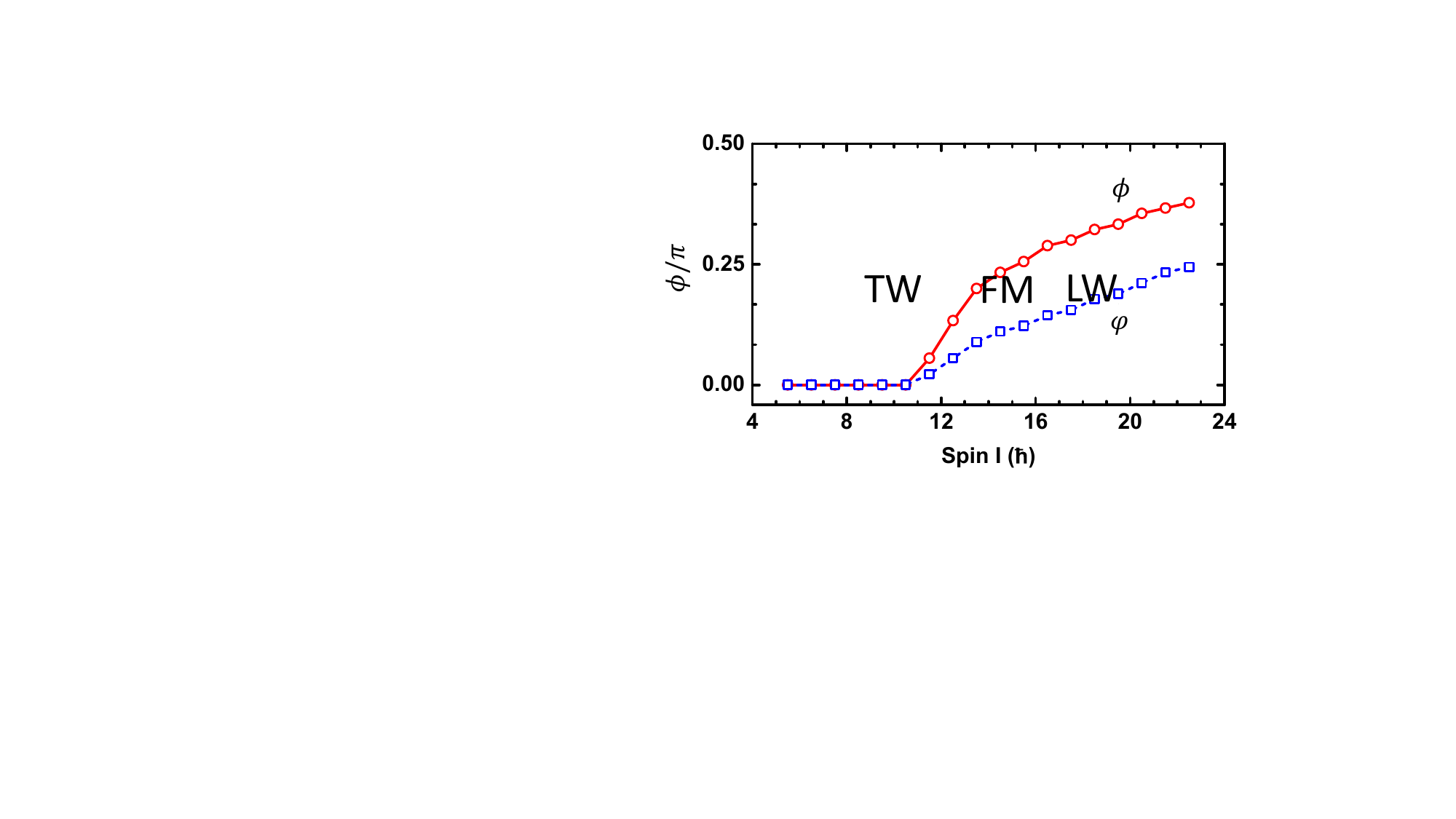}}
  \caption{\label{f:135Pr-angles} Location $\phi_0$ of the minima of the adiabatic classical energy $E_{ad}(\theta,\phi)$ in $^{135}$Pr on the line $\phi=\pi/2$ in Fig. \ref{f:135PrEad} (marked by the
  white crosses).
 The pertaining angles of $\mathbf{j}$ are $\vartheta=\pi/2$ and $\varphi$.
Reproduced with permission and adapted from Ref. \cite{Chen_Frauendorf2022EPJA}.}
\end{figure}

\begin{figure}[h]
\center{\includegraphics[width=\linewidth,trim=0 0 0 0 ,clip]{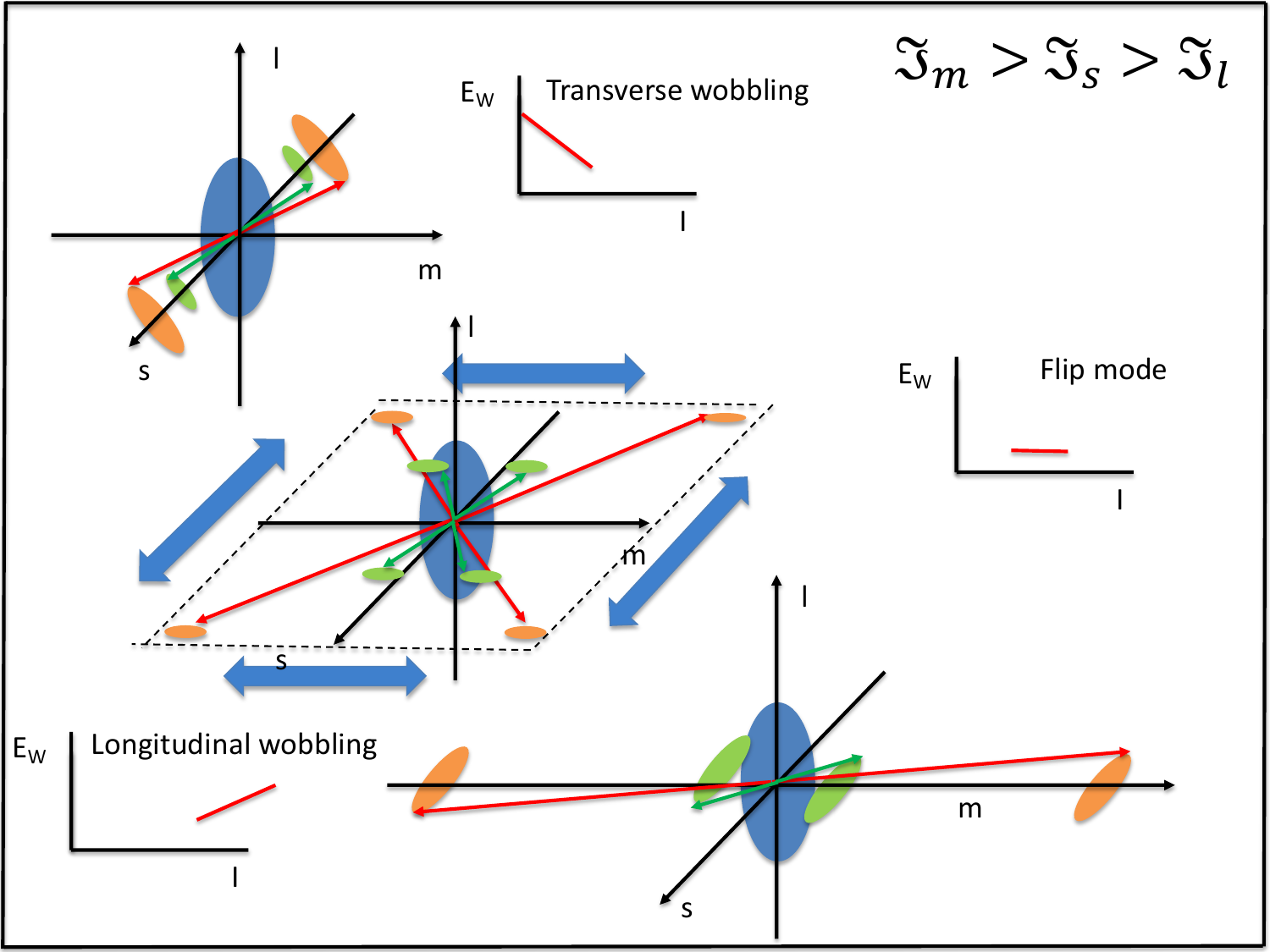}}
 \caption{\label{f:TW-FM-LW} Schematic illustration of the collective wobbling modes.  
 Green arrows show the 
 precession cones of the particle $\mathbf{j}$ and red arrows the precession cones of the total angular momentum $\mathbf{J}$.
 }
\end{figure}

The $I$ dependence reflects the competition between the Coriolis force, which tries to keep the angle between $\mathbf{J}$ and $\mathbf{j}$ small, 
the triaxial potential, which tries to align  $\mathbf{j}$ with the s-axis and the energetic preference of the $m$-axis with
 the maximal MoI. Fig. \ref{f:135Pr-angles} shows the angles $\phi_0$ and $\varphi$ of the minimum of the adiabatic energy.
 
At low $I$ the classical orbits revolve the s-axis, which is the TW regime. 
At the critical  angular momentum of $J_c=10.5$ the $\phi=0$ minimum changes into a maximum and two minima at $\vert \phi\vert>0$ emerge. 
The rotation about the s-axis becomes unstable.
The TW regime changes into the flip mode (FM), which becomes well established at $J=14.5$. The FM represents flipping between the narrow precession cones around the
two tilted axes. Quantum mechanically, the $n=0$ state is an even  and the $n=1$ state  an odd superposition of two states
 localized near the tilted axes. As the two minima at $\phi_0$
approach $\pm \pi/2$ the orbits merge with the ones at $\pi\mp\phi_0$ such that the precession cones revolve around the $m$-axis, 
which means the region of LW has been reached.

The flip regime appears around  $I=14.5$ where $E_W$ has its minimum  in Fig. \ref{f:135PrEwob}. At larger $I$ values $E_W(I)$ increases with $I$.
As seen in Fig. \ref{f:135Pr-angles}, the angle $\varphi$ of the particle $\mathbf{j}$ remains far below $\pi/2$ in this region. Therefore the authors of Ref.  \cite{Chen_Frauendorf2022EPJA}
suggested a generalization of the TW-LW classification of Ref. \cite{Frauendorf2014PRC}, which is based on the topology of the classical orbits that correspond to the quantal states: 
\vspace*{0.5cm}

\begin{table*}
\begin{tabular*}{12.5cm}{|c|c|c|c|c|}
\hline
\hline
Classification  of the &$\mathbf{J}$ revolves &$\mathbf{j}$ revolves &$E_W(I)$&$\frac{B(E2,I\rightarrow I-1)_{out}}{B(E2,I\rightarrow I-2)_{in}}$ \\
lowest bands&around axis&around  axis&&\\
\hline
transverse wobbling (TW) & short or long & short or long& decreases$^*$ & order one\\
longitudinal wobbling (LW)& medium& short, long, tilted& increases &order one\\
flip mode (FM) & tilted& tilted & constant & order one\\
signature partner (SP)& short, medium, long&short, medium, long&increases& small\\
\hline
\hline
\end{tabular*}
\vspace*{0.5cm}\\
$^*$ Depending on $I_C$ there may be  a slight increase at low $I$.\\
\end{table*}

The scheme is quite simple from the experimental point of view and has been used in Refs. \cite{Frauendorf2014PRC,135pr-2015PRL,135pr-2019PLB} in classifying the experimental results.
Fig. \ref{f:TW-FM-LW} illustrates the angular momentum geometry in a schematic way.

The energy difference between the $n=0$ and $n=1$ states of the flip mode reflects the mixing of the states localized near the tilted axis. It disappears when the coupling goes to zero.
This limit represents uniform rotation about a tilted axis, which breaks the signature symmetry and leads to a $\Delta I =1$ band (see discussion of the rotating mean field symmetries in Ref. \cite{SFrmp}).

\begin{figure}[t]
\includegraphics[width=\linewidth]{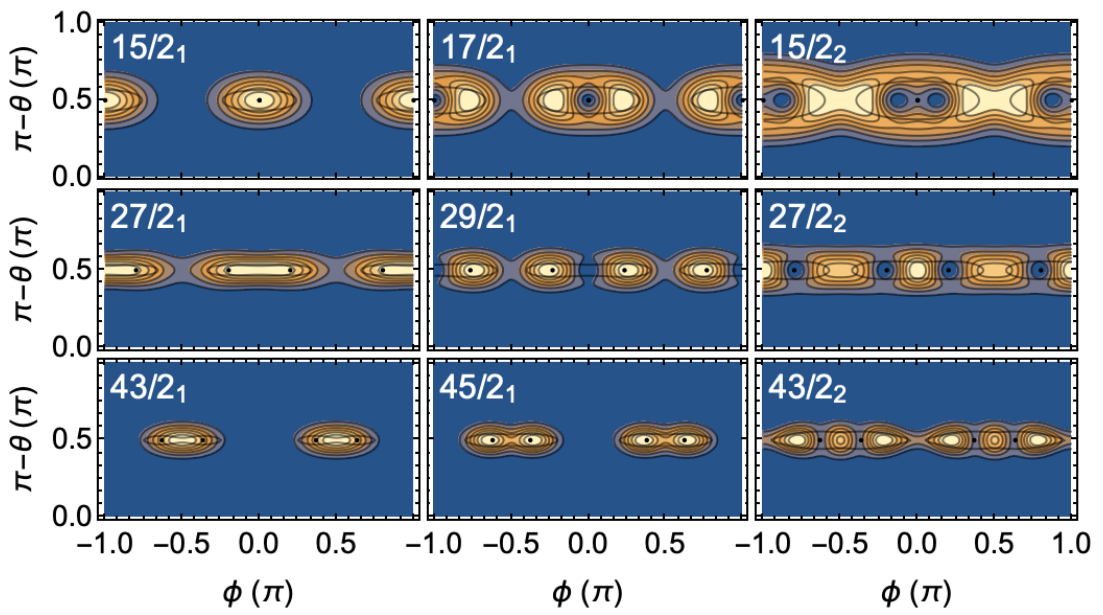}
 \caption{\label{f:135PrSCS} Probability distribution of the angular momentum orientation $ P(\theta \phi)_{I\nu} $  
 generated from the reduced density matrix of the PTR calculation  for $^{135}$Pr in Ref. \cite{Frauendorf2014PRC}. 
 The three rows show the SCS maps for the TW, FM, and LW regimes. The black curves show the classical orbits obtained by setting
  the adiabatic classical energy equal to the quantal energy of the states.
  The author thanks Q. B. Chen for preparing the figure. }
\end{figure}

Fig. \ref{f:135PrSCS} shows the SCS probability distribution of the $\mathbf{J}$ with respect to the principal axes for the lowest three bands $n=$0, 1, 2.
 The classical orbits (as shown in  in Fig. \ref{f:135PrEad}) for the quantal energies of the states are included.

 The SCS maps in the upper row show rims that are centered at the s-axis, which are the hallmark of TW. 

The middle row 
shows the  four spots near the tilted axes between the FM  flips. The $27/2_1$ state represents the even superposition (large $P(0)$) and the $29/2_1$ state the odd superposition  ($P(0)=0$)
of the flip states. The  $27/2_2$ state flips between the s- and $m$-axes, which are both unstable. The structure is analog to the  flip mode of the unstable s-axis of the TR  in "Discussion of the Triaxial Rotor Dynamics".

The lower row displays the LW regime. The very elongated orbits enclose the $m$-axis at $\pm \pi/2$, which is the topological characteristics of LW.  
 They differ substantially from the HFA-LW limit, because the minima of $E_{ad}$ are still away from the $m$-axis (see Fig. \ref{f:135Pr-angles}). 
 
\begin{figure}[t]
\center{\includegraphics[angle=0,width=\linewidth,trim=20 0 0 0 ,clip]{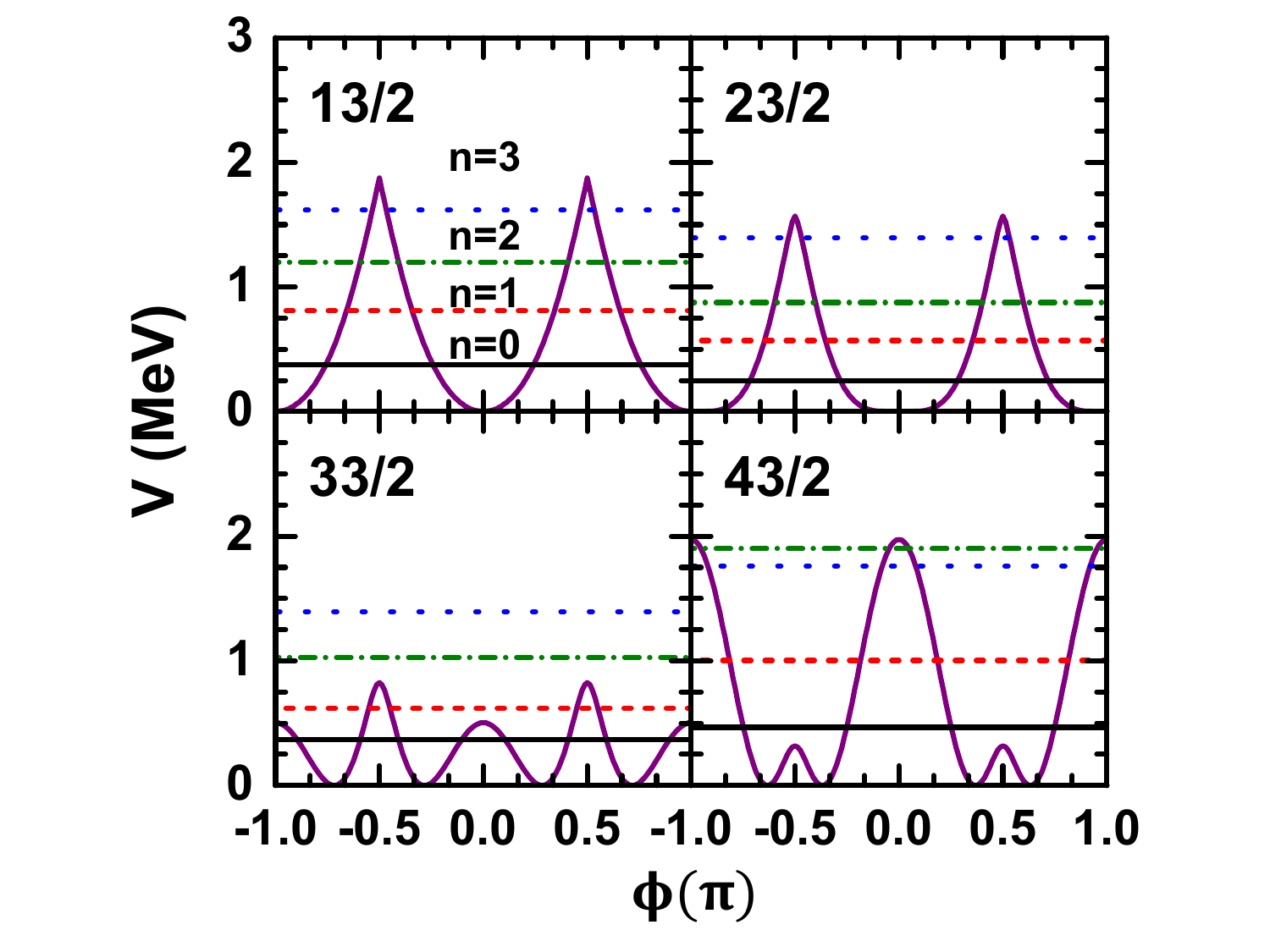}}
 \caption{\label{f:Vad} Classical adiabatic energy  of $^{135}$Pr 
 $V(\phi)=E_\textrm{ad}(\pi/2,\phi)$ for selected spins. 
 The horizontal lines show the PTR energies  $E_n-E_{ad}(\phi_0)$
 relative to the  minimum of $E_{ad}$ at $\phi_0$. For the two values of $n$ that do not appear for  $I$
 the line is located at \mbox{$(E_n(I-1)+E_n(I+1))/2$}.
Reproduced with permission and adapted from Ref. \cite{Chen_Frauendorf2022EPJA}.}
\end{figure}

\begin{figure}[h]

\center{
\includegraphics[angle=0,width=\linewidth,trim=0 0 0 0 ,clip]{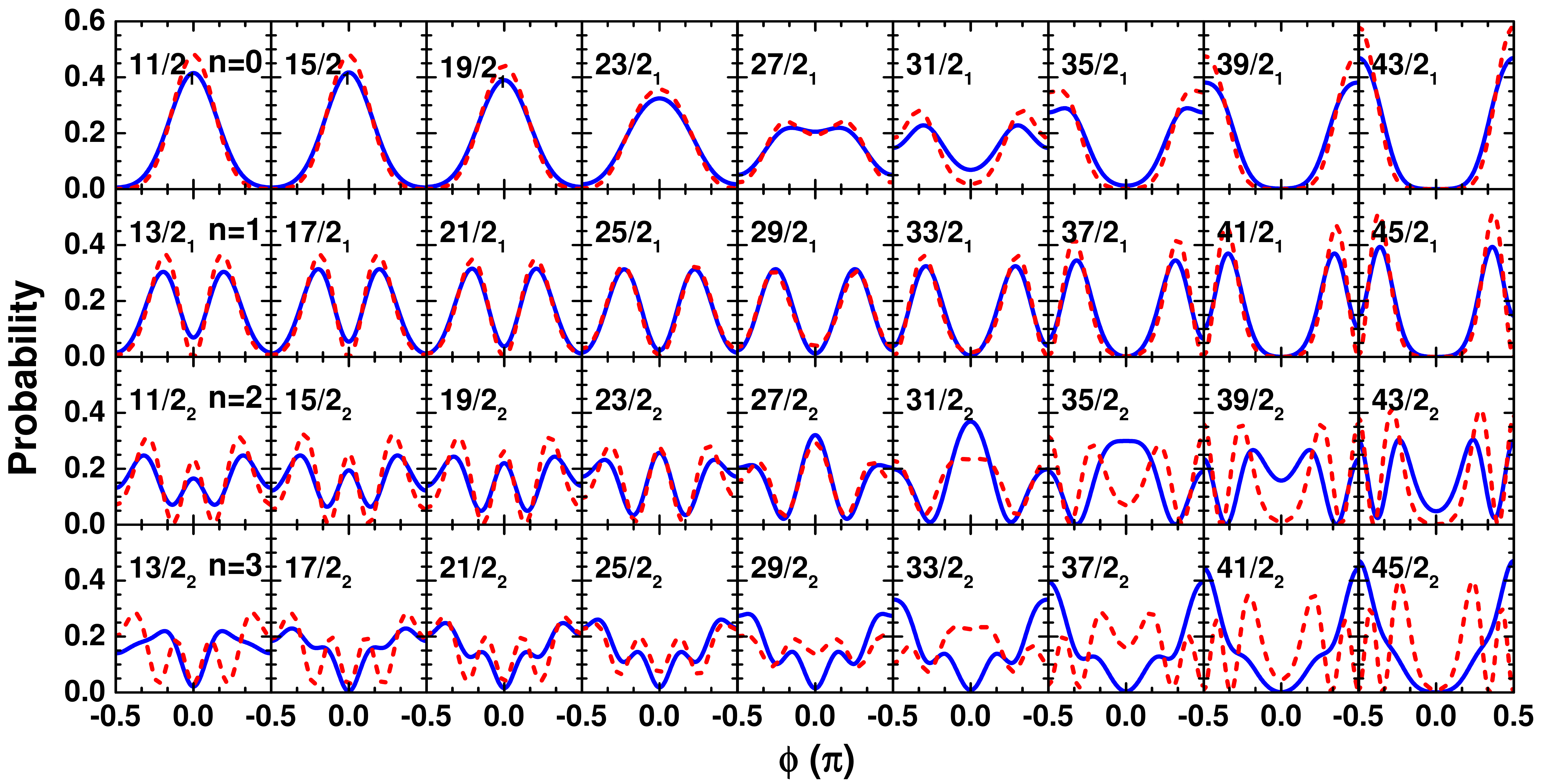}}
 \caption{\label{f:135PrSSS}
 Full curves show the probability density of the SSS for the $n$ = 0, 1, 2, 3 states calculated by PTR for $^{135}$Pr, 
while the dashed curves display the ones from the collective Hamiltonian (CH). Reproduced with permission Ref. \cite{Chen_Frauendorf2024}.}
\end{figure}

For the classical TR the angular momentum component $J_3$ and the angle of $\mathbf{J}$ with the $s$-axis in the s-$m$- axes plane represent
the momentum and coordinate of a pair of canonical variables. This provides a  complementary perspective on the structure of the PTR states.
It employs the correspondence between the classical motion in a potential and the probability density of the corresponding quantal Hamiltonian,
which is familiar from the textbooks. The classical adiabatic energy at $E_{ad}(\pi/2,\phi)$ represents the potential  $V(\phi)$. For not too high energy
one can expand the classical adiabatic energy up to second order in $\theta-\pi/2$, which gives the classical adiabatic Hamiltonian

\begin{equation}\label{eq:Had}
H_{ad}=T+V(\phi),~~T=\frac{J_3^2}{2B(\phi)},~~\frac{1}{B(\phi)}=\frac{1}{I^2}\frac{\partial^2 E_{ad}(\theta,\phi)}{\partial \theta^2}\Bigr\rvert_{\pi/2}.
\end{equation}

The potential  is shown in  Fig. \ref{f:Vad} together with bars at  the PTR energies relative to the  minimum of $V(\phi)$. The difference represents the kinetic energy T. If $T>0$ the wave function
oscillates with a wavelength that decreases with $T$. If $T<0$ the wave function decreases exponentially with  $\vert T\vert$. The value of $T$ defines the TW-FM-LW classification.
 In the TW regime ($I$=13/2, 23/2) the s-axis is in the range $T>0$, and in the LW regime ($I$=43/2) the $m$-axis is in the range $T>0$.  In the FM the $n$ =0 and $n$=1 states are localized 
 around $\phi\approx \pm \pi/4$. The mode flips  between rotation about the two orientations of the tilted axis.

\subsection{Spin squeezed states}
Chen and Frauendorf introduced the  Spin Squeezed State (SSS) representation 
of the reduced density matrix \cite{Chen_Frauendorf2024}. 
\begin{align}\label{eq:PSSS}
 P(\phi)_{\nu}= \frac{1}{2 \pi}
  \sum\limits_{K,K'=-I}^Ie^{-i(K-K')\phi}\rho^{(\nu)}_{KK'},
\end{align}
which  is real because the density matrix is symmetric. 
The parameter  $\phi$ is the the angle of angular momentum $\bm{J}$  with the $s$-axis in the $s$-$m$-plane.
The over-complete, non-orthogonal basis comes as close to a continuous wave function as possible  for the finite Hilbert space of dimension $2I+1$.  
It is generated by rotating the localized state 
\begin{equation}\label{eq:LSSS}
\vert II0\rangle=(2I+1)^{-1/2} \sum\limits_{K=-I}^I \vert IIK\rangle
\end{equation}
 around the 3-axis. It has a smaller width of $\sim 3/2I$ as compared to the 
width $\sim1/\sqrt{2I}$ of the generating state of the SCS basis. Fig. \ref{f:135PrSSS} shows the SSS plots of some states  for $^{135}$Pr, some of which are displayed in the form 
of SCS maps in Fig. \ref{f:135PrSCS}. 

The TW states $I<25/2$  have the profile of oscillations within the potential centered at the s -axis.
The $n=0$ states have a maximum  at $\phi=0$.
The $n=1$ states have a minimum at $\phi=0$, and the $n=2$ states have two minima at $\phi=\pm \pi/6$. If the states were pure, the minima would be the zeros of the collective wave function that
characterize the oscillations.
There is a certain degree of incoherence caused by the deviations from the adiabatic approximation, which lead to a finite probability density at the minima (see discussion in 
subsection "Approximate solutions of the PTR model" paragraph "Collective Hamiltonian").

The LW states $I>37/2$ have the profile of oscillations within the potential centered at the $m$-axis: 
The $n=0$ states have maxima at $\phi=\pm \pi/2$.
The $n=1$ states have minima at $\phi=\pm \pi/2$, and the $n=2$ states have two minima near $\phi=\pm 4/6 \pi$. 

The FM states around $I=31/2,~33/2$ have flip character. The  $n=$0 and 1 are the respective even and odd linear combinations of the two states localized at $\phi\approx\pm\pi/4$.
The $n=2$ states are standing waves with probability maxima at the tops of the barriers where the classical motion is slowest.

The SSS curves are similar to the profile of the SCS maps along the line $\theta=\pi/2$. This reflects the fact that the SCS states along this path are generated by rotating the localized 
state $\vert I I K_s=I\rangle$ around the $l$-axis instead of the SSS basis state, (\ref{eq:LSSS}), which is also centered at the s-axis just being more narrow.
The over-completeness of the SCS and SSS basis sets results in a certain ambiguity in visualizing of the structure.

\subsection{Longitudinal wobbling}\label{s:LWND}

Biswas {\it et al.} \cite{133LaLW} reported the first case of LW for $^{133}$La  based on their observation of large $\frac{B(E2,I\rightarrow I-1)_{out}}{B(E2,I\rightarrow I-2)_{in}}$
ratios  and increasing wobbling energy.
Subsequently evidence for LW was reported  for $^{187}$Au by Sensharma {\it et al.} \cite{187AuLW}
and $^{127}$Xe\cite{127XeLW} by Chakraborty {\it et al.}.
 \begin{figure}[t]
\center{\includegraphics[angle=0,width=\linewidth,trim=0 0 0 0 ,clip]{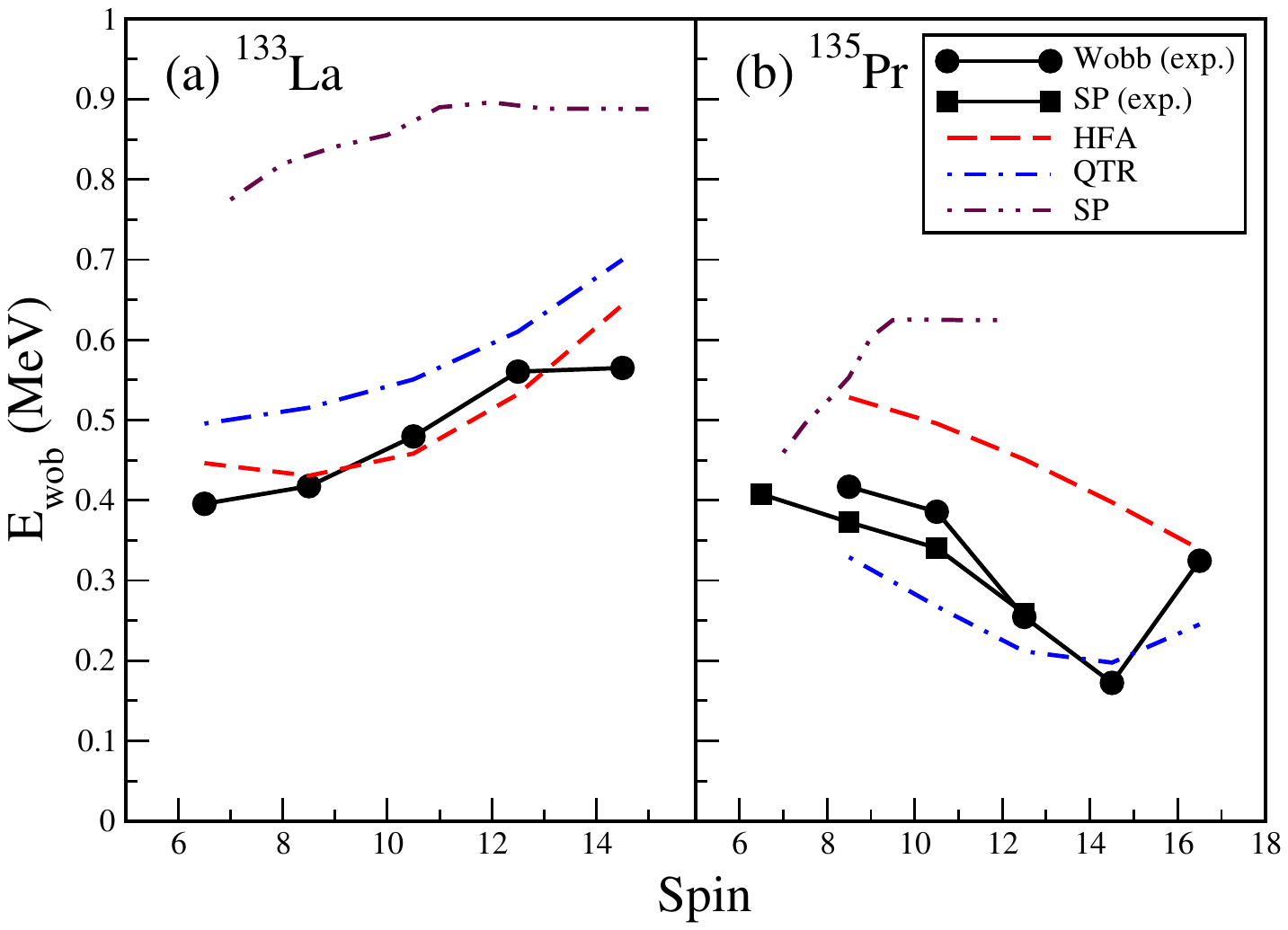}}
 \caption{\label{f:133La} Wobbling and SP band energies in $^{133}$La and $^{135}$Pr compared with QTR calculations. Reproduced with permission from  Ref. \cite{133LaLW} .
}
 \end{figure}

The wobbling energies  of $^{133}$La in Fig. \ref{f:133La} increase with $I$, which is the hallmark of LW. Biswas {\it et al.} \cite{133LaLW} measured the mixing ratios and found that
 $\frac{B(E2,I\rightarrow I-1)_{out}}{B(E2,I\rightarrow I-2)_{in}}$ ratios were collectively  enhanced.
 The right panel shows the isotone $^{135}$Pr with the discussed TW behavior.  As in both nuclei the  proton occupies the bottom of the $h_{11/2}$ shell, TW ought to appear in $^{133}$La.
 The authors of Ref. \cite{133LaLW} explained the appearance LW as follows.   TW in  $^{135}$Pr is the result of  ${\cal J}_s<{\cal J}_m$,  which is seen as a rapid increase of the rotational
  frequency with $I$ along the yrast band. In $^{133}$La a much slower increase is observed, which is caused by the gradual alignment of the angular momentum
   of a pair of positive parity protons with the s-axis.
   The difference  ${\cal J}_m -{\cal J}_s(I)$ becomes quickly small, such that rotation about the 1-axis remains energetically favorable, and the wobbling energy increases with $I$.
   The authors  identified the role of the proton pair by comparing with the rotational response of $^{134}$Pr where the alignment of the proton pair is blocked. 
 The QTR calculations in Fig. \ref{f:133La} were carried out in the framework of the quasiparticle-core-coupling model (see Sec. \ref{s:soft}) using  a TR core with the MoI's
 ${\cal J}_s=(9.12+0.66~ R)~ \hbar^2/MeV$ for $^{133}$La and  ${\cal J}_s=(9.34+0.10 ~R) ~\hbar^2/MeV$ for $^{135}$Pr and for both nuclei ${\cal J}_m=21~ \hbar^2/MeV$, ${\cal J}_l=4~ \hbar^2/MeV$
 and a coupling strength corresponding to $\varepsilon=0.16,~\gamma=26^\circ$. 
 
 The TPSM calculations (see Fig.\ref{f:TPSM_1335pr-133la-131cs_EW} in Sec. \ref{s:TPSM}) reproduce  the change  from TW to LW for, respectively, $Z=61,~59$ in the $N=74$ isotones using nearly
 the same triaxial mean field, which confirms the microscopic origin of the change. The energies of the $h_{11/2}$ bands in the $Z=57,~N=74$ isotone
  indicate TW again, which is reproduced by the TPSM. It would be interesting to substantiate this by measuring the mixing ratios.
  
Based on their highly collective $\frac{B(E2,I\rightarrow I-1)_{out}}{B(E2,I\rightarrow I-2)_{in}}$ ratios, Sensharma {\it et al.} \cite{187AuLW}  assigned LW to the $h_{9/2}$ 
pair of bands in $^{187}$Au.
They carried out QTR calculations using irrotational-flow MoI's with $\gamma=23^\circ$, which correspond to an asymmetry of $\kappa=0.37$ of the inertia
ellipsoid. Good agreement with experimental energies and the  $\frac{B(E2,I\rightarrow I-1)_{out}}{B(E2,I\rightarrow I-2)_{in}}$ ratios was obtained. The TPSM calculations
in Ref. \cite{SJB24} (See  Figs. 12.11 and 12.14.)
 also gave LW with $\frac{B(E2,I\rightarrow I-1)_{out}}{B(E2,I\rightarrow I-2)_{in}}$ ratios very close to the experimental
ones reported by Sensharma {\it et al.} \cite{187AuLW} 

Guo {\it et al.} \cite{187Au-Guo} challenged the LW assignment for $^{187}$Au. They disagree with the authors 
of Ref. \cite{187AuLW} on which of the two minima of the $\chi^2$ fit should be accepted in determining the mixing ratios $\delta$. Guo {\it et al.} \cite{187Au-Guo}
chose the one with small $\vert \delta\vert$ at variance with Sensharma {\it et al.} \cite{187AuLW}, who chose the one with large $\vert \delta\vert$. 
This led to small $\frac{B(E2,I\rightarrow I-1)_{out}}{B(E2,I\rightarrow I-2)_{in}}$ and large  $\frac{B(M1,I\rightarrow I-1)_{out}}{B(E2,I\rightarrow I-2)_{in}}$ ratios,
 which  indicate SP instead of LW
for the excited band. Guo {\it et al.}  \cite{187Au-Guo} carried out QTR calculations using irrotational-flow MoI's with $\gamma=12^\circ$. 
 The weak asymmetry of $\kappa=0.97$ of the inertia ellipsoid is the reason that their QTR calculations assigned a signature partner pair to the two $h_{9/2}$ bands in
 good agreement with the experimental results based  their choice of $\vert \delta\vert$.

 The authors of Ref. \cite{187AuLW} are preparing an extended publication of their experimental results, which will address the controversy.
 The relevant experimental information  is already published in Ref. \cite{Sensharma_Theses}.
In my view, the TPSM calculations, for which the MoI's cannot be adjusted to the data but are fixed by the microscopic structure, represent clear evidence that supports the LW assignment.

 \subsection{Approximate solutions of the PTR model}\label{s:PTRapprox}

\subsubsection{Frozen alignment approximation}
Approximate solutions of the PTR problem have been constructed in order to elucidate the physics of the coupled system. The HFA approximation  was introduced in Ref. \cite{Frauendorf2014PRC}
to define the TW-LW classification. Budaca \cite{Budaca18}  extended the HFA beyond the critical angular momentum $I_c$, given by Eq. (\ref{eq:Ic}).  
 Above $I_c$  the minimum of the FA energy lies at the finite angle 
\begin{equation}
\cos\phi_0=\frac{2A_1j}{(2I-1)(A_1-A_2)}
\end{equation} 
in the case that $\mathbf{j}$ is aligned with the 1-axis. (The alignment with other axes is discussed as well.) The yrast states correspond 
  to uniform rotation about an  axis tilted by $\phi_0$ into the 1-2-plane. 
 The author applied the harmonic approximation with respect to the tilted axis and presented simple algebraic expressions for the energies and 
 reduced $E2$ and $M1$ transition probabilities for $I>I_c$, which complement  the ones of Ref. \cite{Frauendorf2014PRC}. 
 Fig. \ref{f:HFA} shows the HFA wobbling energies  for $j=11/2$ and irrotational flow MoI's at several $\gamma$ values. At the critical angular momentum $I_c$
 the wobbling energy turns zero and then increases, which is the hallmark of LW. The model fails near the instability $I_c$. Above $I_c$  the ellipses around the axes with the tilt angles $\phi_0$
 and $\pi-\phi_0$   soon overlap (see Fig. \ref{f:135PrEad}). It remains to be seen how well the harmonic wobbling excitations about a tilted axis describe the FW regime. 

\begin{figure}[t]
\center{\includegraphics[angle=0,width=\linewidth,trim=0 0 0 0 ,clip]{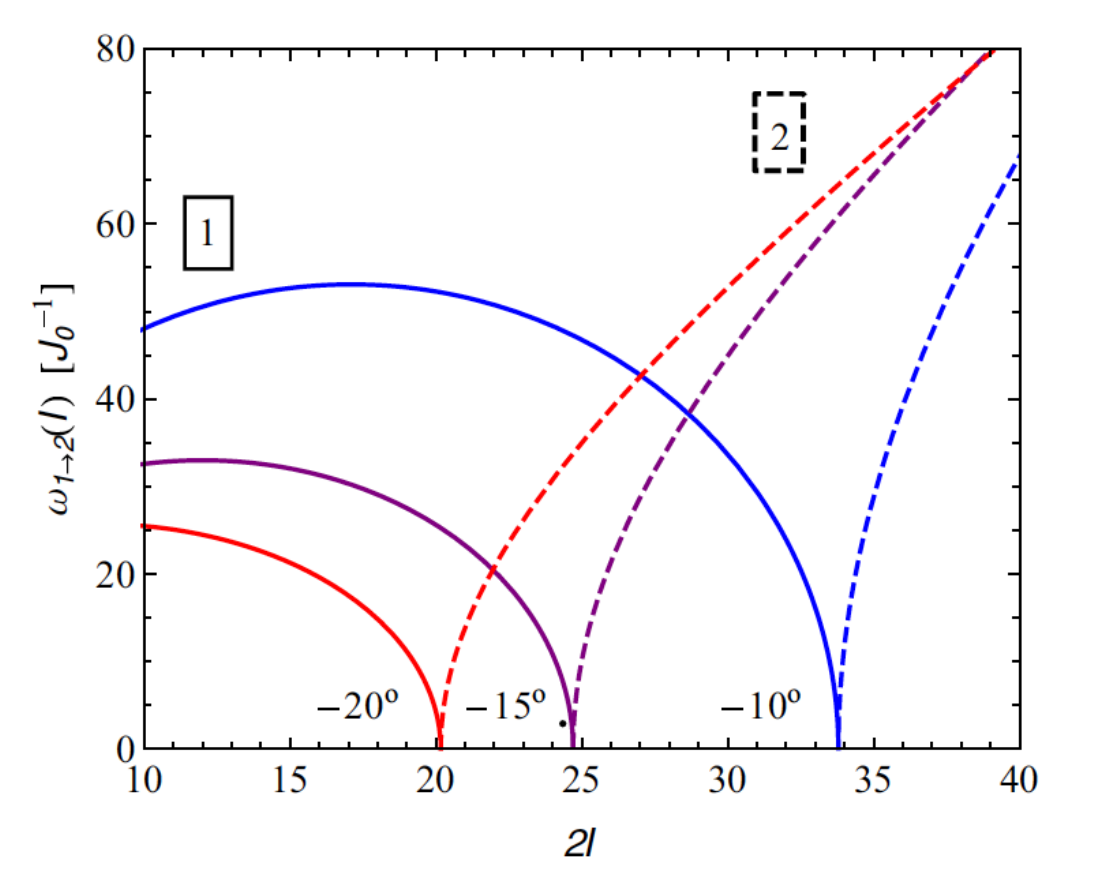}}
 \caption{\label{f:HFA}  Wobbling energies in HFA approximation for different values of the triaxiality parameter $\gamma$. Reproduced with permission  from Ref. \cite{Budaca18}.}
 \end{figure}

\begin{figure}
\center{\includegraphics[angle=0,width=\linewidth,trim=0 0 0 0 ,clip]{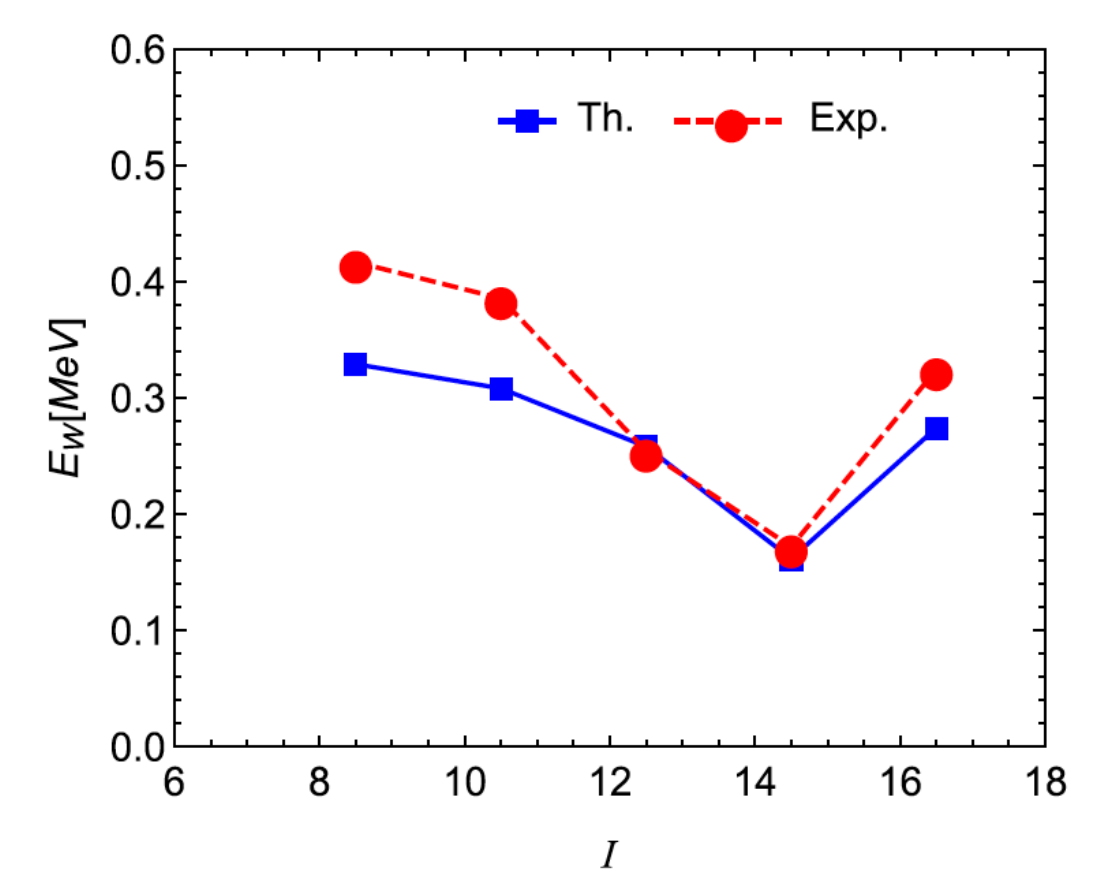}}
 \caption{\label{f:135PrHFA} Wobbling energy adjusted to the experimental energies of $^{135}$Pr. Reproduced with permission from Ref. \cite{Budaca18}.
}
\end{figure}

 The model is a handy tool for a quick exploration of experimental results. The problems around $I_c$ may not not become too serious when the distance  between the discrete $I$ points is large enough,
 like in the case of $^{135}$Pr displayed in Fig. \ref{f:135PrHFA}. The pertaining values of  $\frac{B(E2,I\rightarrow I-1)_{out}}{B(E2,I\rightarrow I-2)_{in}}$ ratios of 0.2 - 0.3 
 deviate substantially from the experimental and QTR values but confirm the collective enhancement.
 
 The model has three parameters: $\gamma$, which
 controls the ratios of the irrotational flow MoI's,  ${\cal J}_0$, which sets the scale of the  rotational energies and $j$ the particle angular momentum aligned with one of the principal axes.
 The author assumes that $j$ remains frozen to the axis, which is not realistic as demonstrated by Fig. \ref {f:135Pr-angles}. This suggests  
 considering a fixed angle  of $\mathbf{j}$ with the 1-axis as another parameter. In Ref. \cite{Budaca23} the authors studied this generalization. Unfortunately, no algebraic 
 expressions for a handy application are given. Their derivation appears straightforward. The only complication is  the solution of the transcendental   equation
 for the angle of the minimum of the classical energy, which could be tabulated or displayed.
 
 \subsubsection{Small amplitude approximation}
 
 Tanabe and Sugawara-Tanabe \cite{Tanabe-energy}  developed an algebraic approximation to the PTR. 
 They applied the Holstein-Primakoff boson expansion to both operators $\mathbf{\hat J}$ and $\mathbf{\hat j}$ 
 and kept the terms up to forth order in the boson number. In second order the PTR Hamiltonian represents two coupled harmonic oscillators. The eigen modes are 
found by solving a biquadratic equation, which has a simple analytic solution. Analytic expressions for the energies and transition probabilities in terms of the three MoI's and the strength
$\kappa$ of the triaxial potential are given. The fourth-oder terms are taken into account by first order perturbation theory in the  eigenmode basis. The expression for the additional
terms are given as well but are rather complex. 
The bands can be classified by the oscillator quantum numbers of the two eigenmodes.

Without coupling,  there  is a pure  $\mathbf{J}$ oscillation, which is the wobbling  mode and a pure $\mathbf{ j}$ oscillation, which is the cranking or SP mode.  
The coupling mixes the modes. In the TW regime the mixed modes retain their character as being predominantly wobbling or cranking, and the $n_{wobb}$ and  $n_{crank}$ 
oscillator quantum numbers can be assigned to the bands.  The HFA represent the limit of a very stiff (large $\kappa$) cranking mode. 
The harmonic approximation breaks down when the minimum of the classical adiabatic energy at the angles $\theta=\pi/2$ and $\phi=0$ changes into a maximum. This the angular momentum
where the angle $\phi$ deviates from zero in Fig. \ref{f:135Pr-angles}.

The authors applied the model to the TSD wobbling bands in the Lu and Tm isotopes assuming rigid body MoI's corresponding to the equilibrium deformation. 
As the respective  ratios  are similar to the ones in Ref. \cite{hamamoto2}
comparable good agreement with the experimental  $\frac{B(E2,I\rightarrow I-1)_{out}}{B(E2,I\rightarrow I-2)_{in}}$ and  $\frac{M1,I\rightarrow I-1)_{out}}{B(E2,I\rightarrow I-2)_{in}}$ ratios 
was obtained. For  rigid body MoI's  the short axis has the largest MoI. Because this implies LW, their wobbling energies  increase with $I$. 
In order to remove the discrepancy with the observed decrease
they made the scale of the MoI's increasing with $I$ as
\begin{equation}
{\cal J}_0(I)={\cal J}_0\frac{I-0.60}{I+23.5}.
\end{equation}
They attributed this increase to a gradual decrease of the pair correlations along the bands. The TDS bands cover the range of $I=13/2 - 97/2$.  
The  scaling gives ${\cal J}_0(97/2)/{\cal J}_0(13//2)=3.14$. However, the experimental ratio of ${\cal J}^{(1)}$ is 73/65=1.12 and of  ${\cal J}^{(2)}$ is 60/70=1.16 \cite{Hagemann2004}.
Obviously, the pairing hypothesis is not convincing. The experimental change of the MoI's cannot revers the $I$ dependence of the wobbling energy. As discussed at the beginning
of Section \ref{s:TR}, the rigid body MoI's ratios contradict the general principle of spontaneous symmetry breaking. 
The conclusions have been discussed at the beginning of subsection "Transverse wobbling in triaxial normal deformed nuclei".

In Ref. \cite{135Pr-tanabe1} the authors applied the model to $^{135}$Pr  assuming irrotational flow MoI's and a triaxialty  parameter $\gamma=26^\circ$ 
as used in Ref. \cite{Frauendorf2014PRC}.
They found that the harmonic vibrations become unstable for  $I>13/2$ and claimed that the TW mode does not appear. Their claim has been refuted by Frauendorf \cite{Frauendorf2018PRC}
(see also their reply in Ref. \cite{135Pr-tanabe2}) for the following reasons. As discussed in the beginning of subsection
 "Transverse wobbling in triaxial normal deformed nuclei", the PTR calculation with irrotational flow MoI's for 
$\gamma=26^\circ$ result in an early transition  of the TW into the FM around $I=17/2$ (See Fig. 16 of Ref. \cite{Frauendorf2014PRC}). 
In the small-amplitude approximation the TW  becomes 
unstable when the minimum of the classical energy at $\phi=0$ disappears. This happens at a smaller  angular momentum than the transition to the FM in the exact PTR calculation.
(Compare $J$=10.5 in Fig. \ref{f:135Pr-angles} with $I=25/2$ in Fig. \ref{f:135PrEwobb}). Hence the very early instability of the TW mode found in Ref. \cite{135Pr-tanabe1} results from
the combination of using irrotational flow MoI's with the near maximal inertia asymmetry of $\kappa=-0.17$ and the harmonic approximation. It cannot be used to question the existence of the TW mode
in normally deformed triaxial nuclei.

Raduta {\it et al.}  \cite{Raduta2017,Raduta2018} (See Chapter 8) re-expressed the PTR Hamiltonian in the basis of the SCS, which becomes a continuous function of  two variables that represent a canonical pair of momentum and coordinate.
They  approximated the Hamiltonian  by expanding it up to second order in momentum and coordinate and solved the pertaining eigenvalue problem on the classical and quantal level. In essence,
the approach is the two-oscillator model of Ref. \cite{Tanabe-energy} (no  fourth order terms). The authors studied $^{163,165,167}$Lu. They 
assumed the rigid-body ratios of the MoI's for the equilibrium deformation and adjusted the scale of the MoI's and the strength of the deformed
potential to the observed energies of the TSD bands. As one can expect, the results were similar to the ones of Ref. \cite{Tanabe-energy}. The model well describes the 
$\frac{B(E2,I\rightarrow I-1)_{out}}{B(E2,I\rightarrow I-2)_{in}}$ and  $\frac{M1,I\rightarrow I-1)_{out}}{B(E2,I\rightarrow I-2)_{in}}$ ratios but gives wobbling energies which increase with $I$.
Hence, my  remarks to the work \cite{Tanabe-energy} apply accordingly.

In Ref. \cite{Raduta2020} Raduta {\it et al.}  suggested that the TSD2 band can be interpreted   as uniform rotation about the $s$-axis like the TSD1 band,  where in the former case 
the core angular momentum $R=$3, 5, 7, ... and in the latter $R=$2, 4, 6, ... In Eq. (25) of Ref. \cite{Raduta2020}, they assigned to both bands the energies $E_{classical}(I)+\hbar\omega_W(I)/2$ with $I=$13/2, 17/2, 21/2, .... for TSD1 and $I=$15/2, 19/2, 23/2, .... for TSD2.  The two terms $E_{classical}(I)$ and $\hbar\omega_W(I)$, derived in Refs. \cite{Raduta2017,Raduta2018},  
are smooth functions of $I$. Therefore the  model
merges the two bands into one $\Delta I=1$ sequence. In other words, the  wobbling energies are zero in stark contrast to the experiment. (A plot of the published energies confirms this.
Private communication of the accurate energies by A. A. Raduta  is acknowledged.)
In Ref. \cite{Raduta2021} Raduta {\it et al.}  applied  the same approach to  $^{135}$Pr with the modification assuming a fixed orientation of the particle's $\mathbf{j}$, 
which gave zero wobbling energies as well.
In my view, the  characterization  of the wobbling mode discussed in Refs. \cite{Raduta2020,Raduta2021}  on the basis of this approach cannot be trusted.

\begin{figure}[t]
\center{\includegraphics[angle=0,width=\linewidth,trim=0 0 0 0 ,clip]{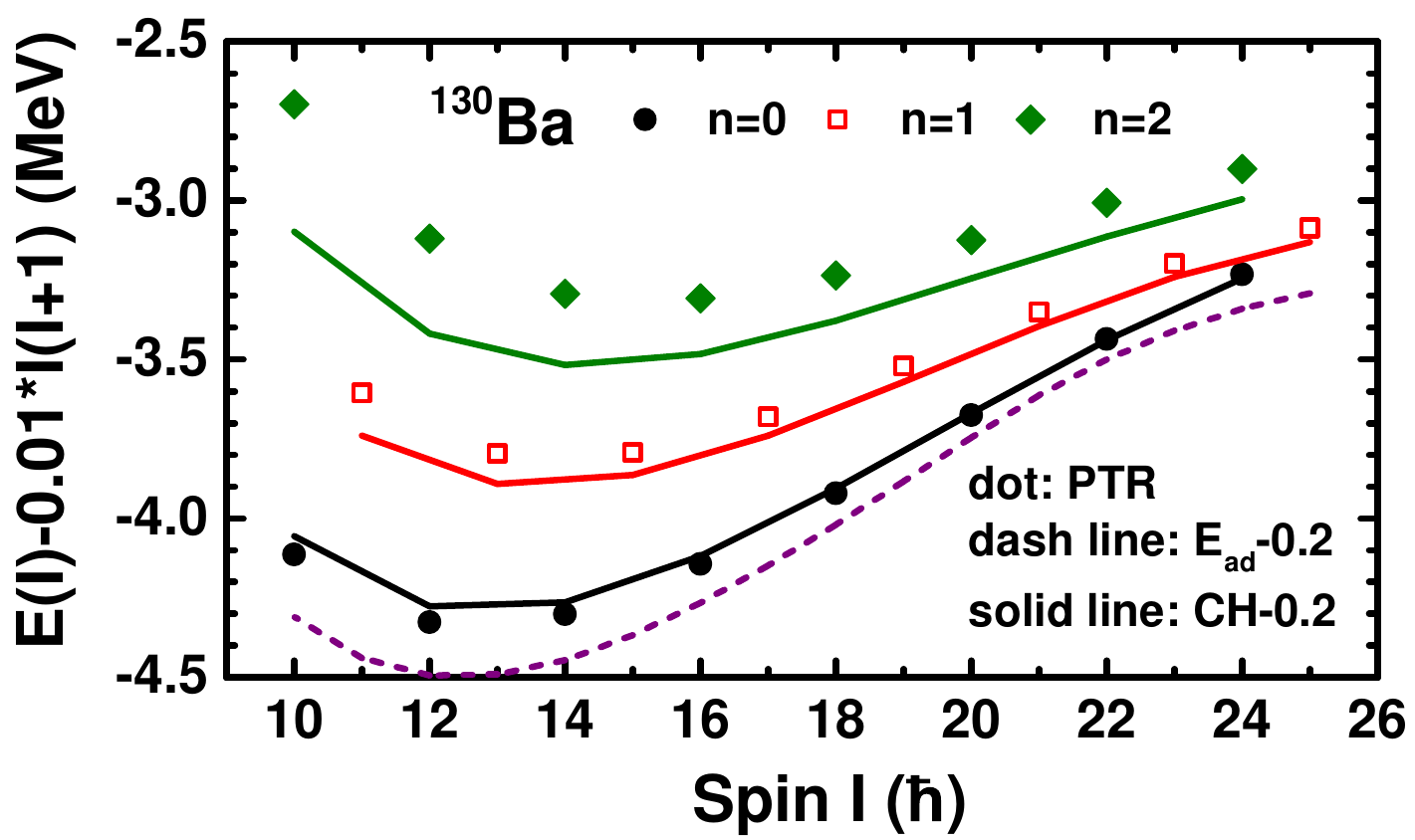}}
\center{\includegraphics[angle=0,width=\linewidth,trim=0 0 0 0 ,clip]{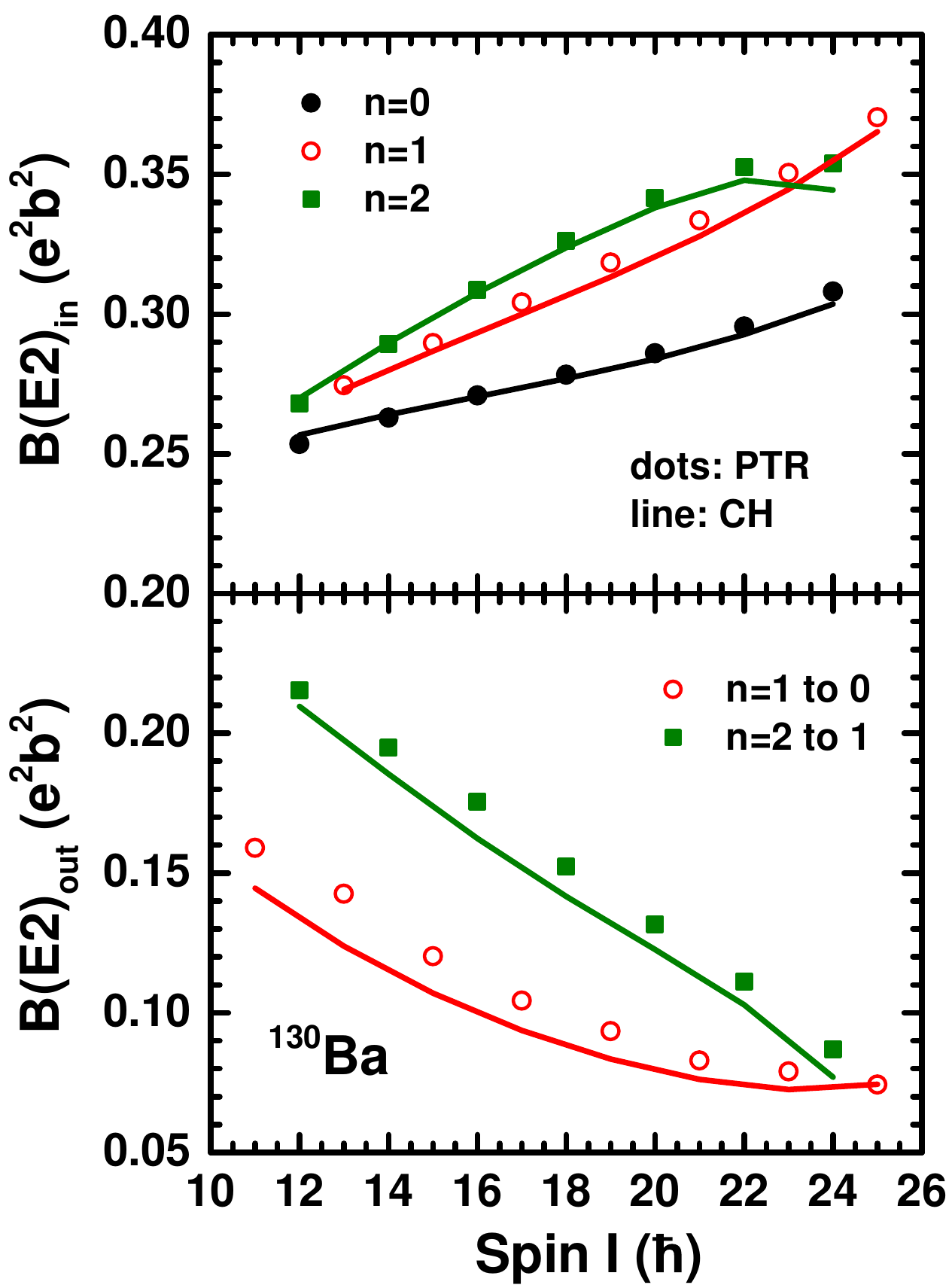}}
 \caption{\label{f:130BaPTR-CH} Comparison of the energies (left panel) and $B(E2)$ values (right panel)
 of $^{130}$Ba calculated by means of the full PTR model (symbols) and the CH (solid curves). 
 The adiabatic classic energy is shown by the dashed curve. All the energies are down shifted by 0.2 MeV.
 Reproduced with permission from Ref. \cite{Chen_Frauendorf2024}.}
\end{figure}

\begin{figure}[h]
\center{\includegraphics[angle=0,width=\linewidth,trim=0 90 0 90 ,clip]{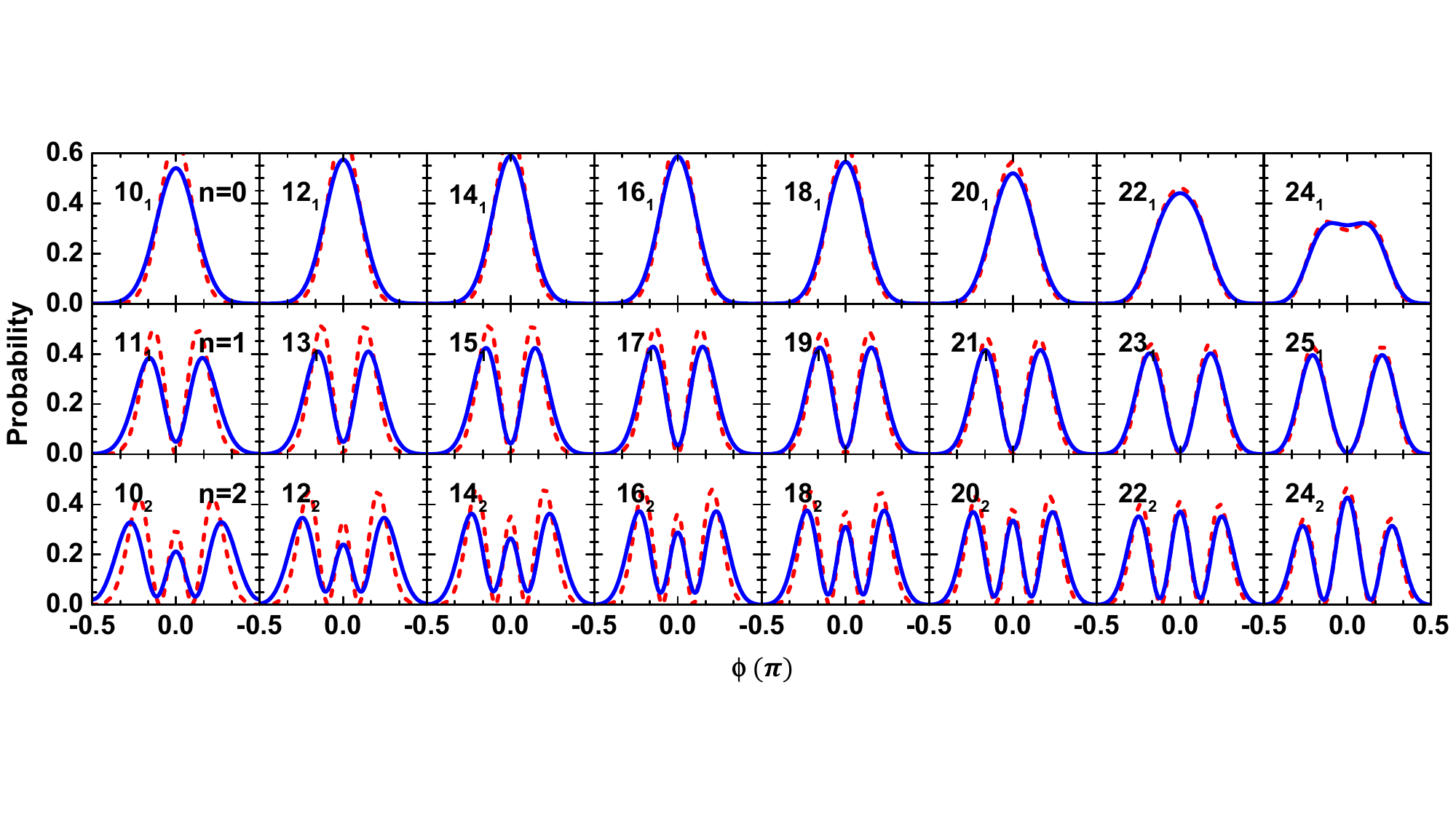}}
 \caption{\label{f:130BaSSS} Full curves show the probability density of the SSS for the $n$ = 0, 1, 2, 3 states calculated by PTR for $^{130}$Ba, 
 while the dashed curves display the ones from the collective Hamiltonian (CH). Reproduced with permission from Ref. \cite{Chen_Frauendorf2024}.
}
\end{figure}

\subsubsection{Collective Hamiltonian}
Frauendorf and Chen \cite{Chen_Frauendorf2024} quantized the classical adiabatic Hamiltonian (\ref{eq:Had}) using the Pauli prescription for the replacing the classical 
kinetic energy $T$ by the corresponding operator $\hat T$. The resulting collective Hamiltonian (CH) was diagonalized in the basis of the axial rotor states $\vert IMK\rangle$.
As the CH is invariant with respect to the $D_2$ group, all representations of the point group appear as eigenstates. Only the ones that 
are  similar to the full PTR states are accepted. 

Fig. \ref{f:130BaPTR-CH}  compares the energies and $B(E2)$ values calculated by means of the PTR and of the CH for $^{130}$Ba. The CH reproduces the exact PTR values rather well.

The CH is based on the familiar concept of a continuous coordinate $\phi$, the orientation of $\bm{J}$ in the 1-2-plane and its conjugate momentum $\hat J_3$.
A sharp localization of the orientation of $ \bm{\hat J}$ is not possible within the finite Hilbert space of $-I\leq K\leq I$.
The SSS (\ref{eq:LSSS}) are  the closest proxies of the collective wave functions that can be generated from the
PTR states. Fig. \ref{f:130BaSSS} compares the SSS probabilities (\ref{eq:PSSS}) generated from the reduced density matrices of the PTR solutions with
the ones generated from the corresponding eigenstates of the CH with the density matrices Eq. (\ref{eq:dmTR}) for pure states. 
The CH densities for the $n=$0, and 1 states agree very well with the PTR densities.
 At the PTR minima, they become zero, which characterizes an oscillating pure wave function. The PTR density has only minima because the coupling to the two particles generates some 
degree of incoherence. The degree of incoherence increases with $n$, which is seen as a progressive washing out of the wave-function pattern. The pattern changes from
the oscillations of the lowest Hermit polynomials in the harmonic TW region near $I$=10 to the onset of the flip regime near $I=$24.

 For $^{135}$Pr, the second example studied,
the agreement of the observables between CH and PTR is comparable. As seen in Fig. \ref{f:135PrSSS}, the incoherence increases more rapidly with $n$. 
This is expected, because one $h_{11/2}$ proton is weaker
  bound to the triaxial core than a pair of them. The $n=$0 and $n$=1 SSS-CH plots correlate  
  very well with the SSS-PTR plots through the TW-FM-LW transition, which allows one to associate the structures with partially coherent quantum oscillations. 
 For the states $n=2, ~I\geq 39/2$ and $n=3$ the CH plots seriously deviate  from the PTR plots, which indicates the break down of the adiabatic approximation.

There is no problem in generating the numerical solutions for the PTR Hamiltonian. There are several codes available, both with the reduced high-j and the full 
version  the triaxial potential in the particle Hamiltonian $h_p$.
One motivation to approximate the PTR model  by a CH is the intuitive interpretation in terms of  wave functions derived from a one-dimensional Hamiltonian of the form $T+V$.
A second is that the study provides the proof of principle that one can construct a realistic CH Hamiltonian from the energy surface generated by a Tilted Axis Cranking calculation.
The TAC approach is widely used for the interpretation of high-spin experiments. The TAC model accounts only for uniform rotation about the tilted axis, which defines the minimum of the  
energy.  A CH constructed from microscopic TAC calculations will allow one to study excitations with wobbling nature in addition to the TAC solutions with $n$=0 nature.
In the preceding work \cite{Chen2014,Chen2016} a  CH was constructed with $V$  being the Routhian at a fixed rotational frequency and $T$ estimated by the HFA applied at 
the minimum of the Routhian, which also provided promising results. The feasibility of extending the TAC along this avenue should  be investigated.

\subsection{Virtues and limits of the PTR}

The PTR model is simple from the conceptual point of view and has lead to the intuitive pictures of the underlying physics, which have been discussed so far.
The computational effort is negligible. However there is the question of how to choose the input parameters. The triaxiality of the shape determines the potential that
 binds the particles to the rotor and the ratios between the MoI's, the interplay of which produces  the wobbling behavior. Microscopically the two elements are tied together.
 On the PTR  level they are to some extend independent, which has the advantage to achieve a good description of the experiment by adjusting them. 
 On the other hand, the freedom of the parameter choice may lead to qualitative different interpretations as discussed in the preceding sections. 
 The parameter choice is restricted by the fundamental principle of spontaneous symmetry breaking, which requires the irrotatotional flow-like oder ${\cal J}_l<{\cal J}_s< {\cal J}_m$.
 The expressions (\ref{eq:momiIF}) with the triaxility parameter $\gamma$ taken from some  mean field calculations  should be used with  caution, because the equilibrium
 values  usually represent  the minimum of a shallow energy surface and often differ between various versions of the mean field approaches. The asymmetry parameter $\kappa$ of the
 MoI's depends sensitively on $\gamma$ between $20^\circ$ and $30^\circ$ (see Fig. \ref{f:kap-ga}), which introduces an element of  uncertainty.
 Adjusting  the MoI's  to reproduce the energies of observed wobbling bands turned out to be the best strategy for interpreting specific experiments. 
  
  In the  microscopic
 approaches to be presented in the next section the deformed mean field determines both elements in a unique way, which reduces the number of parameters. 
 In addition they avoid a basic problem of the PTR, which assumes that the collective orientation angles of the TR are completely independent of the particle degrees of freedom.
 Microscopic calculations by means of the cranking model show that this is only correct  to some extend. 

\section{Microscopic descriptions of wobbling}
There is extended work on the rotational excitations of triaxial nuclei. In this article I only review studies, which explicitly address the wobbling mode.
  \subsection{The pairing-plus-quadrupole-quadrupole Hamiltonian}

The microscopic descriptions of wobbling is based on the pairing-plus-quadrupole-quadrupole many-body Hamiltonian
\begin{eqnarray}
\label{H2} 
\hat H_{PQQ}  &=& 
\sum_{\tau =\pi,\nu}
 [\,\,\hat{h}^\circ_{\tau}   
-G_\tau (\hat{P}^\dagger_\tau \hat{P}_\tau)  
- \lambda_\tau \hat{N}_\tau  \,\,]
 \nonumber\\
&&\quad\quad -\frac{\kappa_{_0}}{2} 
\sum_{m=-2,2} (-1)^{m}\hat{Q}_{m}\hat{Q}_{-m}. 
\end{eqnarray} 
The operator $\hat{h}^\circ_{\tau}$ is the spherical part of the Nilsson Hamiltonian, where the
isospin index $\tau=\pi,\nu$ distinguishes the neutron and proton contributions, respectively, and
 $\hat{P}^\dagger_\tau$ and $\hat{P}_\tau $ are the familiar monopole pair operators.
As pairing is treated in BCS approximation the terms $\lambda_\tau \hat{N}_\tau$,  containing the particle
number operators $\hat{N}_\tau$, are introduced to attain the 
average particle numbers.  
The following term in Eq.\,(\ref{H2}) is the  quadrupole-quadrupole interaction, where $\hat Q_{m}=\sum_{\tau =\pi,\nu}\hat Q_{m}(\tau)$.
The BCS approximation is applied to $H_{PQQ}$, which leads to the quasiparticle Hamiltonian
\begin{eqnarray}\label{htac}
\hat{h}_\tau=\,\,\hat{h}^\circ _\tau -
\Delta_\tau (\hat{P}^\dagger_\tau+ \hat{P}_\tau ) - \lambda_\tau \hat{N} _\tau\qquad\qquad\qquad\qquad\nonumber\\
- \,\hbar\omega_{_0} \frac{2}{3}\,\varepsilon \,\left(\,\cos{\gamma} \hat Q_{_0} 
 -  \frac{\sin{\gamma}}{\sqrt{2}}(\hat Q_{_2}+\hat Q_{_{-2}})\,\right)
.\quad
\end{eqnarray}\label{eq:SC}
and the selfconsistency relations
\begin{eqnarray}\label{scepsga}
\kappa_{_0}\langle \hat Q_{_0}\rangle=\quad 2/3\,\hbar\omega_{_0} \,\varepsilon \cos{\gamma},\nonumber\\
G_\tau\langle P^\dagger_\tau\rangle=\Delta_\tau,~~~~~
\langle\hat N_\tau\rangle=N_\tau,
\end{eqnarray}
where $\hbar\omega_{_0}=41A^{-1/3}$\,MeV.
The average $\langle ...\rangle$ is taken with the quasiparticle configuration of interest.
In the applications the deformation parameters $\varepsilon$, $\gamma$ and the pairing gaps $\Delta_\tau$ are taken as input, which
fix by the relations (\ref{scepsga}) the coupling constants $\kappa_0$ and $G_\tau$. The last equation is fulfilled  by the 
appropriate choice of the Fermi energy $\lambda_\tau$. 

 \begin{figure}[t]
\center{\includegraphics[angle=0,width=\linewidth,trim=0 0 0 0 ,clip]{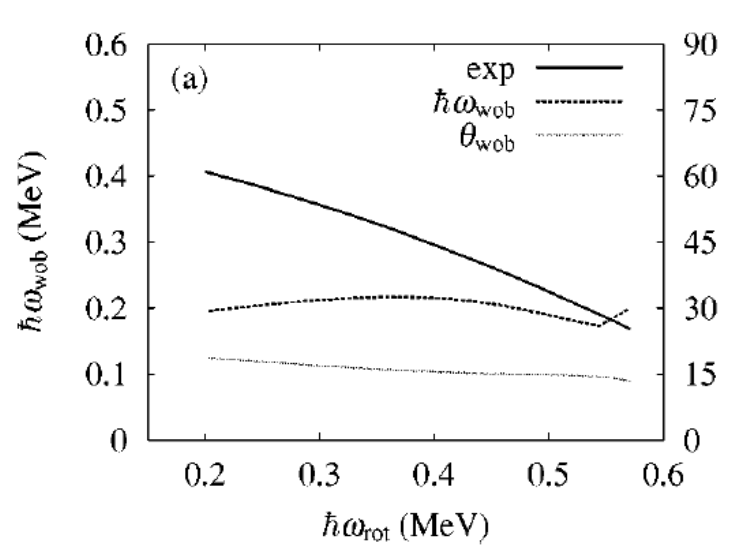}}
 \caption{\label{f:RPA_EW_mat}  Wobbling energies in RPA approximation for $^{163}$Lu.   The opening angle ot the precession cone around the s-axis is $\theta_{wob}$
 with the scale (in degree) on the right axis. 
 Reproduced with permission  from Ref. \cite{Matsuzaki2002}.}
 \end{figure}
 \begin{figure}
\center{\includegraphics[angle=0,width=\linewidth,trim=0 0 0 0 ,clip]{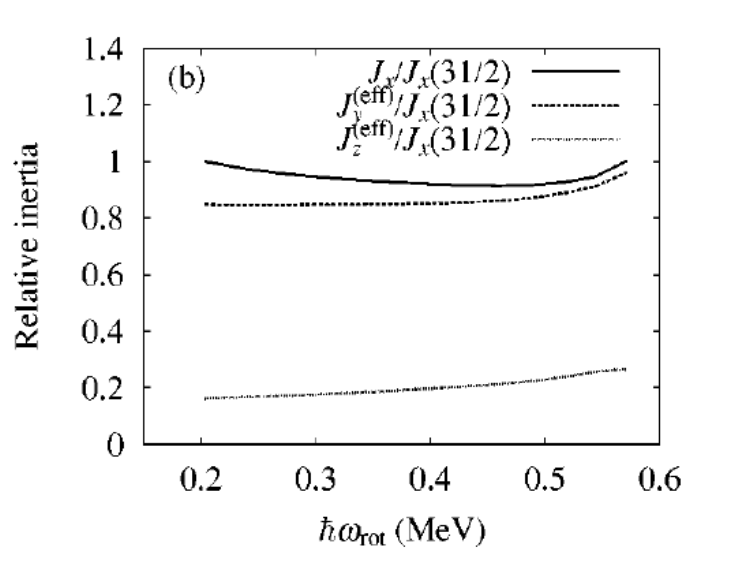}}
 \caption{\label{f:RPA_J_mat} The moments of inertia (\ref{eq:Jdisperse}) in RPA approximation for $^{163}$Lu. The labels x, y, z correspond to 1, 2, 3.
Reproduced with permission  from Ref. \cite{Matsuzaki2002}.
}
\end{figure}

 \begin{figure}[h]
\center{\includegraphics[angle=0,width=\linewidth,trim=0 0 0 0 ,clip]{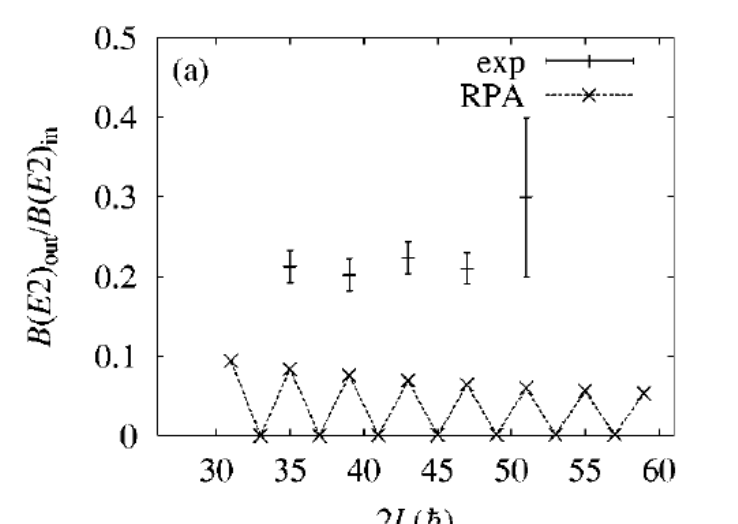}}
 \caption{\label{f:RPA_BE2_mat}  The $\frac{B(E2,I\rightarrow I-1)_{out}}{B(E2,I\rightarrow I-2)_{in}}$  ratios in RPA approximation for $^{163}$Lu.  
 Reproduced with permission  from Ref. \cite{Matsuzaki2002}.}
 \end{figure}
\begin{figure}
\center{\includegraphics[angle=0,width=\linewidth,trim=0 0 0 0 ,clip]{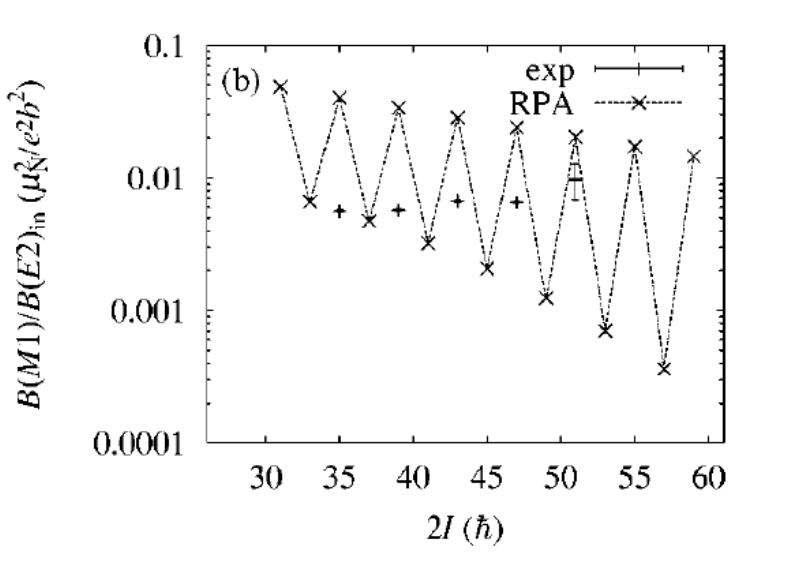}}
 \caption{\label{f:RPA_BM1_mat} The $\frac{B(M1,I\rightarrow I-1)_{out}}{BE2,I\rightarrow I-2)_{in}}$  ratios in RPA approximation for $^{163}$Lu. 
 Reproduced with permission  from Ref. \cite{Matsuzaki2002}.
}
\end{figure}

\subsection{Random Phase Approximation}\label{s:RPA}

The random phase approximation (RPA) to the wobbling mode has been developed by Michailov and Janssen \cite{MichailovRPAwobb,JanssenRPAwobb} and Marshalek \cite{Marshalek!979}.
For the TW regime the RPA starts with finding the mean field solutions to the TAC Routhian
\begin{eqnarray}  
\label{eq:H'} 
\hat H'_{PQQ}=\hat H_{PQQ}- {\omega_1} {\hat{J}_1} - {\omega_2} {\hat{J}_2}- {\omega_3} {\hat{J}_3},
\end{eqnarray}
with $\omega_1=\omega_{rot},~\omega_2=\omega_3=0$. The solution represent uniform rotation about the $s$-axis, which is the yrast band, where
the rotational frequency is chosen such $\langle J_1\rangle=I +1/2$. The solutions are given as configurations of quasiparticles generated by the quasiparticle Routhian
\begin{eqnarray}  
\label{eq:h'} 
\hat h_\tau'=\hat{h_\tau}- {\omega_{rot}} {\hat{J}_1}, ~~~[\hat h_\tau',\alpha^\dagger_m]=E'_m\alpha^\dagger_m,
\end{eqnarray}
where $E'_k$ are the quasiparticle energies  in the  rotating frame (quasiparticle routhians) and $\alpha^\dagger_m$ the pertaining quasiparticle creation operators.
(For details of the TAC approach see e .g. Ref. \cite{SFrmp}.)

Then it is assumed that $\omega_2(t)$ and $\omega_3(t)$ execute small harmonic oscillations, the wobbling mode. 
The solutions are found in first order of time-dependent perturbation theory. The wobbling frequency is given by the expression 
\begin{equation}\label{eq:EWRPA}
\omega_W=\omega_{rot}\left[\frac{({\cal J}_1-{\cal J}_2(\omega_w)({\cal J}_1-{\cal J}_3(\omega_w)}{{\cal J}_2(\omega_w){\cal J}_3(\omega_w)}\right]^{1/2},
\end{equation}
 which has the form of the original expression (\ref{eq:omW-harm}) from Ref. \cite{BMII}. The expressions for the transition propabilities are the same as well.
 The microscopic MoI's are similar to the ones from the cranking model
 \begin{equation}\label{eq:Jdisperse}
 {\cal J}_1=\frac{\langle J_1\rangle}{\omega_{rot}},~~~ {\cal J}_{i=2,3}=
 4\sum_{k,m}\frac{\vert \langle \vert J_1\alpha^\dagger_k\alpha^\dagger_m\vert\rangle\vert^2(E'_k+E'_m)}{(E'_k+E'_m)^2-\omega_W^2},
 \end{equation}
 where the indices $k,~m$ run over all quasiparticle excitations of the yrast configuration $\vert\rangle$.  As both sides depend on $\omega_W$, Eq. (\ref{eq:Jdisperse}) is 
 the RPA dispersion relation, which must be solved numerically.
 
 Matsuzaki {\it et al.} \cite{Matsuzaki2002, Matsuzaki2004} applied the method to the wobbling mode in $^{163,167}$Lu. They used the deformation $\varepsilon=0.43, ~\gamma=20^\circ$
 and weak pairing of $\Delta_n=\Delta_p=0.3$ MeV. The one-quasineutron configuration $\alpha^\dagger_{i_{13/2}}\vert0\rangle$ was the yrast band, which
  implies the replacement $E_m\rightarrow -E_m, ~ \alpha^\dagger_m\rightarrow \alpha_m$ for $m=i_{13/2}$ in the sum (\ref{eq:Jdisperse}). Their results are compared with the experimental
  ones in Figs. \ref{f:RPA_EW_mat} - \ref{f:RPA_BM1_mat}.
  
 As the wobbling energies are constant with a slight downward trend and the $\frac{B(E2,I\rightarrow I-1)_{out}}{B(E2,I\rightarrow I-2)_{in}}$  ratios are collectively enhanced, the results
 represent the first example for a microscopic description of TW. The authors explained the angular momentum dependence of $E_w(I)$  in the same way as   Frauendorf and D\"onau
 when they introduced the TW-LW classification based on the HFA approximation
 \cite{Frauendorf2014PRC}. The linear increase $\propto \omega_{rot}$ is overcompensated by the decrease of the factor ${\cal J}_1-{\cal J}_2$ in Eq. (\ref{eq:EWRPA}), which is caused by the 
 decrease of ${\cal J}_1$. The origin of the latter is the presence of the particle's $\mathbf{j}$ being aligned with the s -axis. In such a case the angular momentum 
 $J_1=j+\omega_{rot}{\cal J}_1^{(2)}$ and $\bar {\cal J}_1=J_1/\omega_{rot}={\cal J}_1^{(2)}/(1-j/J_1)$, which agrees with the $\bar A_1$ in Eq. (\ref{HFAham}).
 Assuming that $j=13/2$, one can estimate from the value of ${\cal J}_1$ in Fig. \ref{f:RPA_J_mat} that the TR MoI is ${\cal J}_1^{(2)}=0.67$ in the units of the figure at $\hbar \omega_{rot}$=0.2 MeV.
 The RPA calculations  confirm the qualitative requirement  ${\cal J}_m>{\cal J}_s> {\cal J}_l$ in a quantitative way.
 
  Compared with the experiment, the wobbling energy is too small at low angular momentum and stays almost constant.
  The $\frac{B(E2,I\rightarrow I-1)_{out}}{B(E2,I\rightarrow I-2)_{in}}$ are by a factor of two 
  too small and the $\frac{B(M1,I\rightarrow I-1)_{out}}{B(E2,I\rightarrow I-2)_{in}}$  ratios by a factor 5 - 10 too large.
  
  The wobbling motion represent oscillations of $\mathbf{J}$ with respect to the principal axes of the triaxial shape. In the laboratory frame $\mathbf{J}$ stands still, and the triaxial shape 
  oscillates. For small amplitudes  wobbling is a harmonic oscillation of the mass quadrupole moment $Q_{1-}(t)=(Q_1(t)-Q_{-1}(t))/\sqrt{2}$.  
  Application of the time-dependent perturbation
  theory leads to the well known RPA for separable interactions, which is equivalent with  the just discussed RPA in the principal axes system \cite{Marshalek!979}.

  In Ref. \cite{Frauendorf2015RPA} Frauendorf and D\"onau  re-investigated $^{163}$Lu by means of the standard RPA, because the available computer code allowed them to include other separable interactions. They kept  $\kappa_{_0}$ constant and  adjusted it to obtain $\varepsilon\approx 0.4$ over the range of
  $\omega_{rot}$, which gave $\gamma\approx 10^\circ$ from the self-consistency condition for $Q_2$. Weak pairing gaps of $\Delta_n=0.35,~\Delta_p=0.45$ MeV  and standard effective charges  of $e_p=e(1+Z/A)$ and $e_n=eZ/A$ charges were used. The results are labelled  as $\mathrm{QQ}_{\mathrm{sc}}$ in Figs. \ref{f:RPA_BEW_FD} and \ref{f:RPA_BM1_FD}. 
  The results are quite similar to the ones of Matzusaki {\it et. al.} with the exception that the wobbling mode becomes unstable around  $\hbar\omega=0.45$ MeV.
 
  Using the deformations $\varepsilon\approx 0.4$ and $\gamma\approx 20^\circ$ from Matzusaki {\it et al.} marginally changed the results. Including the isovector $QQ$ 
  and spin-spin interactions changed the results  insignificately. However a current-current interaction of the type $\mathbf{L}\cdot\mathbf{L}$ did.
  As displayed by the curves $\mathrm{QQ}_{\mathrm{sc}}+\mathrm{LL}$  in Figs. \ref{f:RPA_BEW_FD} and \ref{f:RPA_BM1_FD}, the wobbling energies agree very well
  with the experimental values, and the $\frac{B(M1,I\rightarrow I-1)_{out}}{BE2,I\rightarrow I-2)_{in}}$  ratios are reduced to the observed ones, while the 
   $\frac{B(E2,I\rightarrow I-1)_{out}}{BE2,I\rightarrow I-2)_{in}}$  ratios did not change much.
   
  The $\mathrm{LL}$ interaction
  couples the wobbling mode to the scissors mode, which is the isovector version of the wobbling mode:
   the proton system and the neutron system precess with anti-phase around the total angular momentum $\mathbf{J}$. The RPA calculation for the neighbor $^{164}$Yb with the same
   coupling strength give the position and $B(M1)$ strength observed for the scissors mode in the nuclei of the region. A more detailed investigation of the relation 
   between the scissors and wobbling modes seems interesting because the PTR calculations systematically 
    overestimate the $\frac{B(M1,I\rightarrow I-1)_{out}}{BE2,I\rightarrow I-2)_{in}}$  ratios.

  \begin{figure}[t]
\center{\includegraphics[angle=90,width=\linewidth,trim=0 50 70 10 ,clip]{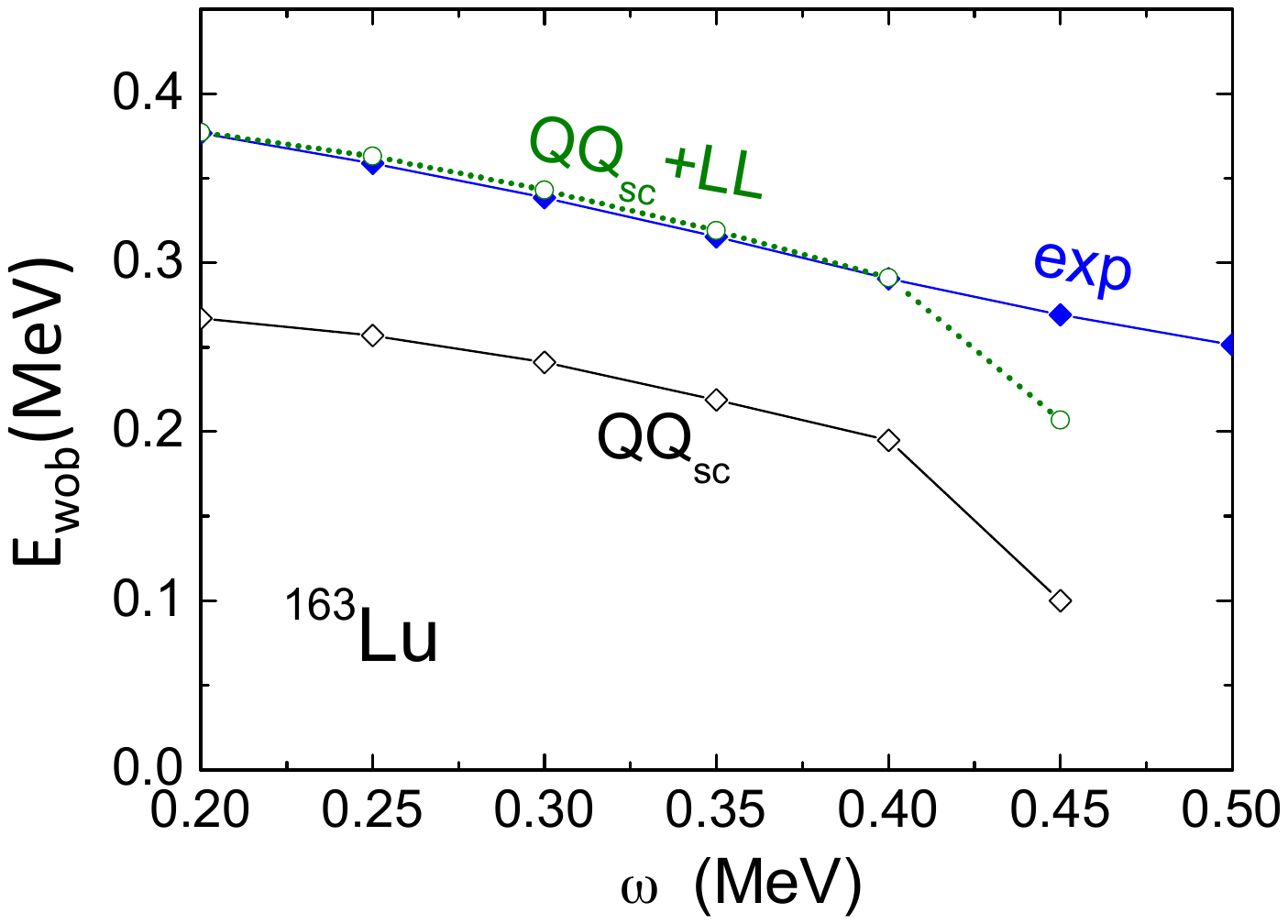}}
 \caption{\label{f:RPA_BEW_FD}   Wobbling energies in RPA approximation for $^{163}$Lu.  Reproduced with permission from Ref. \cite{Frauendorf2015RPA}.}
 \end{figure}
\begin{figure}
\center{\includegraphics[angle=90,width=\linewidth,trim=50 50 50 20 ,clip]{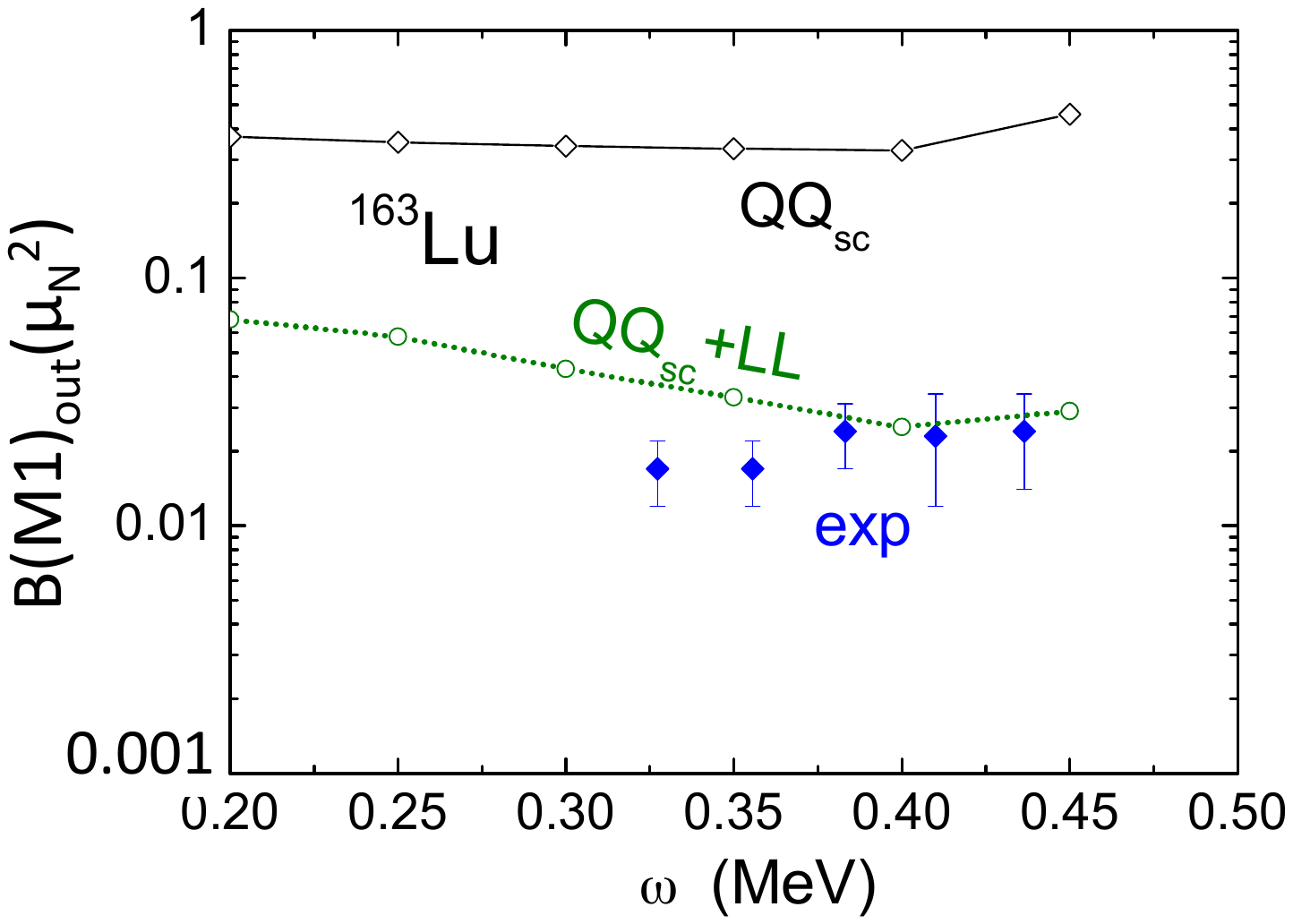}}
 \caption{\label{f:RPA_BM1_FD} The $\frac{B(M1,I\rightarrow I-1)_{out}}{BE2,I\rightarrow I-2)_{in}}$  ratios in RPA approximation for $^{163}$Lu. 
 Reproduced with permission from Ref. \cite{Frauendorf2015RPA} .
}
\end{figure}

\subsection{Triaxial projected shell model}\label{s:TPSM} 

The triaxial projected shell model (TPSM) \cite{tpsm} has been the other microscopic approach  for studying the wobbling mode. 
The TPSM diagonalizes the pairing-plus-quadrupole-quadrupole many-body Hamiltonian (\ref{H2}) (augmented by a quadrupole pairing interaction) 
in the basis of quasiparticle configurations which are projected onto good angular momentum,
\begin{eqnarray}\label{eq:basis}
\hat P^I_{MK}\left|0\right>,& \hat P^I_{MK}~a^\dagger_{p_1}  \left|0\right>, \nonumber\\
 \hat P^I_{MK}~a^\dagger_{p_1} a^\dagger_{p_2} \left|0\right>,& \hat P^I_{MK}~a^\dagger_{n_1}  \left|0\right>,  \nonumber\\
\hat P^I_{MK}~a^\dagger_{n_1}a^\dagger_{n_2} \left|0\right>, & \hat P^I_{MK}~a^\dagger_{p_1}a^\dagger_{n_1}a^\dagger_{n_2} \left|0\right>,\nonumber \\
\hat P^I_{MK}~a^\dagger_{p_1} a^\dagger_{p_2} a^\dagger_{n_1}a^\dagger_{n_2} \left|0\right> , &\hat P^I_{MK}~a^\dagger_{n_1} a^\dagger_{p_1} a^\dagger_{p_2} \left|0\right>,\nonumber \\
..... &\hat P^I_{MK}~a^\dagger_{p_1} a^\dagger_{p_2} a^\dagger_{p_3}a^\dagger_{n_1}a^\dagger_{n_2}  \left|0\right>,\nonumber \\
 &\hat P^I_{MK}~a^\dagger_{n_1} a^\dagger_{n_2}a^\dagger_{n_3} a^\dagger_{p_1} a^\dagger_{p_2} \left|0\right>,\nonumber \\
&.....
\label{basis}
\end{eqnarray}
where $\left| 0\right>$  represents the triaxial quasiparticle
vacuum state, and $ \hat P^I_{MK}$ projects the quasiparticle configuration on to good angular momentum states $\vert IMK\rangle$. In the majority
of the nuclei, the near-yrast spectroscopy up to$ $I=20 is well described using the above basis space. The extension is straight forward. 

The input  parameters are the pairing gaps and the deformation parameters $\varepsilon,~\varepsilon'=\varepsilon \tan\gamma$. The PQQ coupling constants are fixed by means of the selfconsistency
relations (\ref{scepsga}). Usually $\varepsilon$ is taken from experiment or a mean field calculation and $\varepsilon'$ is adjusted to the experimental band energies. 

 \begin{figure}[h]
\center{\includegraphics[angle=0,width=\linewidth,trim=0 0 0 0 ,clip]{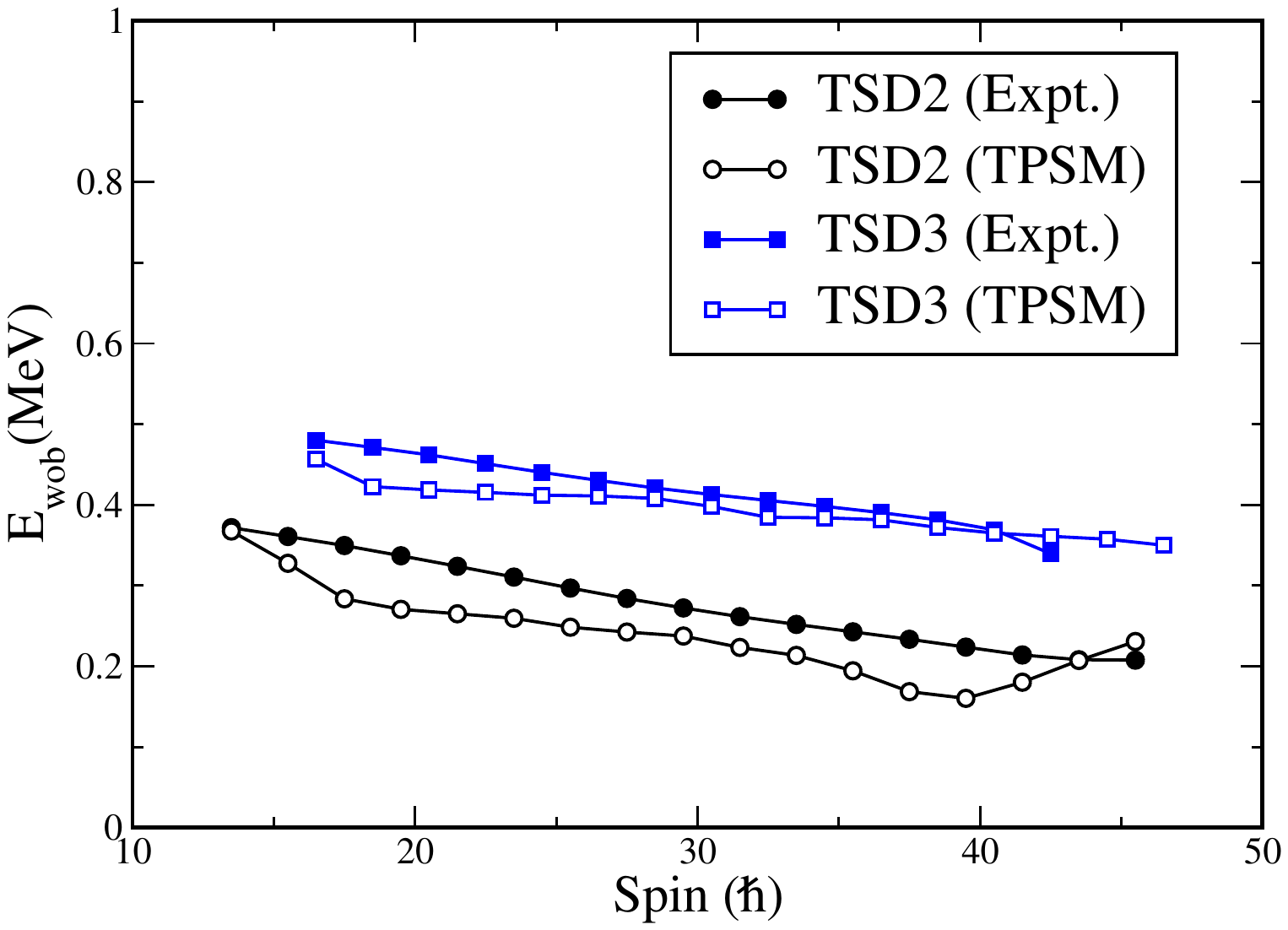}}
 \caption{\label{f:TPSM_163Lu_EW}   Wobbling energies in TPSM approximation for $^{163}$Lu.  From Ref. \cite{SJB24}, 
 private communication by G. Bhat is acknowledged.}
 \end{figure}
\begin{figure}
\center{\includegraphics[angle=0,width=\linewidth,trim=0 0 0 0 ,clip]{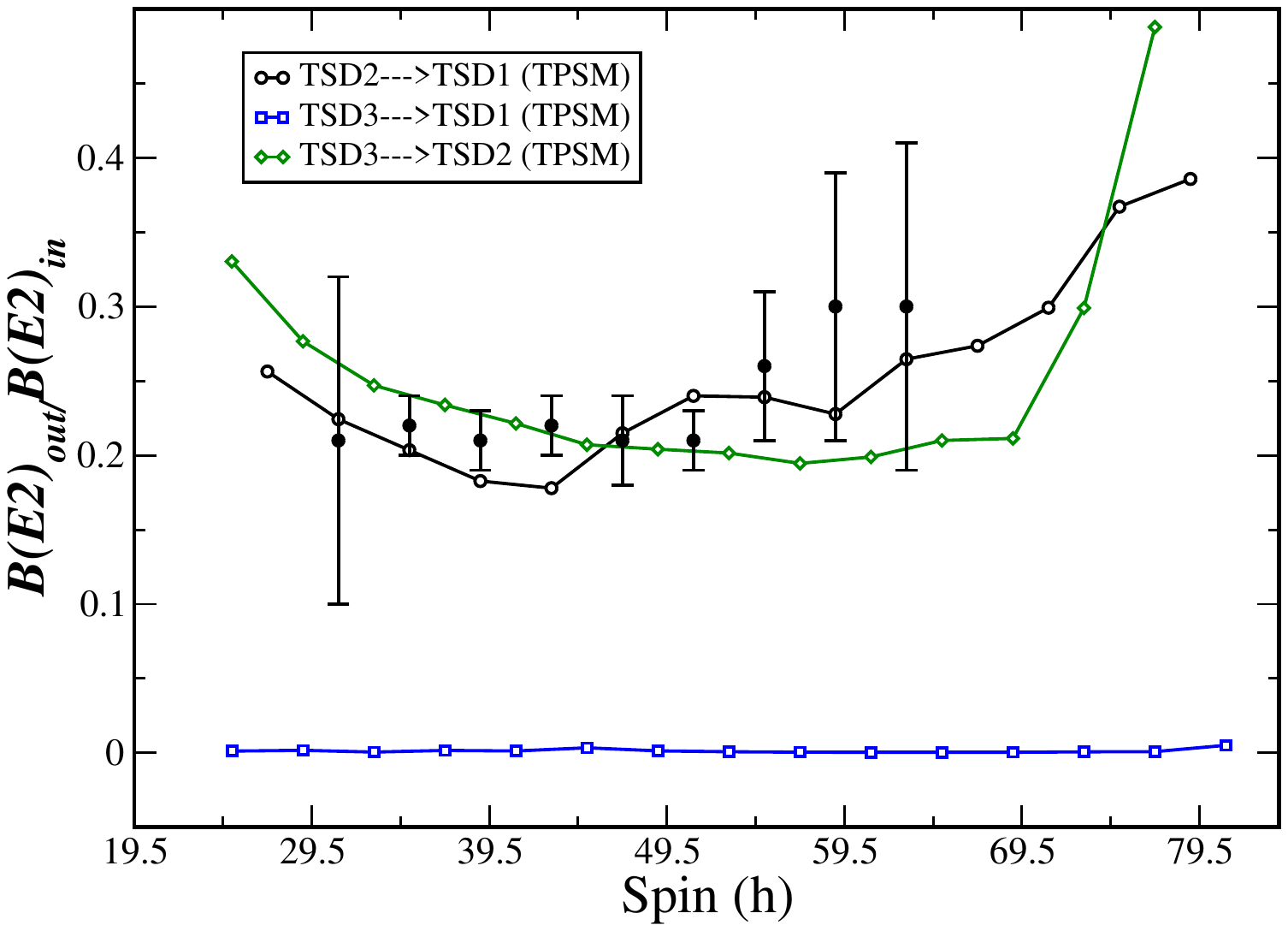}}
 \caption{\label{f:TPSM_163Lu_BE2iout/in} The $\frac{B(E2,I\rightarrow I-1)_{out}}{BE2,I\rightarrow I-2)_{in}}$  ratios in TPSM approximation for $^{163}$Lu.
 From Ref. \cite{SJB24}, 
 private communication by G. Bhat is acknowledged.
}
\end{figure}

  \begin{figure}[h]
\center{\includegraphics[angle=0,width=\linewidth,trim=0 0 0 0 ,clip]{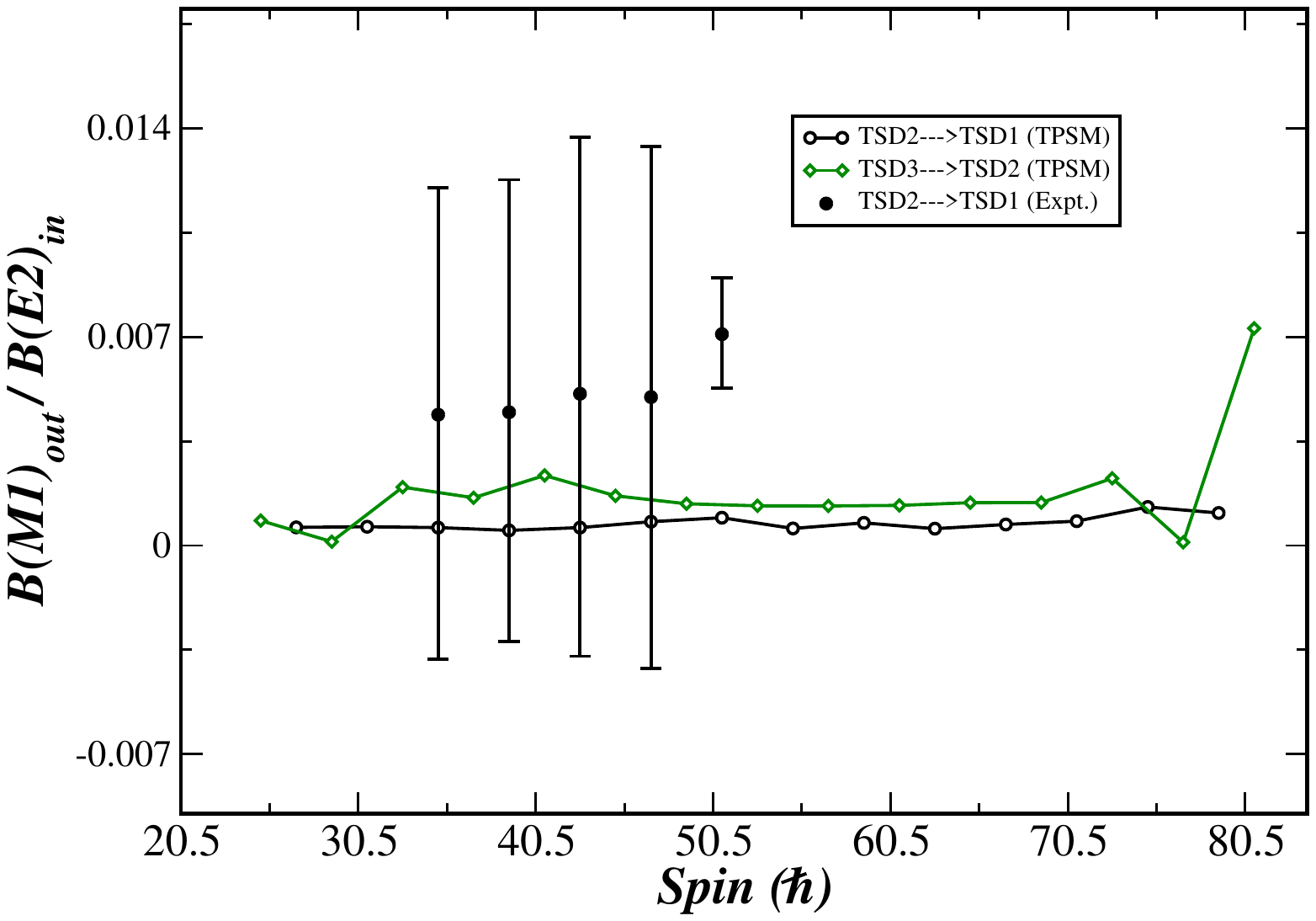}}
 \caption{\label{f:TPSM_163Lu_BM1}   The $\frac{B(M1,I\rightarrow I-1)_{out}}{BE2,I\rightarrow I-2)_{in}}$  ratios in TPSM approximation for $^{163}$Lu.  From Ref. \cite{SJB24},  
 private communication by G. Bhat is acknowledged.
}
 \end{figure}
\begin{figure}
\center{\includegraphics[angle=0,width=\linewidth,trim=0 0 0 0 ,clip]{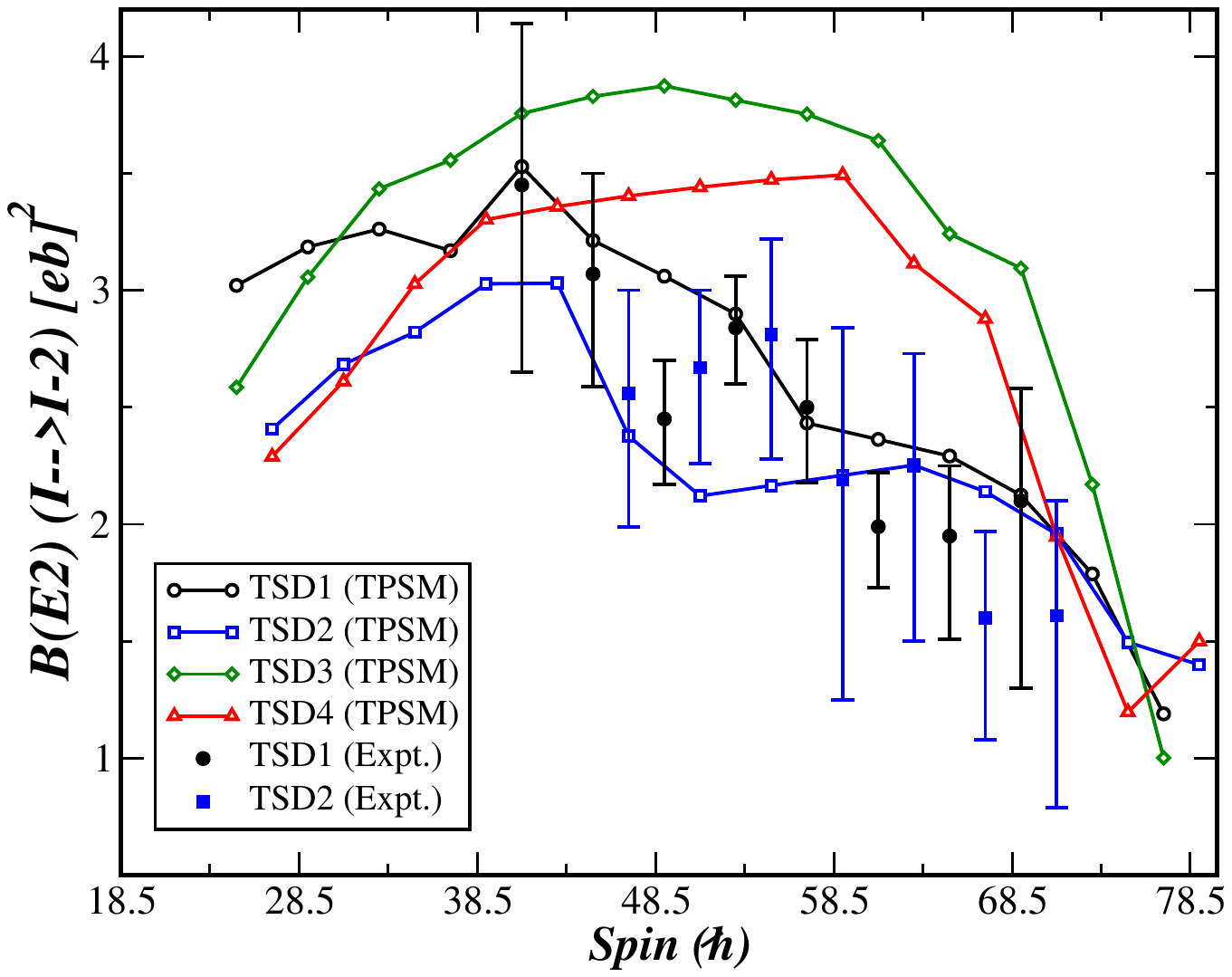}}
 \caption{\label{f:TPSM_163Lu_BE2in}  Intra band $BE2,I\rightarrow I-2)_{in}$ values in TPSM approximation for $^{163}$Lu. From Ref. \cite{SJB24}, 
 private communication by G. Bhat is acknowledged.
}
\end{figure}

Sheikh, Jehangir and Bhat \cite{SJB24} carried out an extended TPSM  study of nuclei with candidates for  chiral and wobbling bands.
 They found systematic good agreement of the  TPSM results with the experimental energies and available transition probabilities.
From their work I discuss only few examples which demonstrate
 the new aspects  of the microscopic TPSM as compared to the semi-phenomenological PTR approach.
The authors studied wobbling in the $^{163}$Lu and the other Lu isotopes. They used $\Delta_n=0.7527$ MeV, $\Delta_p=0.9637$ MeV and 
$\varepsilon=0.40$ and $\gamma=15^\circ$, which was adjusted to 
the experimental value of $E_W(21/2)$. The results  displayed in Figs. \ref{f:TPSM_163Lu_EW} - \ref{f:TPSM_163Lu_BE2in} agree very well with 
the experimental data. The bands TSD1, TSD2, TSD3 are assigned to the bands with $n$=0, 1,  2 wobbling quanta. The band TSD4 contains the $h_{9/2}$ proton.
The close-to-harmonic TW picture is  substantiated. The wobbling energy decreases with the correct average slope. There are wiggles, which are not seen
in experiment. The $\frac{B(E2,I\rightarrow I-1)_{out}}{BE2,I\rightarrow I-2)_{in}}$ ratios for the $n\rightarrow n-1$ transitions 
show the collective enhancement of collective wobbling. The transition $n=2\rightarrow 0$ is suppressed.

Remarkably, the $\frac{B(M1,I\rightarrow I-1)_{out}}{BE2,I\rightarrow I-2)_{in}}$  ratios in Fig. \ref{f:TPSM_163Lu_BM1} agree with the experimental values.
These ratios are systematically overestimated in the PTR calculations. As discussed in section "Random Phase Approximation", this is also the case for the RPA calculations
based on the standard  pairing-plus-quadrupole-quadrupole Hamiltonian (\ref{H2}). Only the inclusion 
of the coupling to the scissors mode  reduced  the ratios to their experimental values. It would be interesting to see if there is a connection between coupling to the scissors mode
and the mutli-quasiparticle states of the TPSM basis (\ref{eq:basis}), which include scissors configurations.

  \begin{figure}[t]
\center{\includegraphics[angle=0,width=0.8\linewidth,trim=0 0 0 0 ,clip]{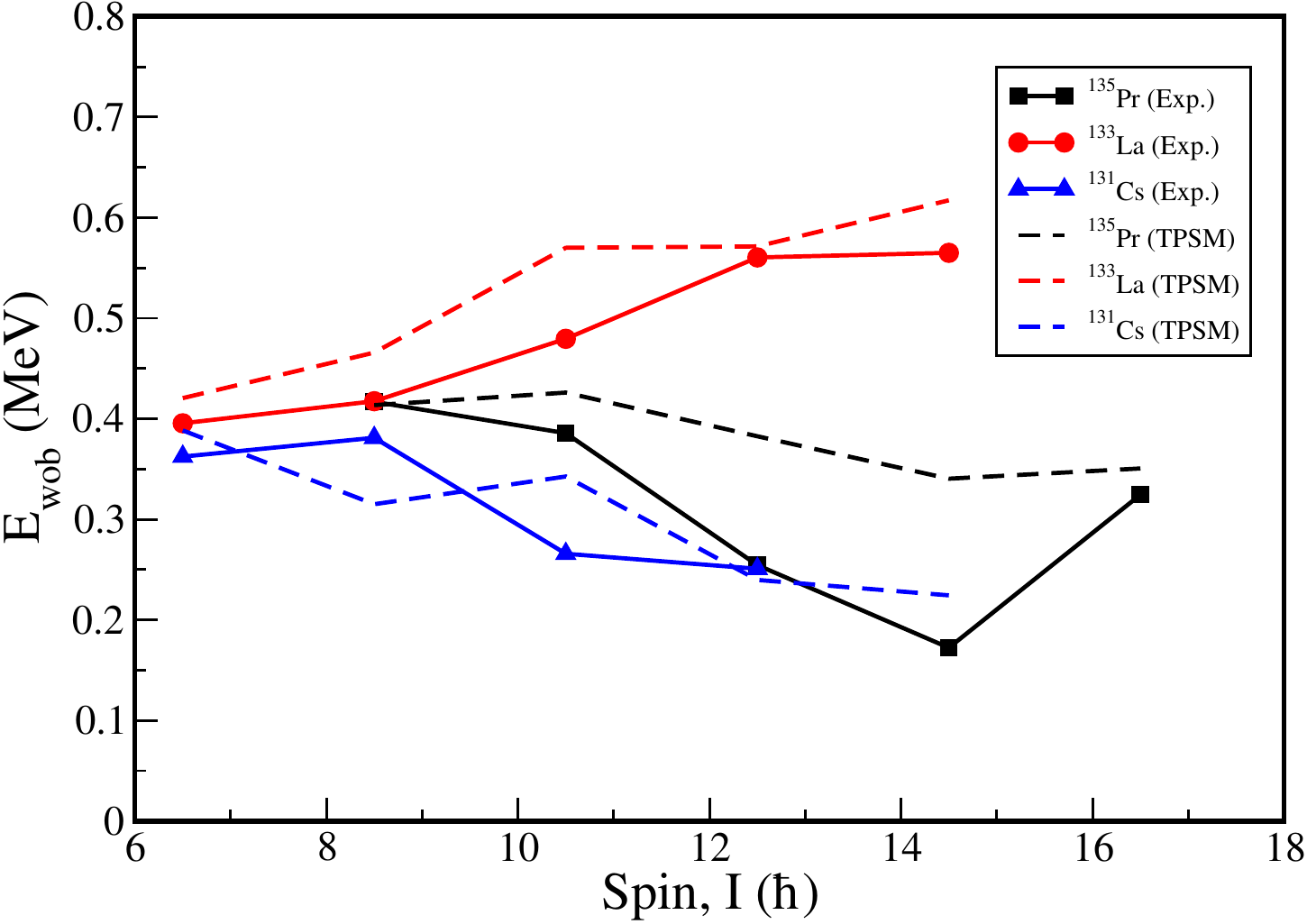}}
 \caption{\label{f:TPSM_1335pr-133la-131cs_EW}   Wobbling energies in TPSM approximation for $^{131}$Cs, $^{133}$La, $^{135}$Pr.  From Ref. \cite{SJB24}, 
 private communication by G. Bhat is acknowledged.}
 \end{figure}
\begin{figure}
\center{\includegraphics[angle=0,width=0.8\linewidth,trim=0 0 0 0 ,clip]{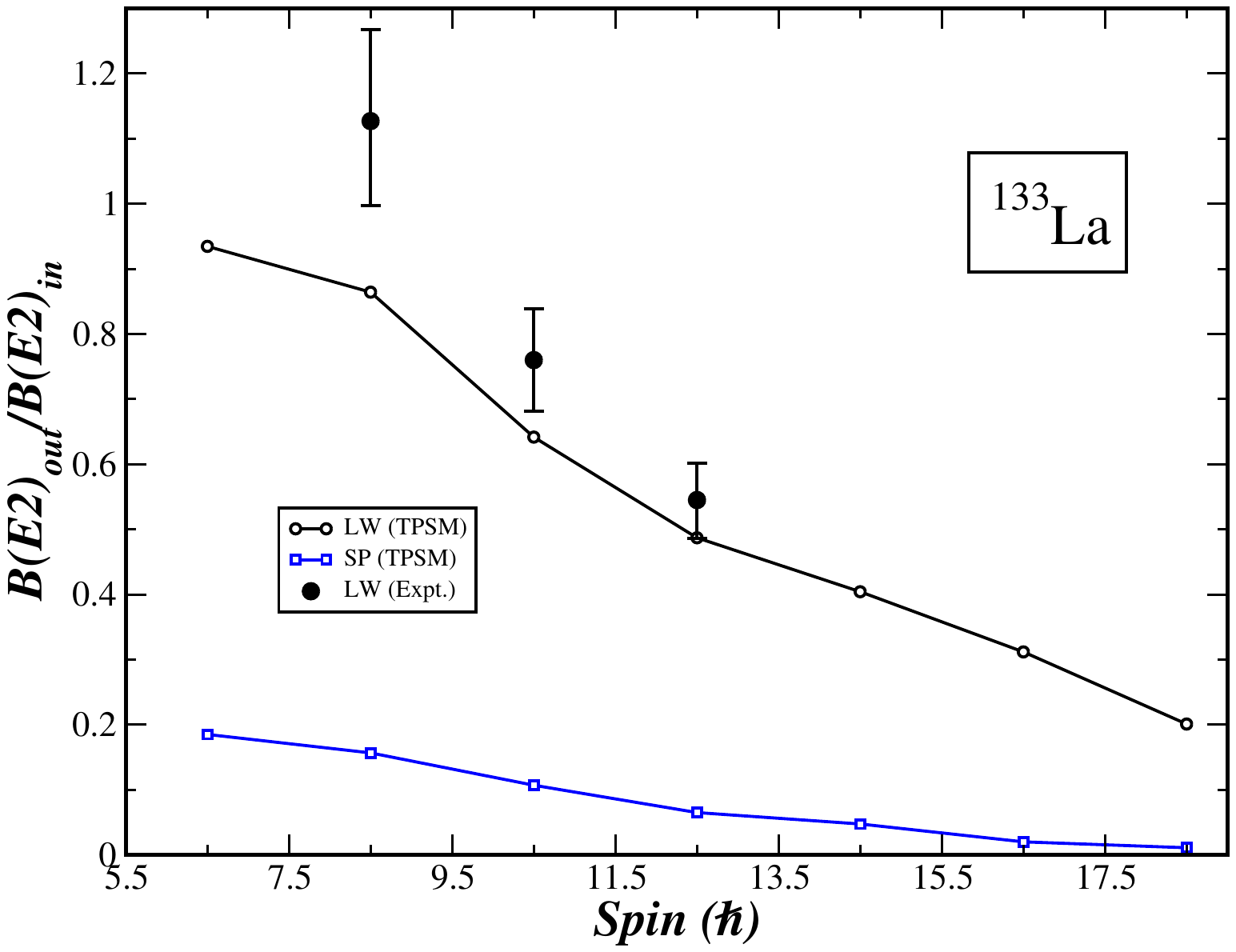}}
 \caption{\label{f:TPSM_133La_BE2iout/in} The $\frac{B(E2,I\rightarrow I-1)_{out}}{BE2,I\rightarrow I-2)_{in}}$  ratios in TPSM approximation for $^{133}$La.
 From Ref. \cite{SJB24}, 
 private communication by G. Bhat is acknowledged.
}
\end{figure}

Fig. \ref{f:TPSM_1335pr-133la-131cs_EW} compares  the experimental wobbling energies of the three $N=74$ isotones $Z=$57, 59, 41 with the TPSM
calculations, which used, respectively, $\varepsilon=$0.14, 0.15, 0.16 and  $\varepsilon'$=0.10 (corresponding to $\gamma=35^\circ,~33^\circ,~32^\circ$) as input.
The fact that the TPSM  with nearly constant $\varepsilon, ~\varepsilon'$ accounts for the observed  TW-LW-TW sequence along the isotone chain
 indicates that
the change originates from the change of the quasiparticle  orbitals  at the  Fermi level. The details are discussed in section  "Longitudinal wobbling".
Fig. \ref{f:TPSM_133La_BE2iout/in} and \ref{f:135PrBE} demonstrate the collective enhancement of the $n\rightarrow n-1$ transitions between the wobbling band
in $^{133}$La and $^{135}$Pr, respectively.

 Figs. \ref{f:133La} displays two examples for a general problem: The QTR model predicts the SP band too high when locating the wobbling bands 
 at the experimental energies.
 As seen in Fig.  \ref{f:135PrEwob}, TPSM locates both the SP and wobbling bands close to their experimental positions. 
 The  treatment of the TR core and the of the odd proton on equal footing in the microscopic TPSM seems to be decisive.

  \begin{figure}[h]
\center{\includegraphics[angle=0,width=0.8\linewidth,trim=0 0 0 0 ,clip]{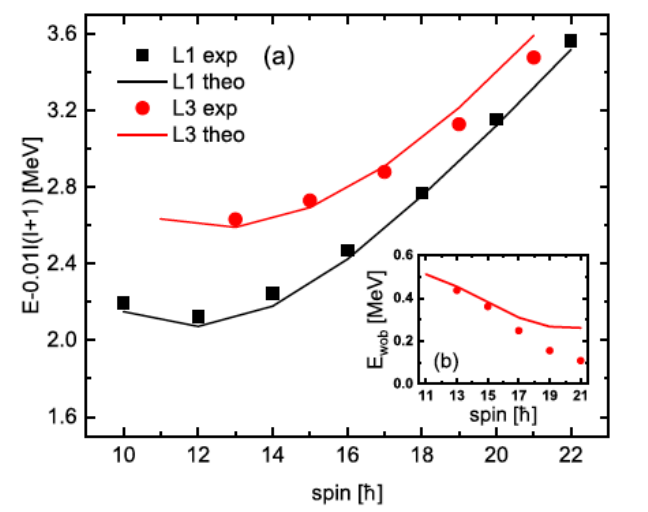}}
 \caption{\label{f:TPSM_136Nd_EW}   Energies in TPSM approximation for the $n$=0 (L1) and $n$=1 (L3) TW candidate bands in $^{136}$Nd.
 The inset shows the wobbling energies.
 Reproduced with permission from Ref \cite{Fang-Petrache2021}.}
 \end{figure}
\begin{figure}
\center{\includegraphics[angle=0,width=\linewidth,trim=20 0 45 30 20 ,clip]{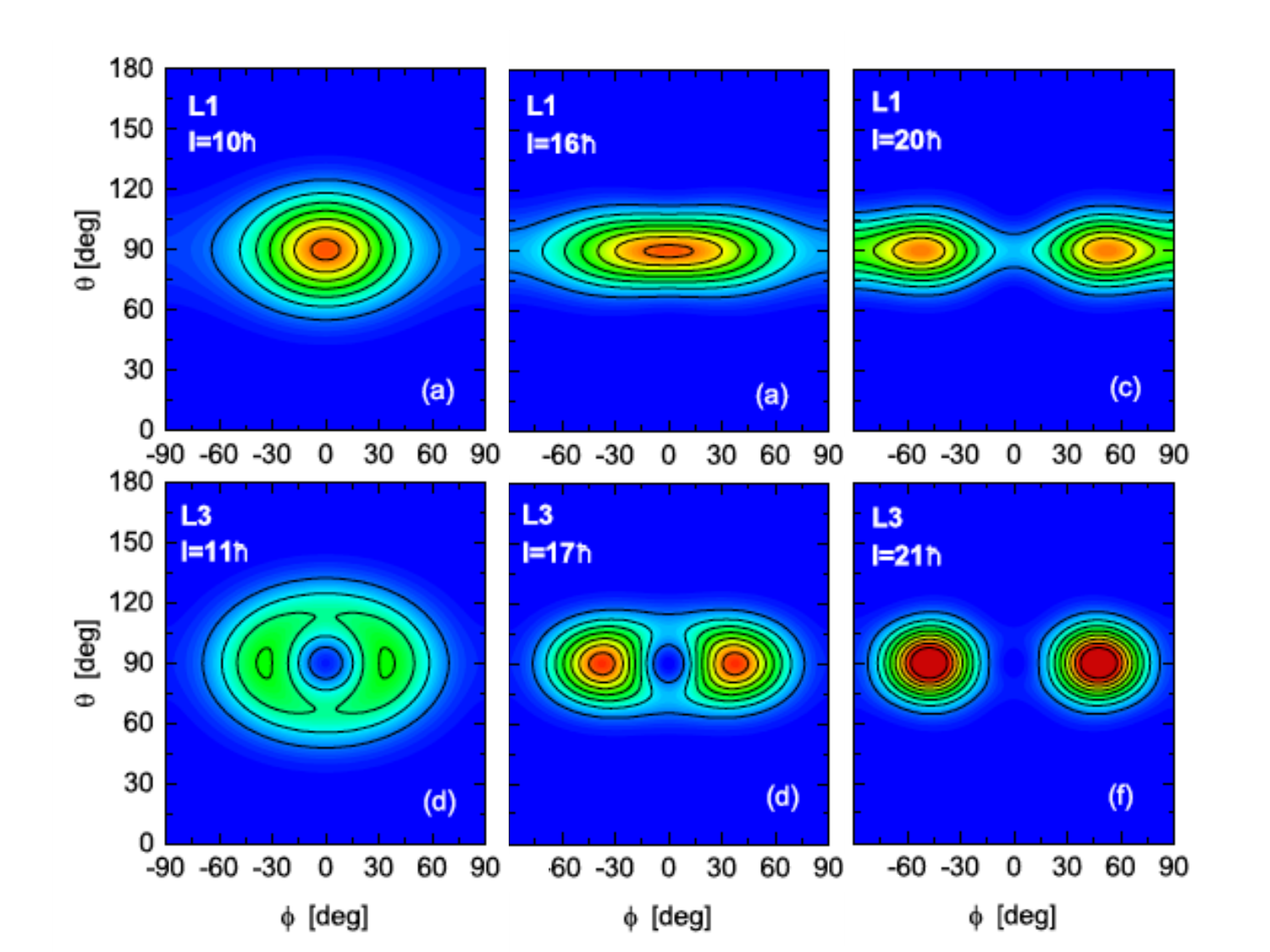}}
 \caption{\label{f:TPSM_136Nd_SCS}  SCS probability distributions for $^{136}$Nd. Reproduced with permission and selected from Fig. 3 of Ref. \cite{Fang-Petrache2021}.
}
\end{figure}

 Chen and Petrache \cite{Fang-Petrache2021}   carried out TPSM calculation for the two-quasiproton configuration $(h_{11/2})^2$ in $^{136}$Nd. 
 Fig. \ref{f:TPSM_136Nd_EW} compares the TPSM energies with the ones of bands L1 and L3, which are interpreted as the $n$=0 and $n$=1 TW bands.
  Fig. \ref{f:TPSM_136Nd_SCS}
 displays a selection of the SCS probability distributions  generated from their TPSM results. For $I=10$ one 
 sees the $n$=0 characteristics: uniform rotation around the $s$-axis and the ground state fluctuations around it.
 For $I=11$ the precession cone  enclosing the $s$-axis is seen as the elliptical rim around it, which characterizes the $n$=1 TW state. 
   The  $I$=16 state is topologically  classified as $n=0$ with
 large fluctuations in direction of the $m$-axis. The  $I$=17 state is topologically  classified  as a strongly perturbed $n=1$ TW excitation.
 It encloses the $s$-axis, where 
  the probability around $\phi=0$ is strongly reduced. The approach of  the FM is apparent in form of the two 
  maxima at $\phi=\pm 50^\circ$.  The $I$=20 state has the characteristics 
 of the $n=$0 FM. It flips between $-120^\circ,~-60^\circ,~60^\circ,~120^\circ$ with no phase change, which  indicated by the finite probability densities  between the blobs. 
 The relatively large probabilities at $\phi=\pm 90^\circ$ herald the transition to the LW regime. The $I=21$ state has the characteristics of the $n=1$ FM
  with phase change between the flips. The approach of the LW mode is not apparent, because the symmetry requires $P(\theta=90^\circ,~\phi=\pm 90^\circ)=0$.

\section{Soft core} \label{s:softcore}
In even-even nuclei triaxiality is indicated by the low energy ratio $E(2^+_2)/ $E($4^+_!)$ ratio and the splitting between the even- and odd- spin members of the quasi $\gamma$ band.
A staggering parameter 
\begin{eqnarray}\label{eq:staggering}
S(I)= \frac{[E(I)-E(I-1)]-[E(I-1)-E(I-2)]}{E(2^{+}_1)}~~~~.
\end{eqnarray}
is defined, which measures the relative distance  of the even- and odd-spin branches. When $S(I)$ is negative for even $I$ the even-$I$ sequence is lower than the odd-$I$ one.
Then the nucleus is  classified  as "$\gamma$-soft", i. e. the triaxiality appears as a large-amplitude oscillation around around a near-axial shape. When $S(I)$ is negative
 for odd $I$ the odd-$I$ sequence is lower than the even-$I$ one. Then the nucleus is  classified as "$\gamma$-rigid, i. e. the shape does not fluctuate much around some finite value of $\gamma_)$. The aspects of $\gamma$-softness are reviewed Ref. \cite{Frauendorf15} and discussed in detail in Ref. \cite{Rouoof2024}.
 
 Most even-even nuclei with a low ratio of  $E(2^+_2)/ $E($4^+_!)$  are $\gamma$-soft. The few $\gamma$-rigid cases include the  examples for 
wobbling discussed in section 
"Experimental evidence for triaxial rotor states in even-eve- nuclei". In these cases $S(I)$ is equal to  two times the difference between the single and double wobbling bands 
divided by the energy of the 2$^+_1$ state.
The PTR model describes wobbling by coupling  a rigid TR with one or two excited quasiparticles. This is in contrast with the observation that the "experimental cores",
i. e. the $\gamma$ bands of the even-even neighbors (or the same nucleus), are $\gamma$-soft.  Already Meyer-ter-Vehn was surprised how well his model applied to
soft nuclei \cite{MtV1975}.

 \begin{figure}[t]
\center{\includegraphics[angle=0,width=0.6\linewidth,trim=0 0 0 0 ,clip]{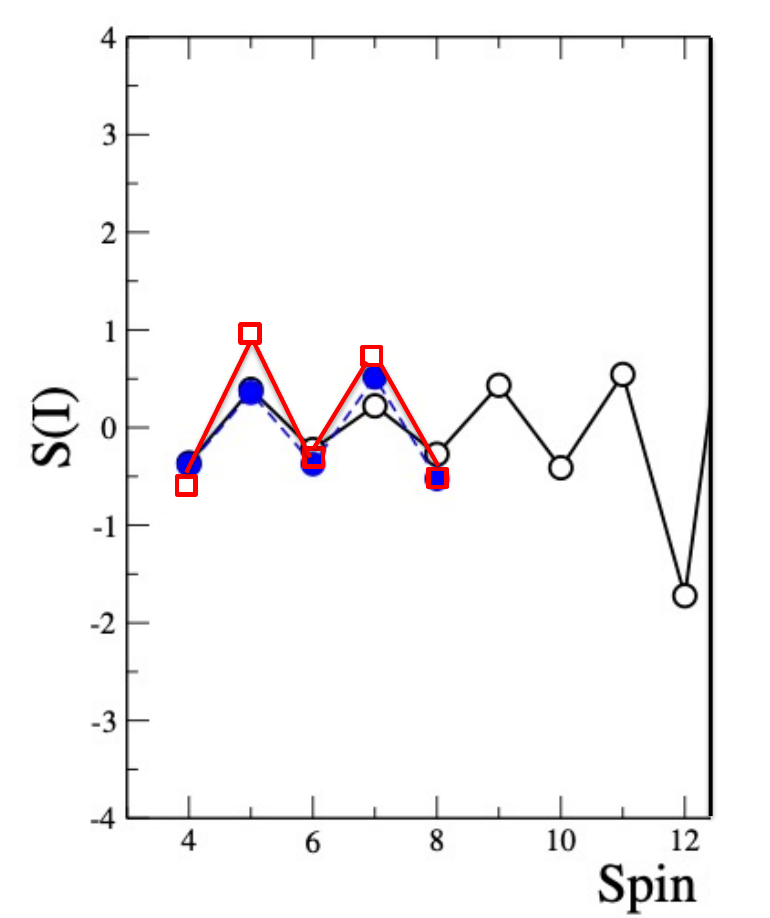}}
 \caption{\label{f:135Pr-stag}  Staggering parameter for $^{134}$Ce. 
Blue circles: experiment, red squares: Bohr Hamiltonian, black circles: TPSM - Private com. by G. Bhat acknowledged. 
 }
 \end{figure}
 \begin{figure}
\center{\includegraphics[angle=-90,width=\linewidth,trim=0 0 0 0 0 ,clip]{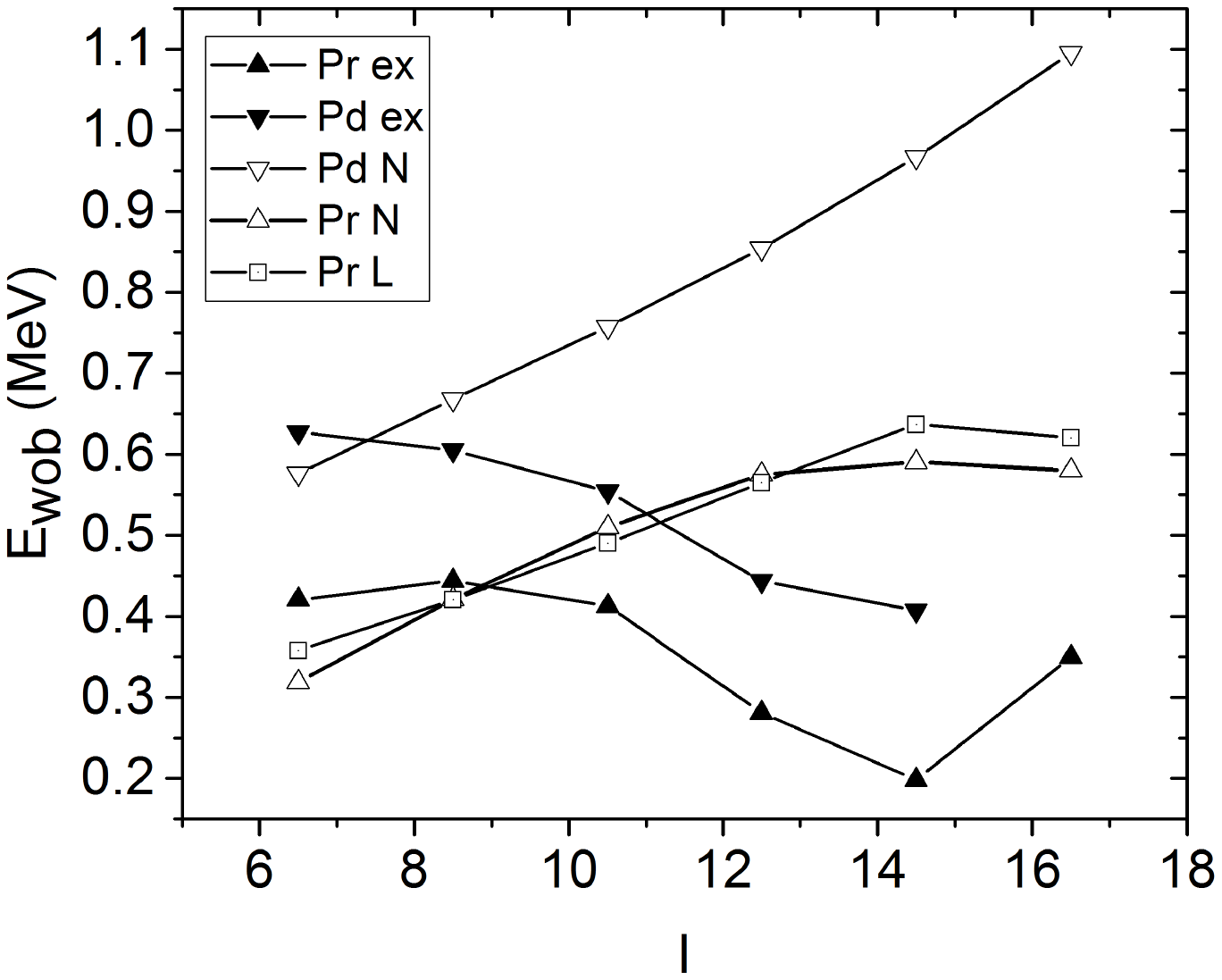}}
 \caption{\label{f:Nomura}  Wobbling energies from $\gamma$-soft cores for $^{135}$Pr (Pr) and $^{105}$Pd (pd)
 calculated by the model of Ref. \cite{Li-thesis} (L) and the model of Ref. \cite{nomura-petrache} (N) compared with the experiment (ex).
}
 \end{figure}
 \begin{figure*}[t]
 \center{\includegraphics[angle=0,width=\linewidth,trim=0 150 0 50 0 ,clip]{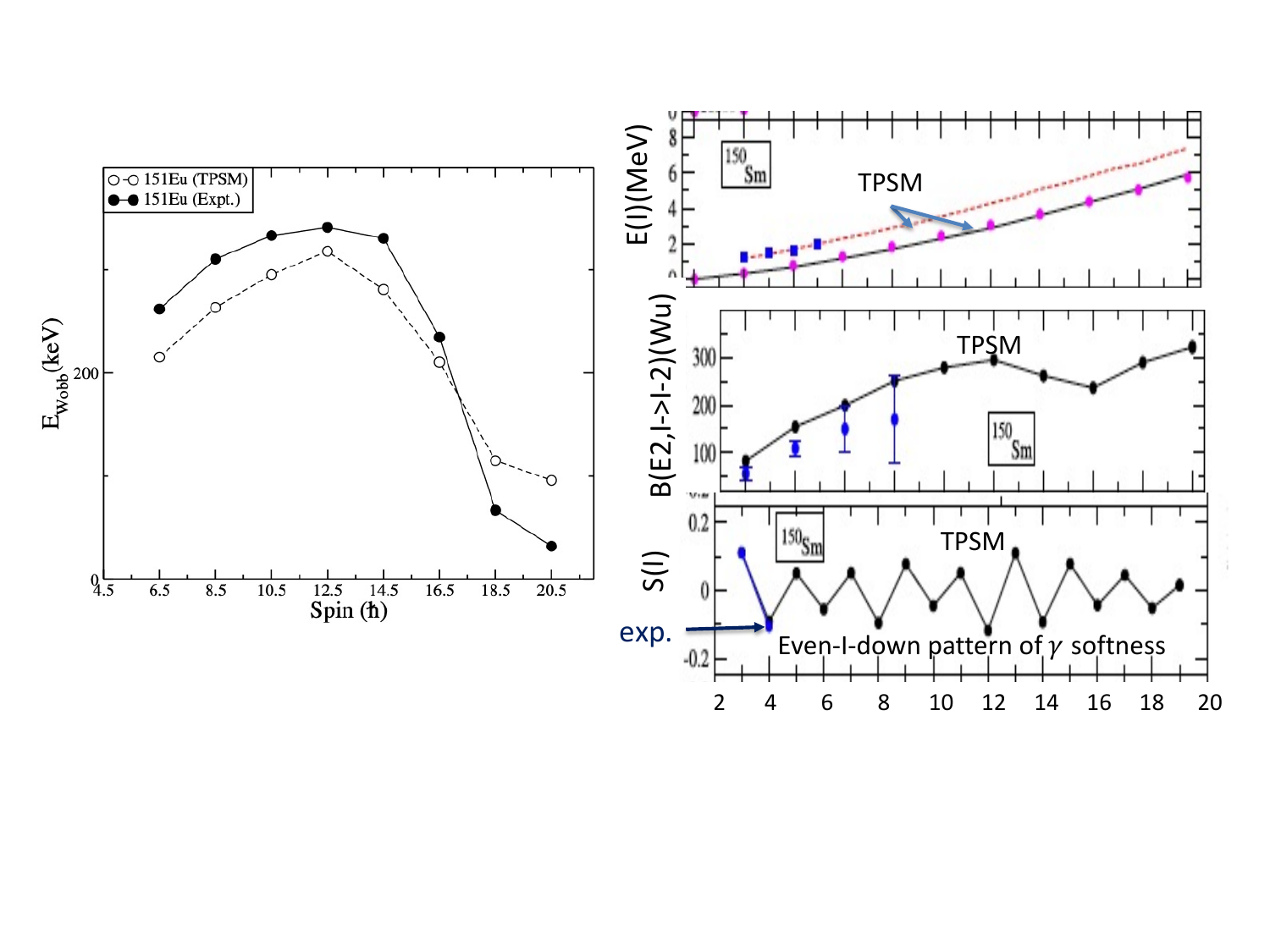}} 
 \caption{\label{f:151Eu} Left panel: Wobbling energies for $^{151}$Eu 
 calculated by TPSM  compared with the experiment \cite{151Eu}. Right panel:  energies, $B(E2)_{in}$ values and staggering parameter of 
 $^{150}$Sm calculated by the TPSM compared with the experiment (blue dots and squares) 
 \cite{Tab2018}. Reproduced with permission from Refs. \cite{151Eu,Tab2018}
}
 \end{figure*}

 For this reason Li and advisors \cite{Li-thesis,Li2022} used the core-quasiparticle-coupling model \cite{CQPCM} for studying the coupling of a high-j
quasiparticle to the soft core of Ref. \cite{MC11}, which incorporates the fluctuations of the $\gamma$ degree of freedom.
The core is a simplified version of the Bohr Hamiltonian, the parameters of which were adjusted to the energies of the lowest collective states in $^{134}$Ce.  
Fig. \ref{f:135Pr-stag} shows that the fitted Bohr Hamiltonian well approximates the  $\gamma$-soft even-$I$-down pattern seen in $^{134}$Ce. The resulting wobbling
energies of the coupled system $^{135}$Pr  in Fig. \ref{f:Nomura} have a  clear LW character, which is in contrast to the experiment. 
The systematic study of the Pd and Rh isotopes  \cite{Li2022}
showed that $\gamma$-soft cores enhanced the increase of the wobbling energy with $I$ as compared to a rigid $\gamma=30^\circ$  core. The $B(E2)$ values did not differ much
while the $B(M1)$ values were larger for the soft cores. 
 
Nomura and Petrache \cite{nomura-petrache} used the IBAF approach, which couples the odd particle to a soft core  described by the Interacting Boson Model, the parameters of which were
obtained by mapping a mean field deformation energy surface onto the Boson space. Fig. \ref{f:Nomura} shows the wobbling energies calculated from the energies published in 
Ref. \cite{nomura-petrache} (Private communication of the accurate values by K. Nomura is acknowledged.), which both for $^{135}$Pr (Pr) and $^{105}$Pd have a clear
 LW character. The contrast with the observed TW behavior raises doubts that the calculations substantiate 
 the claimed "Questioning the wobbling interpretation of low-spin bands in $\gamma$-soft nuclei".

  The studies \cite{Li-thesis,Li2022} and \cite{nomura-petrache} could not obtain the observed TW characteristics by coupling the high-j particle to a phenomenological soft core,
  which appears surprising at the first sight. The microscopic TPSM resolves the problem. In the series of publications \cite{SJ21,Na23,Ruoof2024,Rouoof} the collaboration
  demonstrated that  the energies and $E2$ and $M1$ transition matrix elements of $\gamma$-soft even-even nuclei are very well reproduced by the TPSM, provided 
  the basis contains the quasiparticle configurations explicitly stated in Eq. (\ref{eq:basis}, left). The softness features appear as a consequence of the mixture of basis states with different 
  mean values of $\gamma$. Treating the wobbling nuclides in the same TPSM basis (\ref{eq:basis}, right) incorporates the basis states that are responsible for the fluctuations of $\gamma$.
  Thereby it is important that the TPSM obeys the Pauli exclusion principle between the core and valence particles, which is not ensured in the case of the core-particle coupling models.
  
Fig. \ref{f:135Pr-stag} also shows the staggering parameter $S(I)$ for $^{134}$Ce, calculated by means of the TPSM with the same parameters as used in the TPSM calculation for the neighbor 
$^{135}$Pr. It is close to the experiment and shows the even-$I$-down characteristics of a $\gamma$-soft nucleus. As already discussed, the TPSM  well reproduces the TW 
features of $^{135}$Pr (see Figs. \ref{f:135PrEwob}, \ref{f:135PrBE}) and \ref{f:135PrEwobb}).

The authors of  Ref.  \cite{151Eu} identified $^{151}$Eu as another example of  TW. The left panel of Fig. \ref{f:151Eu} displays the wobbling  energies with the
 characteristic slight increase at low and the decrease at high spin. They are well reproduced by the TPSM calculations with the parameters $\varepsilon=0.20$ and
 $\gamma=27^\circ$. The right panel compares the results of a TPSM calculation with the parameters $\beta=0.20$ and $\gamma=22^\circ$ for the 
 even-even neighbor $^{150}$Sm \cite{Tab2018} with the experiment. The TPSM very well accounts for the  energies and $B(E2)_{in}$ values, where the staggering parameter
$S(I)$ signifies $\gamma$-softness. 
 
 The two examples demonstrate that adding a high-j orbital drastically changes the behavior of the core, which may be modeled by a rigid TR although the 
 neighbors look $\gamma$ soft. Treating the core as an system of its own and taking only into account its polarization  by the particle in the frame work of 
 the core-particle-coupling or IBFA approaches does not work. The success of the TPSM points to the importance of the exchange terms between the valence and core nucleons.
  The ratios $\frac{B(E2,I\rightarrow I-1)_{out}}{BE2,I\rightarrow I-2)_{in}}\sim0.2$ indicate that the breaking of the axial symmetry is weaker 
 than the breaking of the rotational symmetry. This suggest that the Pauli exclusion principle is more import for wobbling than for ordinary rotation. To investigate its
 role in more detail  seems to be interesting.

\section{Conclusions}

The appearance of the wobbling mode in strongly  and normal deformed nuclei is well established by 
the observation of collectively enhanced $E2$  between the bands carrying $n+1$ and $n$ wobbling quanta. 
For even-even nuclei the softness of the triaxial shape complicates the interpretation. Although the TPSM quite well 
accounts for the experimental energies and  relevant transition rates, the relation to the simple wobbling mode of the TR remains to be clarified.
The evidence for wobbling is clear in the presence of one or two high-j quasiparticles. 
There are a number of nuclei that show TW. There are less examples for LW. The expected appearance of LW for mid-shell high-j quasiparticles has not yet been demonstrated.

The PTR model has become a major tool for interpreting the data. The most successful strategy is to take the deformation parameters of the triaxial potential 
from mean field calculations and adjust the rotor MoI's to the observed band energies under the restriction that ${\cal J}_m>{\cal J}_s>{\cal J}_l$. The SCS  maps of the PTR states reveal 
 the closest proxies to the corresponding classical orbits. Their topology classifies the wobbling mode: TW orbits enclose the axes perpendicular to the axis  with the largest MoI, 
 which is enclosed by the LW orbits. The TW becomes unstable at a critical angular moment and changes into LW  via the FM, where the angular momentum vector flips between
 four equivalent orientations in the $s$-$m$-plane. The SSS plots provide the closest proxies to the probability densities of wave functions that belong to
 a one-dimensional potential. Which axes the potential encloses classifies the wobbling as TW-LW-FM as well. 
 
 The microscopic RPA and TPSM approaches have only the triaxial deformations as input, which removes ambiguities of the PTR model  due to the choice of the MoI's.
 Being a small-amplitude approximation, the RPA works only for angular momenta far from the instability of the TW. Its successful application to the strongly deformed nuclei
 revealed the TW dynamics. The RPA study of normally deformed nuclei indicated that the coupling to the scissors mode seems to be important for for the 
 for reproducing the strong reduction of the $M1$ transition between the wobbling states.
 
 The TPSM  very successfully accounts for the energies of the wobbling  and SP excitations
as well as for the  transition matrix elements between them. It resolves the apparent contradiction between the appearance of wobbling in the present  of high-j quasiparticles and the 
features of $\gamma$ softness in the even-even nuclei in their absence. The tools for extracting the physics from the results of the truncated shell model diagonalization need to be further developed.

\end{document}